\documentclass[oneside,a4paper,12pt]{book}
\usepackage{graphics, graphicx}
\usepackage{natbib}
\usepackage{amsfonts}
\usepackage{hyperref}
\usepackage{lscape}
\begin{document}
\pagenumbering{arabic} \pagestyle{plain}

\frontmatter
\thispagestyle{empty}
\begin{center}
\LARGE{
Charles University in Prague\\
Faculty of Mathematics and Physics}\\
\vspace{1cm}
{\bf{\Huge{DOCTORAL THESIS}}}\\
\vspace{7mm}
\scalebox{0.5}{\includegraphics*[1mm,55mm][140mm,210mm]{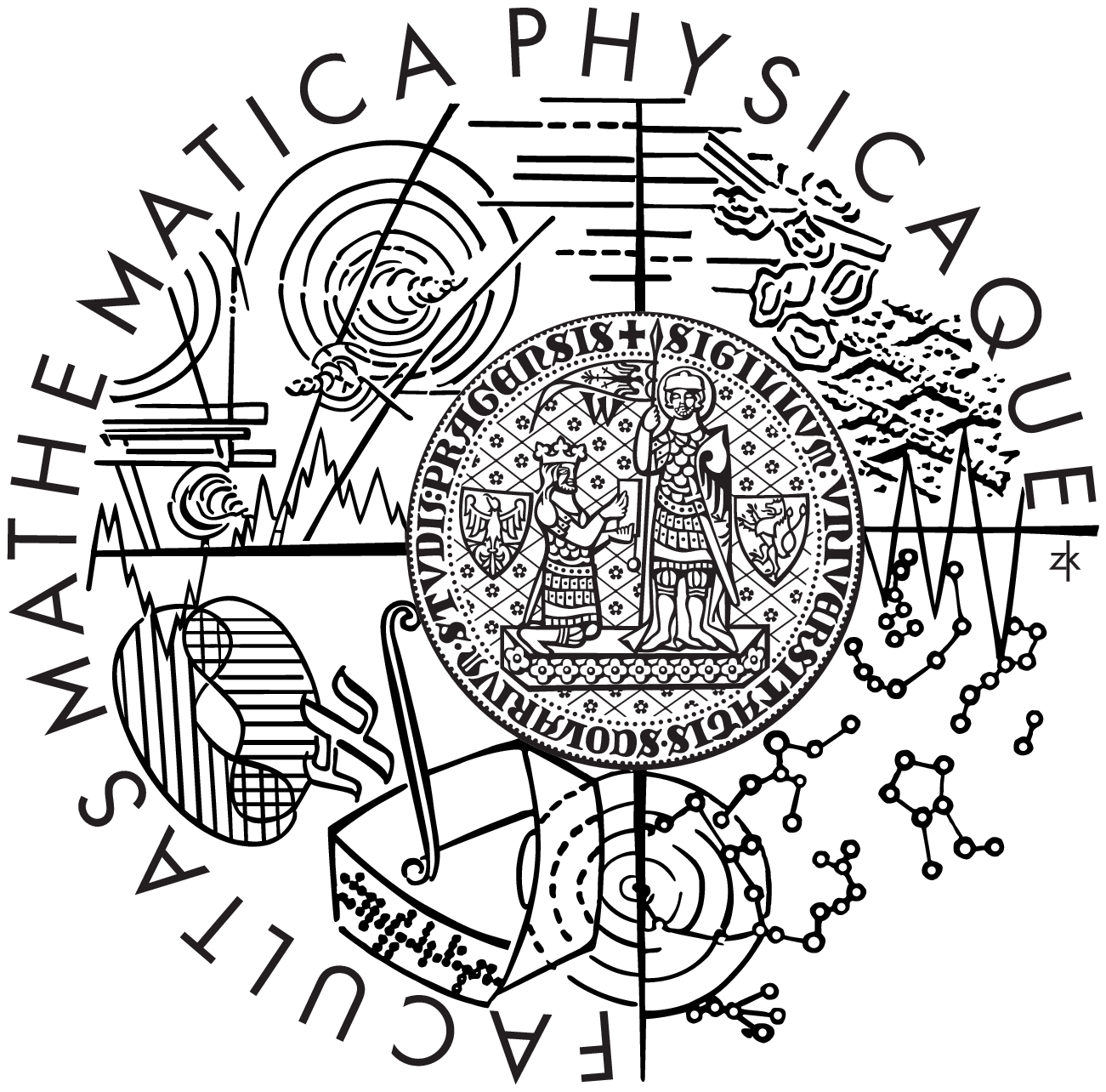}}\\%
\vspace{1mm} {\LARGE Mgr. Petr Zasche\\} \vspace{0.5cm} {\huge \bf
Multiple stellar systems \\ under photometric and astrometric
analysis\\} \vspace{9mm}
The Astronomical Institute of Charles University \\ Prague \\
Supervisor: Doc. RNDr. Marek Wolf, CSc.\\
Study program: F1 -- Theoretical physics, astronomy and astrophysics \\
\end{center}

{{\newpage\thispagestyle{empty} \noindent \vspace{16cm}

\noindent \textbf{\underline{Acknowledgements:}} \vspace{2mm}

\noindent Here at this place I would like to thank sincerely all the people, who helped me with this work and
during my PhD study. Most of all to my supervisor, Doc. RNDr. Marek Wolf, CSc., for his care and leadership
always full of enthusiasm and optimism. My thanks also go to Prof. RNDr. Petr Harmanec, DrSc., RNDr. Pavel
Mayer, DrSc., and Dr. William Hartkopf for their valuable critical notes, remarks, advices and other support. I
should not forget to thank all the members and students of The Astronomical Institute MFF~UK for their attitude
and the unique friendly atmosphere they created and last but not least my friends and family for their help and
support.

\noindent Support from the grants GA\v{C}R 205/04/2063 and
205/06/0217 is also acknowledged.
%
%

\newcommand{\arcs}{$^{\prime\prime}$}
\newcommand{\Mo}{$\mathrm{M_\odot}$}

\newcommand\aj{{AJ}}%
\newcommand\araa{{ARA\&A}}%
\newcommand\apj{{ApJ}}%
\newcommand\apjl{{ApJ}}%
\newcommand\apjs{{ApJS}}%
\newcommand\ao{{Appl.~Opt.}}%
\newcommand\apss{{Ap\&SS}}%
\newcommand\aap{{A\&A}}%
\newcommand\aapr{{A\&A~Rev.}}%
\newcommand\aaps{{A\&AS}}%
\newcommand\azh{{AZh}}%
\newcommand\baas{{BAAS}}%
\newcommand\jrasc{{JRASC}}%
\newcommand\memras{{MmRAS}}%
\newcommand\mnras{{MNRAS}}%
\newcommand\pra{{Phys.~Rev.~A}}%
\newcommand\prb{{Phys.~Rev.~B}}%
\newcommand\prc{{Phys.~Rev.~C}}%
\newcommand\prd{{Phys.~Rev.~D}}%
\newcommand\pre{{Phys.~Rev.~E}}%
\newcommand\prl{{Phys.~Rev.~Lett.}}%
\newcommand\pasp{{PASP}}%
\newcommand\pasj{{PASJ}}%
\newcommand\qjras{{QJRAS}}%
\newcommand\skytel{{S\&T}}%
\newcommand\solphys{{Sol.~Phys.}}%
\newcommand\sovast{{Soviet~Ast.}}%
\newcommand\ssr{{Space~Sci.~Rev.}}%
\newcommand\zap{{ZAp}}%
\newcommand\nat{{Nature}}%
\newcommand\iaucirc{{IAU~Circ.}}%
\newcommand\aplett{{Astrophys.~Lett.}}%
\newcommand\apspr{{Astrophys.~Space~Phys.~Res.}}%
\newcommand\bain{{Bull.~Astron.~Inst.~Netherlands}}%
\newcommand\fcp{{Fund.~Cosmic~Phys.}}%
\newcommand\gca{{Geochim.~Cosmochim.~Acta}}%
\newcommand\grl{{Geophys.~Res.~Lett.}}%
\newcommand\jcp{{J.~Chem.~Phys.}}%
\newcommand\jgr{{J.~Geophys.~Res.}}%
\newcommand\jqsrt{{J.~Quant.~Spec.~Radiat.~Transf.}}%
\newcommand\memsai{{Mem.~Soc.~Astron.~Italiana}}%
\newcommand\nphysa{{Nucl.~Phys.~A}}%
\newcommand\physrep{{Phys.~Rep.}}%
\newcommand\physscr{{Phys.~Scr}}%
\newcommand\planss{{Planet.~Space~Sci.}}%
\newcommand\procspie{{Proc.~SPIE}}%

\newpage
\tableofcontents

\mainmatter  \pagenumbering{arabic}

\def\ps@headings{%
       \def\@oddfoot{}%
       \def\@evenfoot{}%
       \let\@mkboth\markboth
       \def\@evenhead{\parbox{\textwidth}{{ \thepage}
                \hfill{\markfont \leftmark}\smallskip}}%
       \def\@oddhead{\parbox{\textwidth}{{\sl
       \rightmark}\hfill{ \thepage}\smallskip\hrule}}}
\pagestyle{headings}

\chapter{Introduction}

The binary stars are crucial for our knowledge about the universe. Especially eclipsing binaries provide us an
unique insight to the basic physical parameters of the stars, stellar clusters, interstellar medium and
galaxies. They are excellent distance indicators. We are able to learn more about the matter composition of the
stars, about their evolution status, or the presence of planets or other components in these systems.

The very first task is the data acquisition, because only with precise input data is one able to get precise
results. In last few decades mainly due to excellent satellite observatories (and not only in the visible part
of the spectrum) our knowledge of them has rapidly grown.

Regarding the astrometry, there is still decreasing the number of observations of the wide pairs. On the other
hand, due to the new interferometers, which could resolve the milli- and micro- arcsecond angular distances, the
observable semimajor axes of the astrometric binaries are still decreasing. Unfortunately, most of the systems
analyzed below have the angular size of the astrometric orbit from 1 arcsec down to 100 mas, which is beyond the
limits for the modern multi-aperture interferometers. And the lack of recent observations lead to the low
accuracy of the results.

Another approach is photometry and the classical observation of minimum light. Due to the large "baseline" of
observers in our country and the interest of amateur astronomers, the number of these observations is growing
very rapidly and the cooperation between professional and amateur astronomers is very intensive. Many of the
observations of minimum timings used in this study came from amateur astronomers and these measurements are as
accurate as from the professional observatories. Thanks to the large minimum times data set we are able to
analyze many of the eclipsing binary systems for their long-term period variations.

The whole thesis is divided into several parts. In the first one is presented the theory needed for the analysis
of multiple stellar systems by photometric and astrometric techniques, description of such systems and some
limitations which have to be considered. In the second part are introduced several systems which show apparent
period changes in their $O-C$ diagrams. And in the third one are the systems analyzed by photometry and
astrometry simultaneously. Also the catalogue of other suggested systems for simultaneous analysis is included
in this chapter. This is the crucial part of this thesis. The method itself is introduced in the chapter
\ref{Ch1} and the results are in the chapter \ref{Ch3}.

\chapter{Theory} \label{Ch1}
\section{Binaries}

Most stars are found to be members of binary or multiple stellar systems. A recent analysis of a large set of
close binaries \citep{PribRuc2006} indicates that even most of the binaries are in multiple systems. According
to this study about 59 \% of northern-hemisphere contact binaries are members of multiple stellar systems. The
sample of binaries was analyzed very precisely, some distance-independent techniques were used, and the
selection effects were also discussed.

During the last decades, many of the observational techniques have become so effective and precise that it is
possible to discover low-massive stars, brown dwarfs, or even exoplanets in eclipsing binary systems (hereafter
EB).

The astrometric binaries are a special subset of binaries, where the individual components could be resolved
into separate stars, which means the angular separation of the components has to be above a certain limit (a
function of the telescope aperture and technique used). Because the individual components in the system were
discovered at different time epochs, they were marked by different labels. Most common is the use of A-B-C\dots
sequence, which means that the component "A" was discovered as the first one, then was discovered the second one
"B", and after then "C", etc. Sometimes one component was resolved to be a double, so it turns A-B $\rightarrow$
Aab-B. Mostly A component is the brightest one. Here comes the problem with the hierarchy of such a system.
Sometimes it is so-called \emph{hierarchical-type} Aabc-B system (where the B component is far away from the
triple Aabc, where the c component could be far away from the ab double), and sometimes is is so-called
\emph{trapezium-type} Aab-Bab system (two doubles Aab and Bab far away from each other), for the detailed
description see e.g. \citet{Docobo2006}.

In the present analysis another approach was used. It is based on the physical properties and gravitational
bounding of the stars. The numbers 1 and 2 were used for the primary and secondary component of the eclipsing
pair (these are not spatially resolvable) and index 3 as a label for the third, distant, component which is
astrometrically observable. Sometimes an additional component, the fourth, is suggested. In this study only the
hierarchical systems are analyzed. It means that the higher the number of the component, the bigger the
semimajor axis (the distance of the component is larger than the previous one) $a_1 \ll a_2 \ll a_3 \ll \dots$.
Because here we deal only with the relative astrometry (position angles and angular separation) it should be
more precise to write $a_{12} \ll a_{12-3} \ll \dots$.

\section{Principal methods for analyzing the EBs}

Despite the fact that the methods introduced below are the most important and useful ones, only a short
description was presented here. These methods are not the essential ones for this thesis. Two other methods (the
astrometry and the analysis of times-of-minima observations) are more crucial and are described in detail in the
next section.

\subsection{Spectroscopy}

Observing the spectra of stars is perhaps the most time-consuming activity performed at astronomical
observatories all around the world. However, obtaining the spectrum of the star is also the most powerful tool
for deriving the relevant parameters of the star. Thanks to different techniques of spectral analysis
(spectrophotometry, line-profile analysis, disentangling, etc.) one can model the stellar atmosphere, derive the
orbital parameters, or discover the surface structures. For a brief introduction to the topic and overview of
the methods see e.g. \citet{HilditchBook2001}.

Besides spectral classification, the modelling of radial velocity curves (hereafter RV) is one of the oldest
methods of spectral analysis. If only one component of the binary is observable, we deal with a so-called
SB1-type, and if both components are evident, it is called SB2-type binary. From both RV curves, we could derive
many parameters of the relative orbits of the stars in the system (see Table \ref{Entities} for the parameters
and e.g. \citet{WilsonRV1976} for the method). Cross-correlation methods are now often used for analyzing RVs,
see e.g. \citet{McLean1981}.

Spectral disentangling (see e.g. \citealt{HadravaDisentangling1995}) of composite spectra into the separate ones
helps us to model the particular star in the system. Theoretical stellar models could be compared to the
observed ones and one could study the physical conditions in the atmosphere of the star, temperatures,
pressures, rotation, stellar wind or the chemical composition.

Another quite new technique is Doppler profile mapping (see e.g. \citealt{RiceDopplerImaging1989}). With this
technique one is able to discover the dark and bright (cool and hot) spots on the surface of the star, as well
as its rotation, or the evolution of the spots. Doppler tomography (see e.g. \citealt{MarshDopplTomogr1988}) is
able to detect similar structures and effects in accretion discs in interacting binaries. Evidence for discs,
jets, or outflows could be observed in the precisely measured spectrum of the star.

\begin{landscape}
\begin{table}
\caption{The scheme of directly derivable entities.}
\label{Entities} \centering \scalebox{0.85}{
\begin{tabular}{c|c|c|c|c|c|c|c|c|c|c}
 \hline \hline   
                                                                                                                 &            &    Only    &    Only    &     LC     &     LC     &   Only     &   Only     &            &    LITE    &  LITE      \\
                                                                                                                 &    Only    &     RV     &     RV     &     RV     &     RV     &   $O-C$    &   $O-C$    &  Astrom.   &     +      &  Apsid.    \\
                                                                                                                 &     LC     &     SB1    &     SB2    &     SB1    &    SB2     &   LITE     &   Apsid.   &            &   Astrom.  &  Astrom.   \\ \hline
 $a_{1} \sin{i}\,$ or $\,a_{2} \sin{i}$                                                                          &            & \checkmark & \checkmark & \checkmark & \checkmark &            &            &            &            &            \\ \hline
 $a \sin{i}$, $a_1 \sin{i}$, $a_2 \sin{i}$, $M_1 \sin^3{i}$, $M_2 \sin^3{i}$                                     &            &            & \checkmark &            & \checkmark &            &            &            &            &            \\ \hline
 $a$, $a_1$, $a_2$, $M_1$, $M_2$, $R_1$, $R_2$, $\mathcal{L}_1$, $\mathcal{L}_2$, $d$                            &            &            &            &(\checkmark)& \checkmark &            &            &            &            &            \\ \hline
 $P$, $e$, $\omega$, ($\dot \omega$)                                                                             & \checkmark & \checkmark & \checkmark & \checkmark & \checkmark &            & \checkmark &            &            & \checkmark \\ \hline
 $\gamma$                                                                                                        &            & \checkmark & \checkmark & \checkmark & \checkmark &            &            &            &            &            \\ \hline
 $q$                                                                                                             &(\checkmark)&            & \checkmark &(\checkmark)& \checkmark &            &            &            &            &            \\ \hline
 $i$, $\frac{R_1}{a}$, $\frac{R_2}{a}$, $L_1/L_2$, $g_1$, $g_2$, $A_1$, $A_2$, $F_1$, $F_2$, $x_1$, $x_2$, $l_3$ & \checkmark &            &            & \checkmark & \checkmark &            &            &            &            &            \\ \hline
 $T_2$                                                                                                           & \checkmark &     ?      &     ?      & \checkmark & \checkmark &            &            &            &            &            \\ \hline
 $JD_0$, $P$                                                                                                     & \checkmark & \checkmark & \checkmark & \checkmark & \checkmark & \checkmark & \checkmark &            & \checkmark & \checkmark \\ \hline
 $p_3$, $T_0$, $A_3$, $\omega_3$, $e_3$, $f(M_3)$, $M_{3,min}$                                                   &            &            &            &            &            & \checkmark &            &            & \checkmark & \checkmark \\ \hline
 $p_3$, $T_0$, $\omega_3$, $e_3$, $i_3$, $\Omega_3$, $a_3$\arcs                                                  &            &            &            &            &            &            &            & \checkmark & \checkmark & \checkmark \\ \hline
 $M_3$, $D$                                                                                                      &            &            &            &            &            &            &            &            & \checkmark & \checkmark \\
 \hline \hline
\end{tabular}}
\end{table}
\begin{picture}(1,1)(-12,13.5)
\put(-1,96.2){\scalebox{3}{\line(1,0){125}}}     
\put(374,96.1){\scalebox{3}{\line(0,1){46.2}}} 
\end{picture}
\vspace{-1cm}
\begin{list}{}{} \item[]
\small{Some comments: The table shows which entities could be derived when one has measurements of a particular
type. The table is divided into two parts. In the first one (the left top corner) are the parameters and methods
adopted from \cite{KallrathMilone}, page 142. In the second part are the parameters of the third-body orbit
which could be derived from the methods used in this thesis. 'LC' stands for the light curve, 'RV' for the
radial velocity curve, 'SB1' and 'SB2' for the types of the spectroscopic binaries, 'LITE' for the light-time
effect, 'Apsid.' for the apsidal motion and 'Astrom.' for the astrometric orbit, respectively. In the parameters
column $a_1$ and $a_2$ denote the semimajor axis of the primary and secondary component in the EB, $i$ for the
inclination of the EB orbit, $M_1$ and $M_2$ masses of the primary and secondary, $R_1$ and $R_2$ radii of
components, $\mathcal{L}_1$ and $\mathcal{L}_2$ for the bolometric luminosities, $d$ for the separation between
the components, $e$ for the eccentricity of the EB orbit, $\omega$ for the argument of periastron of the orbit,
$\gamma$ is the systemic velocity of the EB pair, $q$ is the photometric or the spectroscopic mass ratio, $L_1$
and $L_2$ monochromatic luminosities in a specified passband, $g_1$ and $g_2$ gravity darkening coefficients,
$A_1$ and $A_2$ albedo coefficients, $F_1$ and $F_2$ rotation parameters, $F_i = \frac{\omega_i}{\omega}$, $x_1$
and $x_2$ limb-darkening coefficients, $l_3$ the third light and $T_2$ temperature of the secondary ($T_1$
fixed), respectively. Only one parameter was added compared with \citeauthor{KallrathMilone}, the apsidal motion
rate $\dot \omega = \frac{\mathrm{d}\omega}{\mathrm{d}t}$, and parentheses mean that the parameter is computable
only if some apsidal motion is presented. Sometimes it is difficult to compute the parameters, or the ability to
compute them depends on the type of the EB. For example the photometric mass ratio could be computed only when
the EB is contact or over-contact type (that is the reason why there are the parenthesis). The question mark
indicates that the temperatures could be derived from the spectrum of the star. The parameters from the second
part are explained in the text, $D$ stands for the distance to the system.}
\end{list}
\end{landscape}

Sometimes also an additional component is observable in the spectra. The spectral lines of the third component
typically remain at a fixed wavelength, while the lines of the binary components move in agreement with the
actual orbital phase of the EB. In the more favourable case the third lines also move very slowly according to
the third-body orbit. In some others only the barycenter of the whole system is moving very slowly in agreement
with the phase of the third-body orbit (see Eq. \ref{RV3} in chapter \ref{LITE}). Such effects are hardly
observable and their final detection is questionable in most of the cases (see \cite{Mayer2004} for the list of
such systems and the subsections \ref{VWCep} and \ref{V505Sgr} below).

\subsection{Photometry}

Modern photoelectric and CCD photometry is a very powerful tool for studying the eclipsing binaries. Thanks to a
very wide network of amateur astronomers with their CCD cameras all around the world, it is nowadays quite easy
to observe the whole light curve of a particular eclipsing binary in the standard filters and perform a detailed
analysis of the system.

From the light curve (hereafter LC) one could derive many useful
parameters of the eclipsing binary itself. The comparison between
the parameters derivable from the LC solution, from the RV fitting
and from the techniques described below is shown in Table
\ref{Entities}.

From this set of parameters, together with the radial velocity
curves, one could obtain a complete set of parameters describing
the orbit of the individual components in the EB, as well as the
basic physical parameters of both components in absolute units.
The masses, the radii, the luminosities and the temperatures could
be calculated, if the precise photometry together with both radial
velocities were carried out (see Table~\ref{Entities}). The
limb-darkening coefficients (see e.g. \citealt{vanHammeLD1993}),
as well as gravity brightening (see e.g.
\citealt{LucyGravDark1967}) and albedos (see e.g.
\citealt{RucinskiAlbedos1969}) of the components could be also
calculated from the model. The most common codes for the EB
modelling are the Wilson-Devinney \citep{WD1971}, {\tt FOTEL}
\citep{Fotel2004}, Linnell's model \citep{Linnell1984}, {\tt
LIGHT} \citep{Hill1979}, etc.

Modern advanced tools for analyzing the light curves of EBs are
very powerful for discovering the surface structures and their
evolution. Dark or hot spots on the star surface could be
implemented into the model (see e.g. \citealt{Poe1985}).

Precise photometry of the binary in different filters could also reveal the third light from a distant component
in the LC solution. The third light $l_3$ could be the indicator or the proof for the presence of a third body
suggested by some independent method.

\subsection{Additional techniques}

Besides photometry and spectroscopy other methods also play a role in our modern knowledge about EBs. The
invisible part of the spectrum is also used to study these objects. Some EBs have been successfully identified
to be radio- or X-ray active. Such systems are mostly the complicated ones, currently in an unstable
evolutionary status, undergoing rapid mass transfer, having an accretion disc or be a member of a cataclysmic
variable.

Another approach is the use of polarimetry (see e.g. \citealt{HallPolariz1949}), or magnetometry (see e.g.
\citealt{MarcyMgPole1984}) to study EBs. These two modern techniques are also very powerful and could reveal
some properties of these systems which are otherwise undetectable, for example a disc or stream in the system,
large coronae of one of the components, or magnetic fields.

\section{$O-C$ diagram analysis}

The periodic behavior of eclipsing binaries can also be studied. Long-term changes of the binary period could be
caused by various effects and these effects could be analyzed thanks to the large database of times-of-minima
observations which cover more than a century in many cases. Especially due to amateur astronomers with their CCD
cameras in the recent few years the number of times of minima is growing rapidly, because these observations are
much easier to obtain than the photometric observation of the whole LC. The accuracy of such observations are
sometimes not very good (mainly the old ones), so the analysis is also problematic. The topic was discussed in
detail for example in \cite{SterkenOC2005}.

The $O-C$ diagram in our case is a special plot, where the x-axis is either the epoch (number of cycles relative
to $JD_0$) or the Julian date,, while the y-axis is the difference \textbf{O}bserved minus \textbf{C}alculated.
Here in the case of the eclipsing binaries this means the difference in times of minima in the particular
system, expected minus predicted moment of minimum light.

The linear ephemeris of the binary indicates the minimum time $JD$
after $E$ cycles ($E$ is the epoch number) since the initial time
of minimum $JD_0$ occured. One could write
$$ JD = JD_0 + P \cdot E, $$
where $P$ is the period of the eclipsing binary. The value $O-C$
is therefore defined as a difference
$$ O - C = JD - JD_0 - P \cdot E.$$

Finding new revised linear ephemeris of the binary could be also
done with these equations. One has to minimize the sum
$$ \sum \limits _{i=1}^N {(O - C)_i}^2 \rightarrow 0 $$
with respect to the parameters ($JD_0$, $P$) over the whole
parameter space and $N$ is the number of times of minima used.

\subsection{Constant period}

One would expect a constant period and therefore a linear trend in the $O-C$ diagrams in most of the eclipsing
binaries. If the assumed period of the EB is lower than the right one, the linear trend in the diagram is
increasing, while if the proposed period is higher than the right one, the times of minima are decreasing in the
diagram. A collection of many $O-C$ diagrams of EBs is for example in Kreiner's Atlas of $O-C$ diagrams, see
\cite{KreinerOCAtlas2001}.

Sometimes $O-C$ diagrams show changes in period which seems to be abrupt. Such "jumps" in the diagram could be
caused by sudden period changes. These are often explained as a mass ejection and/or transfer from one star to
another. Sometimes this explanation is used only due to poor coverage of the abrupt change in the diagram,
sometimes is confirmed by another independent method.

\subsection{Mass transfer}

Another phenomena which could be studied in the $O-C$ diagram is the parabolic behavior of times of minima. If
the data set is large enough, in some binaries it is possible to identify a parabolic (increasing or decreasing)
trend, which means that the period of the binary is steady increasing or decreasing. In $O-C$ diagram it means
$$ JD = JD_0 + P \cdot E + q \cdot E^2, $$
where $q$ is the quadratic term coefficient. It is possible to show that this additional term could be caused by
the mass transfer within the binary.

There are two basic kinds of mass transfer between the components. The first (and the simplest) one is
\emph{conservative mass transfer}. In this case the total mass of the binary as well as its total orbital
angular momentum are conserved in the system. The other one is \emph{nonconservative mass transfer}, where these
quantities are not constant for the whole system. A few mechanisms could cause this latter case: stellar wind,
Roche-lobe overflow, or a sudden catastrophic event. This kind of mass transfer is probably more often in the
nature.

Concerning the former case, from the quadratic term coefficient $q$ also the conservative mass transfer rate
could be derived
$$ \dot{M_1} = \frac{M_1 M_2}{3 (M_1 - M_2)} \cdot \frac{\dot{P}}{P} = \frac{2 q M_1 M_2}{3 P^2 (M_1-M_2)}. $$
A brief introduction to the topic and derivation of the equations for both cases could be found in
\cite{HilditchBook2001}, page 162. The typical values of mass transfer rate are circa $10^{-7} -
10^{-9}$~\Mo/yr.

\subsection{Apsidal motion}

The next effect which could be studied only on the basis of the $O-C$ diagram analysis is the apsidal motion. In
some eccentric binaries the line of apsides of the orbit of such a system is moving in space. One has to take
into consideration two different periods, the sidereal one $P_s$ and the anomalistic one $P_a$, which are in
relation $$ P_s = P_a (1- \dot\omega /{2\pi}).$$ The quantity $\dot\omega$ is the apsidal motion rate.

Such an effect is easily detectable in the $O-C$ diagram of the
binary. Both primary and secondary minima are being periodically
shifted from the linear ephemeris and also against the other one
(the primaries and secondaries are in anti-correlation). The
necessary equations for the apsidal $O-C$ diagram analysis are
presented in \cite{GimenezGarciaPelayo}.

\subsection{Light - time effect} \label{LITE}

Large set of times of minima of the EBs could be also used for discovering the additional component(s) in these
systems. With the light-time effect (hereafter LITE) analysis one can suppose a presence of another component(s)
in the system only by analyzing the times of minima and their long-time behaviour. The motion of the EB around
the common barycenter causes apparent changes of the observed binary period with a period corresponding to the
orbital period of the third body.

\cite{Irwin1959} improved a method by \cite{Woltjer1922} for analyzing the long-term variation of the times of
minima caused by a third body orbiting an eclipsing pair. Very useful comments and limitations were discussed by
\citet{FCH73} and by \citet{Mayer1990}. Nowadays there are several dozens of EBs, where the LITE is certainly
presented or supposed (see e.g. \citealt{BorkovitsHegedus}, \citealt{Albayrak1999}, \citealt{Wolf2004}).

From the numerical point of view the method is a classical inverse
problem. We have $M$ measurements of the times of minima of the
system at certain constant $JD_i$ with the individual
uncertainties $\sigma_{m,i}$. Our task is to find five parameters
describing the orbit of the third body in the system: the period
of the third body $p_3$, the LITE semiamplitude $A$, the
eccentricity $e$, the time of the periastron passage $T_0$, and
the longitude of periastron $\omega_{12}$ for the binary on its
orbit around the common barycenter. We have to compute
simultaneously also two (or three) parameters of the eclipsing
binary itself, namely its linear (or quadratic) ephemeris $JD_0$
and period $P$ (and $q$ for the quadratic one). Altogether, one
has 7 (or 8) parameters to derive from the model fit of the
minimum-time measurements
 \begin{equation}
   \{ (JD_i,\sigma_{m,i}) \}_{i=1,M} \rightarrow ( p_3, A, e, T_0, \omega_{12}, JD_0, P, q ).
 \end{equation}
The least-squares method and the simplex algorithm (see e.g. \citealt{Kallrath}) are used. The basic mathematic
equations are the following. Compute the mean anomaly from the time of the measurement (the subscript $i$ was
omitted for the sake of brevity)
  \begin{equation}
    M = 2 \pi \cdot \frac{(JD - T_0)}{p_3}.
    \label{eq2}
  \end{equation}
Then solve the Kepler equation $M \rightarrow E$ and convert the eccentric anomaly to the true anomaly of the
third body in its orbit
  \begin{equation}
    \nu = 2 \cdot \arctan \left( \sqrt{\frac{1+e}{1-e}} \cdot \tan \frac{E}{2} \right).
    \label{eq3}
  \end{equation}
After then, one can use the formula
  \begin{equation}
\Delta \tau = \frac{A} {\sqrt{1-e^2\cos^2\omega_{12}}}\cdot\left[
{{ \frac{1-e^2}{1+e \cos\nu} \sin(\nu + \omega_{12})+ e
\sin\omega_{12}}} \right]
  \end{equation}
to compute the magnitude of the LITE. Now it is possible to
calculate the difference between the observed and calculated time
of minimum
  \begin{equation}
    (O-C) = JD-JD_0 - P \cdot E - q \cdot E^2 - \Delta \tau  ,
  \end{equation}
where $E$ is the epoch of the $JD$ according to the ephemeris
$JD_0$ and $P$, and $q$ is the quadratic term quotient. The
resultant sum of normalized square residuals is
  \begin{equation}
    \chi^2_{LITE} = \sum_{i=1}^M \left( \frac{(O-C)_i}{\sigma_{m,i}} \right) ^2.
    \label{eq11}
  \end{equation}
Our task is to minimize this value and find the set of parameters
$(p_3, A, e, T_0, \omega_{12}, JD_0, P, q)$ describing the orbit.

The weighting is provided by the uncertainties $\sigma_{m,i}$. These values are obtained from the observations,
or estimated as some typical uncertainty level for the certain kind of measurement provided by specific
instrument. Another way is the following method. If we have information about the type of the observation (the
method by which the measurement was obtained), we could use some weighting scheme $w_i$ instead of uncertainties
(e.g. $w_i=1$ for visual and $w_i=10$ for photoelectric/CCD measurements) and solve the corresponding problem.
From this solution we could find the
uncertainties $\sigma_{m,i}$ 
simply as a differences between the observed and the predicted
values.

The third body in the system also causes variations in gamma
velocities. If we know $v_\gamma$ from different RV investigations
and in different epochs of the system, we could see a variation of
$v_\gamma$ with a period corresponding to the orbital period of
the third body. The variation could be described by
 \begin{equation}
    v_\gamma = K [\cos(\nu+\omega_{12})+ e \cos\omega_{12}], \label{RV3}
 \end{equation}
where $K$[km$\cdot$s$^{-1}$] is the amplitude of such variation
and could be calculated from the LITE parameters $A$[d],
$p_3$[yr], $e$ and $\omega_{12}$ from the equation
 \begin{equation}
    K = \frac{A}{p_3} \cdot
    \frac{5156}{\sqrt{(1-e^2)(1-e^2\cos^2\omega_{12})}}.
 \end{equation}
But the basic limitation is very often the long period $p_3$, which is usually too long to have reliable RV data
for this analysis.

LITE hypothesis could be also applied to other components in the system. The third body may cause LITE$_3$ and
another (fourth) component cause LITE$_4$. The resultant total effect is then simply the sum of the two effects
LITE = LITE$_3$ + LITE$_4$. One necessary condition has to be satisfied. The fourth component has to be more
distant then the third one (and the third one distant from the EB pair). This is the main physical condition,
because the method itself was derived with this condition by the use of von Zeipel's method to the three-body
problem.

If one wishes to include other effects, which could play a role in the problem, additional terms could be easily
added. This means if one wants to describe a system with the LITE caused by the third and the fourth component,
also an apsidal motion is presented and mass transfer together, the variation in $O-C$ diagram could be then
described as $$(O-C) = JD - JD_0 - P \cdot E - q \cdot E^2 - (O-C)_{LITE,3} - (O-C)_{LITE,4} - (O-C)_{apsid}.$$
One has to distinguish between primary and secondary minima in apsidal motion term, which could be computed
according to equations from \cite{GimenezGarciaPelayo}.

\section{Astrometry}

Another method to study binaries and the properties of their orbits is astrometry. The number of visual binaries
with astrometric orbits has grown, but complete phase coverage is often unavailable, due to the long orbital
period involved. Thanks to precise interferometry the observable semimajor axes of astrometric binaries are
still decreasing down to milli- and micro- arcseconds. On the other hand, one has to regret that no recent
astrometric measurements of a wide pair of about 1\arcs~ have been obtained for the systems mentioned below.
Most of the astrometric observations were adopted from The Washington Double Star Catalogue WDS, see
\href{http://ad.usno.navy.mil/wds/}{http://ad.usno.navy.mil/wds/} (\citealt{WDS}). The first astrometric
observations of visual doubles are a few centuries old. Altogether there are about 2000 systems with their
visual orbits known.

The astrometric measurements of binaries consist of a series of measurements of position angle $\theta_i$ and
separation $\rho_i$ secured at different times $t_i~(i=1,N)$. Sometimes also the errors of the individual data
points are available. The weighting scheme is provided by using the uncertainties $\sigma_\theta$ and
$\sigma_\rho$ of the individual observations.

From the astrometric data one is trying to find the parameters of
the relative orbit, defined by 7 parameters: period $p_3$, angular
semimajor axis $a$, inclination $i$, eccentricity $e$, longitude
of the periastron $\omega_3$ for the third body on its orbit, the
longitude of the ascending node $\Omega$, and the time of the
periastron passage $T_0$. One has to solve the inverse problem
  \begin{equation}
    \{ (t_i, \theta_i, \rho_i, \sigma_{\theta,i}, \sigma_{\rho,i}) \}_{i=1,N} \rightarrow ( a, p_3, i, e, \omega_3, \Omega, T_0 ).
  \end{equation}
The least-squares method and the simplex algorithm were used. The basic mathematic equations are the following.
Compute the mean anomaly from the time of the measurement, according to Eq. \ref{eq2}, solve the Kepler equation
$M \rightarrow E$ and after then convert the eccentric anomaly to the true anomaly, according to Eq. \ref{eq3}.
Compute the radius vector in arcseconds (the subscript $i$ was omitted for the clarity)
  \begin{equation}
    r = a \cdot \frac{1 - e^2}{1 + e \cos \nu},
  \end{equation}
and from this equation one can compute the position on the sky,
$\theta$ and $\rho$, respectively:
   \begin{eqnarray}
     \tan (\theta - \Omega) & = & \tan( \nu + \omega) \cdot \cos i \\
     \rho & = & r \cdot \cos( \nu + \omega) \sec(\theta - \Omega).
  \end{eqnarray}
Comparing this theoretical position on the sky with the observed
ones $\theta_0$ and $\rho_0$, one can calculate the sum of
normalized residuals squared
\begin{equation}
  {\chi^2_{astr}} = \sum_{i=1}^N \left[ \left( \frac{\theta_i-\theta_{0,i}}{\sigma_{\theta,i}} \right)^2 + \left( \frac{\rho_i-\rho_{0,i}}{\sigma_{\rho,i}} \right)^2 \right],
  \label{eq7}
\end{equation}
following \citet{Torres2004}. With this $\chi^2_{astr}$ and using
the simplex algorithm (see e.g. \citealt{Kallrath}) one can obtain
a set of parameters $(a, p_3, i, e, \omega_3, \Omega, T_0)$
describing the astrometric orbit.

At this place it is necessary to remark one useful comment. One
has to distinguish between the two angles $\omega_3$ and
$\omega_{12}$. The parameter used in LITE analysis is
$\omega_{12}$, but in astrometry the quantity
$\omega_3~=~\omega_{12}~+~\pi$ is employed. In this thesis the
angle $\omega$ stands for the longitude of the periastron for the
eclipsing binary, i.e. $\omega~=~\omega_{12},$ and the subscripts
will be omitted for clarity.

\begin{figure}[t]
  \centering
  \includegraphics{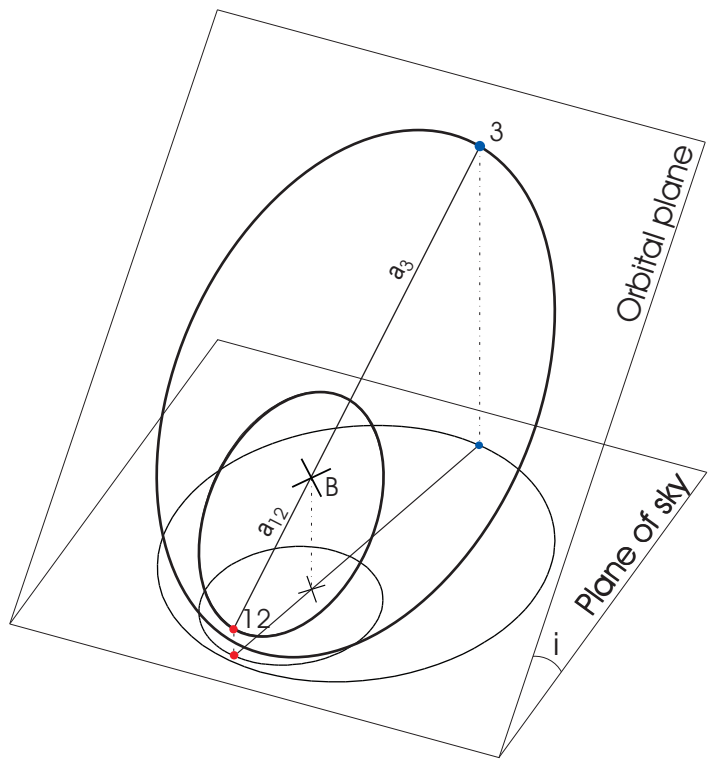}
  \caption{A simplified description of the relative binary orbit on a plane of
  the sky. The true orbit is inclined against the plane of the
  sky (angle $i$). B denotes the barycenter of the whole system,
  12 denotes the eclipsing binary pair and 3 the third
  component. The picture shows the moment, when the bodies are in the
  apocenters on their respective orbits. Also the semimajor axes of
  the third body and EB on the long orbit ($a_{12}$ and $a_3$) are shown.}
  \label{Binary}
\end{figure}

\section{Combining the methods}

Our task is to combine the astrometry and the analysis of times of minima into one joint solution. If one has
$N$ astrometric and $M$ minimum-time measurements, it is possible to merge them together and obtain a common set
of parameters
  \begin{equation}
  (t_i,\theta_i,\rho_i,\sigma_{\theta,i},\sigma_{\rho,i},JD_i,\sigma_{m,i})\rightarrow(A,p_3,i,e,\omega,\Omega,T_0,JD_0,P,q).
  \end{equation}
This set of 10 parameters fully describes the orbit of the eclipsing binary around the common center of mass
with the third unresolved component together with the ephemeris of the binary itself.

One is also able to determine the mass of the third body and the semimajor axis of the wide system because the
inclination is known and one can calculate the mass function of the wide orbit
  \begin{eqnarray}
    f(M_3) = \frac {(a_{12}\sin i)^3} {p_3^2} = \frac {(M_3 \sin i)^3} {(M_1+M_2+M_3)^2} = \frac {1}{p_3^2} \cdot {\left[ \frac
           {173.15 \cdot A} {\sqrt{1-e^2\cos^2\omega}} \right]^3},
    \label{eq13}
  \end{eqnarray}
where $a_{12}$ stands for the semimajor axis of the binary orbit
around the common center of mass and $M_1, M_2, M_3$ are the
masses of the primary, secondary, and tertiary component,
respectively. For more details see e.g. \citet{Mayer1990}.

The only difficulty which remains unclear is the connection
between the angular semimajor axis $a$ and the LITE amplitude $A$.
The quantity $a_{12}$ could be derived from Eq.~\ref{eq13} and
with the masses of the individual components one is able to
calculate also the value $a_3$, i.e. the semimajor axis of the
third component around the barycenter of the system
  \begin{equation}
     a_3= a_{12} \cdot \frac{M_1 + M_2}{M_3}.
  \end{equation}
The total mutual separation of the components is $a_{total} = a_{12} + a_3$ (see Fig.\ref{Binary}). Using
Hipparcos parallax $\pi$ \citep{HIP} one can obtain the distance $D$ to the system. Now it is possible to
enumerate the angular semimajor axis $a$ as a function of $D$ and $a_{total}$
  \begin{equation}
    a~=~\arcsin \left( \frac{a_{total}}{D} \right).
  \end{equation}

The way in which the two different approaches were combined follows a similar approach by \citet{Torres2004}.
From the mathematical point of view both methods are analogous and there is an overlap of the parameters in both
methods. Our task is to minimize the combined $\chi^2$
  \begin{equation}
   \chi^2_{comb} = \chi^2_{astr} +  \chi^2_{LITE},
  \end{equation}
where $\chi^2_{astr}$ and $\chi^2_{LITE}$ are the sums of squares according to Eqs. \ref{eq7} and \ref{eq11}.

There are some circumstances in which the $\chi^2$ values determined by the two methods are inconsistent.
Especially when there are many more data points in one method than the other, the resultant $\chi^2$ would be
much larger and as a consequence this method outweighs the other one. This problem could be eliminated using new
uncertainties $\sigma$ instead of the old ones $$ \sigma_{\theta,i} \rightarrow \sqrt{N} \cdot
\sigma_{\theta,i}, \,\, \sigma_{\rho,i} \rightarrow \sqrt{N} \cdot \sigma_{\rho,i}, \,\, \sigma_{m,i}
\rightarrow \sqrt{M} \cdot \sigma_{m,i} \, .$$

\subsection{Error estimation}

The errors of the output parameters were calculated according to the method described in Numerical Recipes,
pages 684 - 694, see \cite{NumRecipes1986}. Using a confidence limit of 95\%, one could calculate $$1 - 0.95 =
\Gamma (\nu/2, \Delta\chi^2/2),$$ where $\Gamma$ is the incomplete gamma function, $\nu$ is the number of
parameters fitted and $\Delta\chi^2$ defines the boundary around the final solution. The area within this
boundary is scanned and the maximum value of difference between actual parameter and final parameter $\delta
\alpha = (\alpha_i - \alpha_0)$ is taken as an error of the particular parameter.

\section{Distance determination} \label{dst}

Another task is determining the distance to these systems, which
could be also done by combining the astrometry and the LITE
analysis. As was mentioned above, from the LITE analysis it is
possible to derive the quantities $a_{12} \sin i$ and also $M_3
\sin i$, and from the astrometric analysis it is possible to
compute the angle $i$ and determine the semimajor axis and the
third mass in absolute units. Thanks to these values the total
semimajor axis $a_{total}$ (see above) could be determined.
Comparing the value $a_{total}$ with the angular semimajor axis of
the binary on the sky and leaving the parallax of the system as
another free parameter, one could compute this value and derive
the distance to the system.

But this method is useful only in very special cases, where at least one period is covered with sufficient data
points for both methods and both methods give us precise and comparable results. If the methods produce
different results, the method is not useful for the distance determination.

\section{Limitations of the methods}

The methods themselves are very powerful and useful, but one has
to take into consideration also some physical and observational
limitations.

\begin{figure}[t]
  \centering
  \scalebox{0.86}{
  \includegraphics{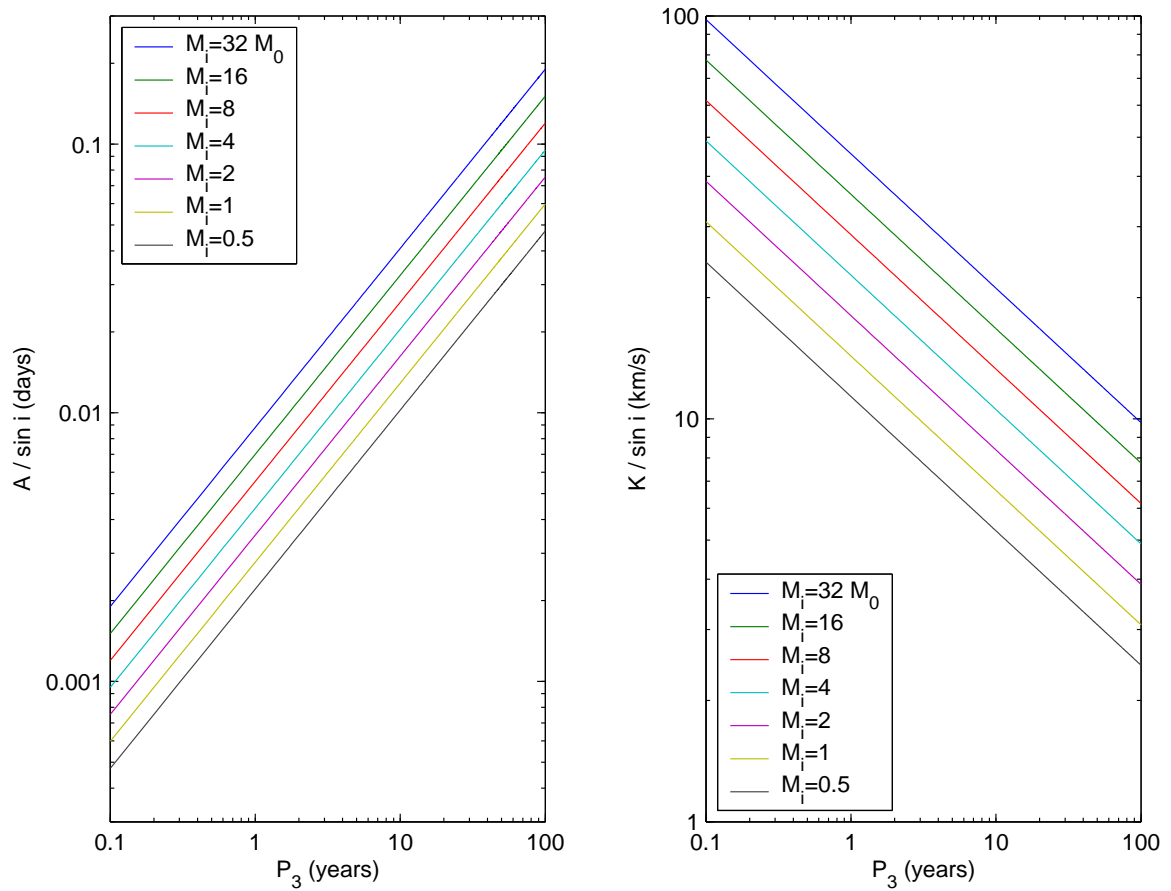}}
  \caption{In the left figure are the physical limits of amplitude
  of LITE as a function of period and mass. The different
  lines represent different masses of the components (for $M_1 = M_2 =
  M_3$). In the right one the same for amplitude of systemic
  velocity variations. The figure adopted from \cite{Mayer1990}.}
  \label{Kepler}
\end{figure}

If one considers only the LITE, the main observational limitation are the amplitude $A$ and period $p_3$. There
are a few often-used methods to determine the time of minimum, the most common being the Kwee-van Woerden method
\citep{Kwee}. Its main advantage is that one can compare the results and the individual errors of the
measurements. It is suitable only for symmetric minima.

For different types of eclipsing binaries we could reach different levels of accuracy of the times of minima.
Using precise photoelectric and CCD detectors one is able to compute the time of minimum light with precision of
about 1--10 seconds or less ($\approx$ 0.0001--0.00001 day). This is only the theoretical value, attainable only
if there are no clouds or moon, and if the observational conditions are very good.

\begin{figure}[t]
  \centering
  \scalebox{0.86}{
  \includegraphics{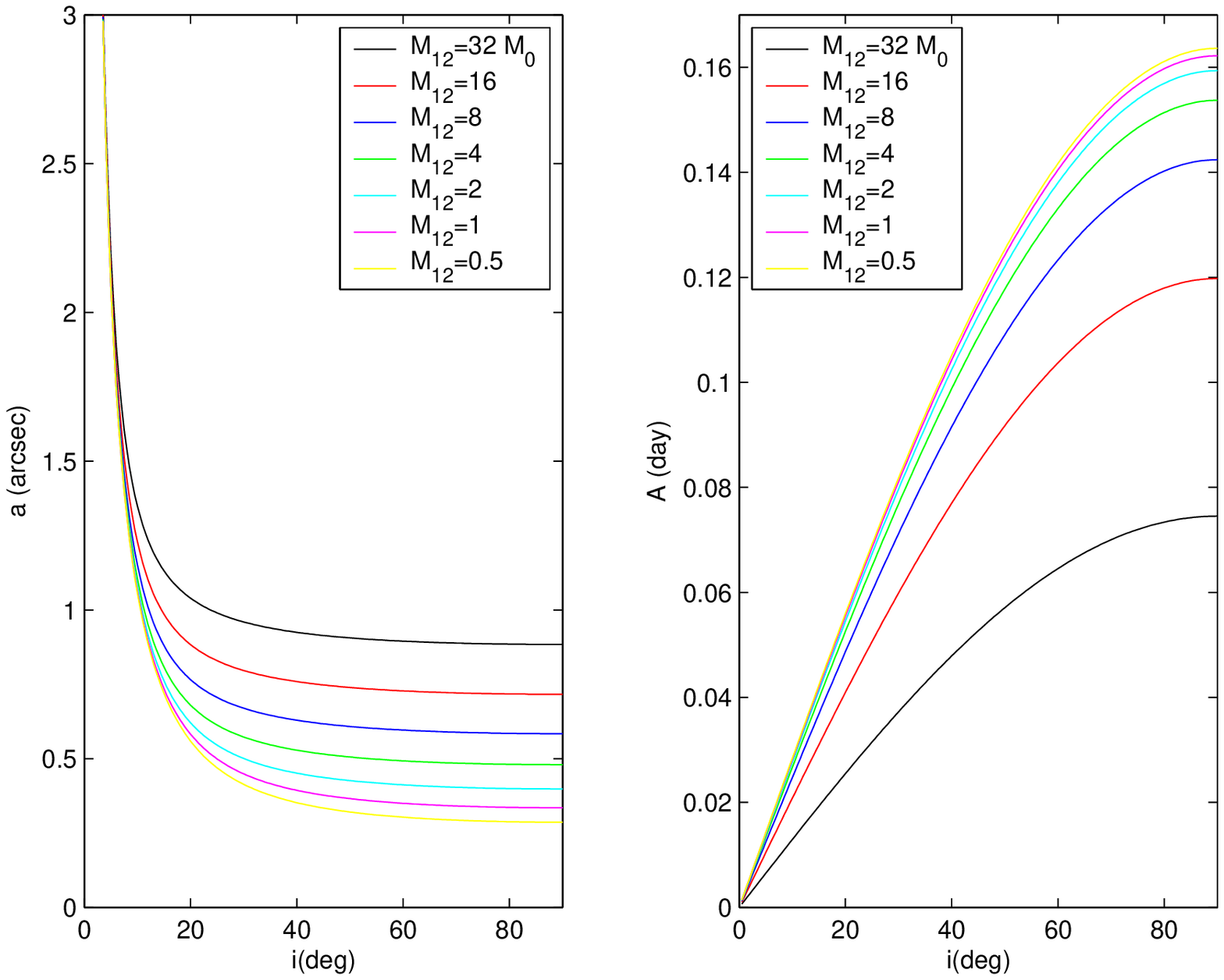}}
  \caption{The principal limits of amplitude of astrometric orbit
  and amplitude of LITE in combined solution. The left figure was
  plotted with fixed LITE amplitude $A=0.03$ d and the period $p_3=20$
  yr. The right one with the same period and the
  amplitude of astrometric orbit $a=1$ arcsec. The value of parallax
  was fixed, $\pi = 35$~mas. The different curves represent the
  different masses of the EB, where the introduced masses
  are $M_{12} = M_1+M_2$. 
  }
  \label{AstrLITE}
\end{figure}

The practice is sometimes quite different. If one compares two observations of the same star and the same
minimum time (after transformation to the heliocentric Julian time), one finds out that there could be a
difference between them of order 0.005 day. So there is a question about the accuracy of the method and the true
error. Especially amateur astronomers have sometimes very precise measurements, but their results (the times of
minima) are not very satisfactory. This could be due to shift of time on their computer (the delay could be from
a few to tens of seconds!), or the wrong method used for determining the time of minimum. Another possible
explanation is, that the time of exposure could be the beginning of the exposure (instead of the middle of
exposure).

Especially due to this reason there is a principal limit of the amplitude of LITE which could be reached.
Amplitudes of LITE below 0.01 day are problematic, but in some systems even lower amplitudes are detectable
(e.g. RT And or RZ Com). It is also necessary to take into consideration that the old measurements are not
photoelectric, but visual or photographic, where the errors of the individual data points are much larger (and
these errors are mostly not available for the analysis).

Considering only LITE, plots of the relative limitations are shown in Fig. \ref{Kepler}. One can see the
dependence of the amplitude and the period on the masses of the individual components (only the case with equal
masses is shown).

If one uses the combined approach, it is necessary to discuss also the astrometric orbit and the accuracy of the
astrometric measurements. A speckle interferometric techniques are very precise and could reach a few
miliarcseconds (mas), but the older data are visual or photographic. In most cases the errors of individual data
points are unavailable.

Figure \ref{AstrLITE} shows the combined solution and the
amplitudes in both methods. They are strongly dependent on the
inclination $i$ of the orbits. In the first one the LITE amplitude
was fixed, while in the second one the astrometric-orbit amplitude
was fixed. The basic properties of the plots in the diagrams are
due to the geometry of the system. The edge-on orbit ($i
\rightarrow 90^\circ$) leads to the binary eclipses, while the
face-on orbit ($i \rightarrow 0^\circ$) leads to the most
pronounced astrometric variation. The parallax (the distance) of
the system was also fixed. One has to consider also the
limitations of the methods described above and the areas in the
diagrams where these limits are satisfied.

\section{Numerics and the strategy to solve the problem} \label{numerics}

The efficiency of the combined method and the computing time
required by the algorithm strongly depend on the initial set of
parameters and the input data. If nothing is known about the
solution, one has to scan a wide range of parameters (eccentricity
$e$ from $0$ to $1$, and the angular parameters from $0^\circ$ to
$360^\circ$, etc.).

The efficiency of the algorithm could be improved if the simplex
is used repeatedly. It can happen that the simplex converges into
a local minimum while the global one is far away. It is therefore
advisable to run the algorithm again, with as large initial steps
as in the previous run, but keeping the values of the parameters
corresponding to the previously found minimum as one vertex.
Repeating this strategy several times over the whole parameter
space, one can judge whether the global minimum was found by
checking whether the sum of squares of the residuals is still
changing or not (see Fig. \ref{RMS} and \ref{RMS2}).

\begin{figure}[t!]
  \centering
  \scalebox{0.75}{
  \includegraphics{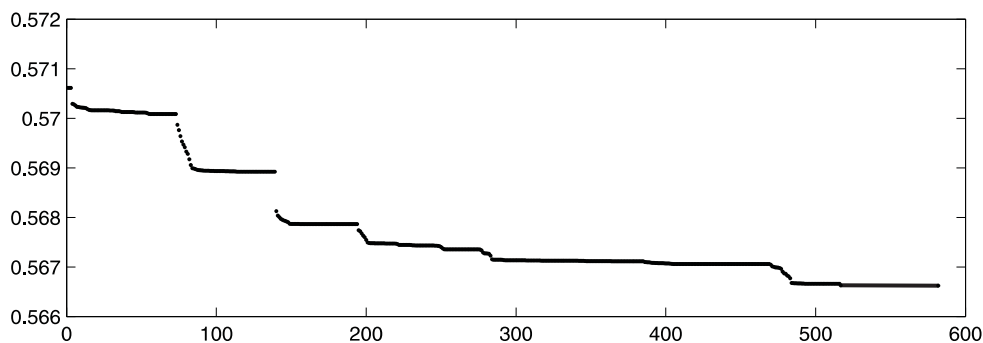}}
  \caption{The sum of squares of residuals as a function of number
  of iterations, for the case VW Cep \emph{Solution I.} (see below).}
  \label{RMS}
\end{figure}

\begin{figure}[t!]
  \centering
  \scalebox{0.65}{
  \includegraphics{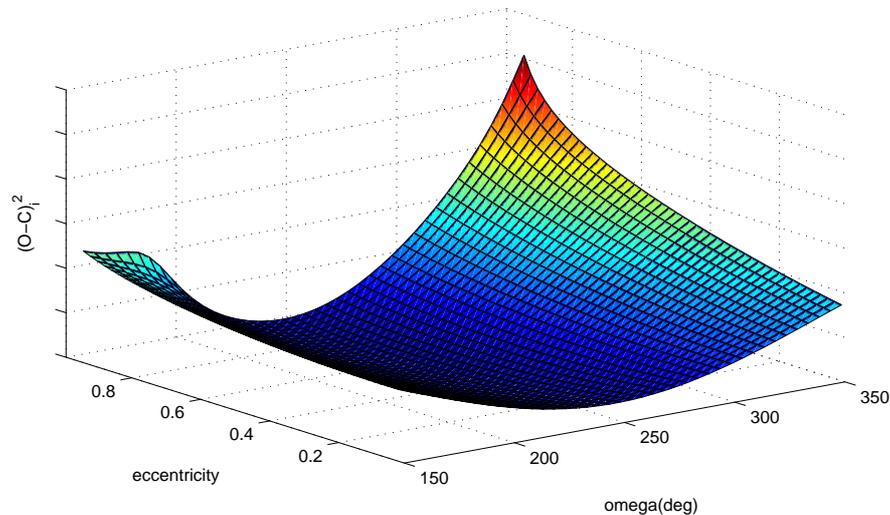}}
  \caption{The sum of squares of residuals as a function of $\omega$
  and $e$ for the third-body orbit.}
  \label{RMS2}
\end{figure}

If one could guess the approximate values of the parameters, it is also recommended to use them. If one does not
have any information about their values, the algorithm itself is able to find the appropriate ones, but these
could be only numerical ones without any physical meaning. It is recommended to set the initial values of
parameters as close as possible to the right values, because the algorithm itself will converge faster and the
probability of the code producing incorrect parameters is lower.

Another problem is the number of parameters used. The number of parameters strongly affects the computing time
required. Regarding the classical LITE problem, one has 7 parameters (2 from ephemeris and 5 from LITE), but
using the quadratic term also one further parameter is necessary to evaluate. One can estimate the ephemeris
parameters ($JD_0, P, q$) in the first step and after then with fixed ephemeris calculate the parameters of LITE
itself. But it is strongly recommended to compute all parameters together, because the LITE could also affect
the ephemeris and change the values slightly. On the other hand sometimes it is better to fix the values of some
parameters for the code to run faster. However, it is necessary to release all of the parameters for the final
fit.

If also another effect appears in the analysis, for example apsidal motion, it is necessary to estimate 3
additional parameters ($\omega_0, e',\dot\omega$). For the combined analysis, there could be even more
parameters. In the most complicated case (VW~Cep, see section \ref{VWCep}), there were 14 parameters which had
to be determined. The strategy which helps the code to converge faster to the minimum could be shown in the case
HT~Vir (see below section \ref{HTVir}). The astrometric orbit is known with high precision, so the astrometric
parameters ($a, p_3, i, e, \omega, \Omega, T_0 $) could be set and fixed for the first step. After the code
returns the values of the ephemeris of the EB, one could release also the astrometric parameters and run the
code once again with fitting all the parameters. This strategy saves a lot of unnecessary computing time.

To conclude, using the strategy presented here and the combined method described above, one gets a satisfactory
result after a large number of iterations. This number and computing time is strongly dependent on the input
data set (the number of observed data) and the number of parameters fitted. For the case VW~Cep with the largest
data set (more than 1600 data points, see below) and also with the most parameters to fit (14 in total) one
reaches the solution, when the sum of squares is not changing significantly, after circa $100\,000$ simplex
steps. This takes about one day on a computer with a 2 GHz processor. But it is only the illustrative example,
the basic condition is the separation between the initial and the final parameters.

There were done also a few tests of the code and its ability to find the same final parameters from different
starting values of parameters. The results were satisfactory, and the code was able to find almost the same
values, but it strongly depends on the separation between the initial and the final parameters. Also these tests
confirmed the fact that setting the initial parameters as close to the right solution spares computing time and
also makes the final solution more reliable.

\section{The program} \label{program}

The code itself was written in the Matlab language. It is available for download via the web pages
\href{http://sirrah.troja.mff.cuni.cz/~zasche/}{http://sirrah.troja.mff.cuni.cz/$\sim$zasche/} The code is
zipped together with the necessary routine files \texttt{*.m}, the sample input data files \texttt{*.epo} and
\texttt{*.dat} and with the initial parameters file \texttt{*.in}. All of these files have to be copied into the
same folder as the code itself. The sample (data files and also the code file) is for the HT~Vir system. The
code is easily modifiable for the user. A brief manual for the user is also available.

The program was initially designed only for the use of this thesis, so the first version was not very
user-friendly, but some modifications were done for the easier use. Some comment lines were also included in the
file.

In the first part of the code (lines 1 - 66) is the input. The files \texttt{*.in}, \texttt{*.dat} and
\texttt{*.epo} include the input parameters, astrometry data and the minimum-time observations, respectively.
The recommended data format is shown in the enclosed files. In the next lines (67 - 121) are some transformation
rules and assigning the values to proper variables. Important here are lines 74-83, where is the input of masses
$M_1$ and $M_2$ and also the parallax and its error.

In the lines 122 - 172 is the relative astrometric orbit of the binary is plotted (before the computation) and
saved as \texttt{HTVir-before.eps}. The next lines (173 - 299) are for plotting of the $O-C$ diagram before the
computation, saved as \texttt{HTVirOC-before.eps}.

Lines 300 to 681 comprise the body of the code, the simplex algorithm. Its initialization is in line 327, where
the user could choose which mode he wants to run. There are three possibilities: 0 for only LITE, 1 for LITE
together with the quadratic term (the mass transfer) and finally 2 for only the quadratic term (this possibility
is not designed for the use of the combined approach). After then there are lines with the input parameters of
the simplex algorithm (l.347 - 387). Here one could change the range of values of the individual parameters
where the simplex works in the first run. The rest of the lines (to l.681) is the simplex itself and has not to
be modified anyway.

The lines from 683 to 1210 are for plotting the resulting $O-C$ diagrams and lines from 1211 to 1268 for
plotting the resulting orbit of the binary. In the lines from 1269 to 1322 is the computation of the errors of
the individual derived parameters. The lines from 1323 to 1476 create the output files \texttt{*.in} and
\texttt{*.txt}, where the output parameters are written.

Generally, the code is ready to be used. The only modification
could be the change of the names of the input and output files in
the code (only find and replace the appropriate file names), and
also the input masses and parallax. The main output file is
\texttt{*.txt}, where are written all of the parameters with their
respective errors. Also the derived quantities with the computed
errors are included in the file.

The computation process of the code is the following. Run the code
\texttt{*.m} in the Matlab program. After two figures ($O-C$
diagram and astrometric orbit) appear on the screen, in the main
Matlab command window the program asks the user whether he wants
to compute also the quadratic term, or only the LITE. After
confirmation which mode one wants to use, there appears another
question about the number of iterations one wants to compute. It
is up to user, but it is recommended to type only a few, because
each of them could take some time. On the other hand more
iterations means better precision of the result. After input these
values on the screen will appear the resulting sum of square
residuals of the problem, which is decreasing after each
iteration. The sum of squares is divided into the two separate
values, the first one from LITE and the second one from the
astrometry. The most usual case is, that one of them is
decreasing, while the other is increasing, but the sum of these
two values has to decrease all the time. When the decreasing
stops, the program terminates. A few plots will appear on the
screen (these are saved as the \texttt{*.eps} files) and also the
output file \texttt{*.txt} is saved to the same folder.

\chapter{Systems with LITE}

There were a lot of studies, where many eclipsing binaries showing period variations of their period were
analyzed. It was decided to present here only a few systems where LITE was not recognized until now, as well as
systems where LITE was supposed to be present, but another (and better) solution for the variations in the $O-C$
diagram has been found. This is not the crucial part of the thesis, and these systems are described only very
briefly.

Altogether there were about 130 systems analyzed during the 3-yrs PhD study for their period changes. In about
one half of these systems the LITE was proposed as a hypothetical explanation. Regrettably, the analysis was
done only on the basis of their times-of-minima observations and in most of the systems no other detailed
analysis was performed. This is also evident in the set of binaries presented here in this chapter.

A few of the analyzed systems were selected for publication in various papers. Namely AD~And, WY~Per and
V482~Per in \cite{Wolf2004}, AR~Aur, R~CMa, FZ~CMa and TX~Her in \cite{ZascheLitomysl}, OO~Aql, V338~Her, T~LMi,
RV~Lyr, TW~Lac and V396~Mon in \cite{ZascheZejdaBratGreece}, EW~Lyr and IV~Cas in \cite{ZascheHungary} and
XX~Leo in \cite{ZascheSvobodaGAIAU}.

\section{Individual systems under LITE analysis} \label{OnlyLITE}

All of the selected systems are Algol-type EBs, and also semidetached ones. According to the recent paper on
period changes in Algols by \cite{Hoffman2006}, there could be a connection between the spectral type of the
secondary component and the nature of the period changes. Systems with spectral types of secondaries earlier
than F5 show $O-C$ variations, which could be caused by the magnetic activity cycles and convective envelopes.
This effect was discussed by \cite{Hall1989}, \cite{Applegate1992}, \cite{Lanza1998}, etc. The role of magnetic
cycles on the period changes is discussed below, but due to lack of information about the systems such analysis
is a difficult task. For some of the systems selected in this thesis the spectral types of secondaries are only
known with a low confidence level, light-curve analysis is missing and spectroscopy has never been done.

The LITE analysis of the systems presented below in this chapter
was also published in \cite{Zasche2007LITE}.

\section{RY~Aqr}

The eclipsing binary system RY~Aqr (AN~125.1908, BD-11~5574, HD~203069) is an Algol-type EB. It was classified
as A3 spectral type (Simbad), A8 \citep{Popper1989}, or most likely as a late A/early F main sequence star and
an early K subgiant \citep{Helt1987}. It is a double-lined SB \citep{Popper1989} with an orbital period of about
2 days. Its relative brightness is about 8.9~mag in V filter.

The variability of RY~Aqr was discovered by \citet{Leavitt1908}
and the ephemeris was firstly derived by \citet{Zinner1913}. Since
then a lot of times of minima observations were obtained.

\begin{figure}[b!]
   \centering
   \scalebox{0.75}{
   \includegraphics{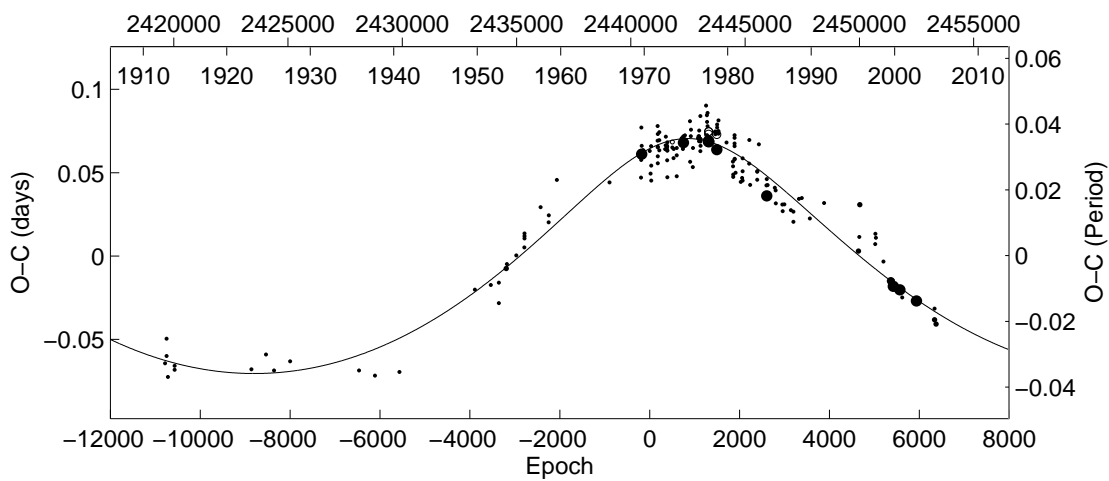}}
   \caption{An $O-C$ diagram of RY~Aqr. The individual observations
   are shown as dots (primary) and open circles (secondary), the
   small ones for visual and the large ones for CCD and photoelectric
   observations, bigger the symbol, bigger the weight. The curve
   represents the predicted LITE variation.}
   \label{FigRYAqr}
\end{figure}

The most detailed analysis was performed by \citealt{Helt1987} on
the basis of the $uvby$ photoelectric photometry together with the
radial velocity measurements by \citeauthor{Popper1989}. This
analysis (using both WINK and Wilson-Devinney programs) results in
a set of parameters, which reveals the nature of the system. The
hotter primary has unusually low mass (about 1.27 \Mo), while the
secondary K subgiant (0.26 \Mo) has undergone a mass-loss from its
initial mass of about 1.9 \Mo. There was observed also an
intrinsic photometric variability which could be caused by the
surface activity of the secondary (the period of this variability
is close to the orbital period of the system). 
RY~Aqr is also a member of a visual binary HU~86, but the distant
body probably does not relate to the system.

There were performed a few period studies of RY~Aqr
(\citealt{Baldwin1974}, \citealt{Mallama1980}) and the most recent
one by \citet{Helt1987}, who suggested that also the LITE could
play a role in this system. Her supposed period of such variation 
(about 70 years) is different from the present one. This new
analysis is based on a set of 178 times of minima published in the
literature. Resultant $O-C$ diagram is plotted in Fig.
\ref{FigRYAqr} and the parameters of the third-body orbit are in
Table \ref{TableRYAqr}.

The minimal mass of the third body ($M_{3,min} = M_3 \cdot
\sin{i_3}$ if $i_3=90^\circ$) was calculated according to the mass
function derived from the LITE hypothesis and total mass of the
eclipsing pair $M_{12} = (1.27 + 0.26)$~\Mo, see
Table~\ref{TableRYAqr}. Computed minimal mass about 1~\Mo~is
rather high and such a star on the main sequence will be evident
because of its luminosity. The mass $M_{3,min}$ leads to the
spectral type around G3 (according to \citealt{Hec1988}), and such
a star would be more luminous than the secondary component of the
EB pair. The third light in the light-curve solution by
\cite{Helt1987} was only estimated. The value about a few percent
(from 2.6\% in \emph{u} to 5.9\% in \emph{y}) was adopted, but not
derived. This value indicates roughly the same third mass, as was
calculated from our analysis.

The radial velocities were analyzed precisely only once
\citep{Popper1989}, therefore no changes in systemic velocities
are available. The systemic velocity of RY~Aqr was derived to be
about $-60~\mathrm{km \cdot s^{-1}}$, which could be caused by the
motion around the common center of mass with the third component.
Precise spectroscopy would probably detect the third body in the
spectrum of the system.

The star has not been measured by Hipparcos, but the distance was
derived from the photometry, see \cite{Helt1987}. The value $(180
\pm 10)$~pc leads to the predicted angular separation of the third
component $a = (169 \pm 10)$~mas, which was calculated using
assumption $i_3 = 90^\circ$. The predicted magnitude difference is
about 3~mag. The companion with such a distance and magnitude
difference is detectable with the modern stellar interferometers.
Only further times of minima, precise spectroscopic and
photometric analysis will reveal the nature of the system.

\begin{table*}[t]
 \caption{The final results: RY~Aqr.}
 \label{TableRYAqr} \centering
 \scalebox{0.87}{
 \begin{tabular}{c c c c c c c c | c c }
 \hline\hline
  Parameter&  $JD_0 $    &      $P$      &  $p_3$   &  $T_0$    & $\omega$ &   $e$    &  $A$      &  $f(M_3)$ & $M_{3,min}$\\
  Unit     & $[$HJD$]$   &   $[$day$]$   & $[$yr$]$ & $[$HJD$]$ & $[$deg$]$&          & $[$day$]$ &  [\Mo]    & [\Mo]    \\\hline
  Value   & $2440824.351$&  $1.9665990$  & $105 $   & $2442300$ & $88$     & $0.35$   & $0.070$   &  $0.165$  & $1.02$    \\
  Error   & $\pm 0.004$  &$\pm 0.0000013$& $\pm 4$  & $\pm 1600$& $\pm 7$  &$\pm 0.05$&$\pm 0.002$&$\pm 0.006$& $\pm 0.03$ \\
  \hline\hline
 \end{tabular}}
\end{table*}

\section{BF~CMi}

BF~CMi is one of the neglected eclipsing binaries, which have been observed only a few times and only very
limited knowledge about it is available. Its period is about 1.18 days and the relative brightness about
10.3~mag in V filter. According to \cite{Svechnikov1990} the star has spectral type A5+K0IV (based only on
photometric indices), mass ratio 0.3, orbital inclination 79$^\circ$ and is a semidetached one.

\begin{table*}[b!]
 \caption{The final results: BF~CMi.}
 \label{TableBFCMi} \centering
 \scalebox{0.87}{
 \begin{tabular}{c c c c c c c c | c c }
 \hline \hline
   Parameter&  $JD_0 $    &      $P$      &  $p_3$   &  $T_0$    & $\omega$ &   $e$    &  $A$      &  $f(M_3)$ & $M_{3,min}$\\
   Unit     & $[$HJD$]$   &   $[$day$]$   & $[$yr$]$ & $[$HJD$]$ & $[$deg$]$&          & $[$day$]$ &  [\Mo]    & [\Mo]    \\\hline
   Value   & $2450789.610$&  $1.1806791$  & $46.3$   & $2447300$ & $170$    &  $0.79$  & $0.040$   & $0.39$    & $2.1$        \\
   Error   &$\pm 0.016$   &$\pm 0.0000026$& $\pm 1.2$& $\pm 200 $& $\pm 21$ &$\pm 0.10$&$\pm 0.018$& $\pm 0.16$& $\pm 0.9$   \\
 \hline \hline
 \end{tabular}}
\end{table*}

Its variability was discovered by \cite{Huruhata1979} and its designation as BF~CMi was presented by
\cite{Kholopov1981}. But the period is still questionable. There were two unsuccessful attempts to observe the
secondary minima - on 2 March 2006 and 4 April 2007. Also \cite{Berthold1981} noted that no secondary minimum is
observable. It means the period could be 2 times longer, about 2.36 days. Two new primary minima were observed
and kindly sent by L. \v{S}melcer from Vala\v{s}sk\'e Mezi\v{r}\'i\v{c}\'i observatory.

\begin{figure}[t!]
   \centering
   \scalebox{0.75}{
   \includegraphics{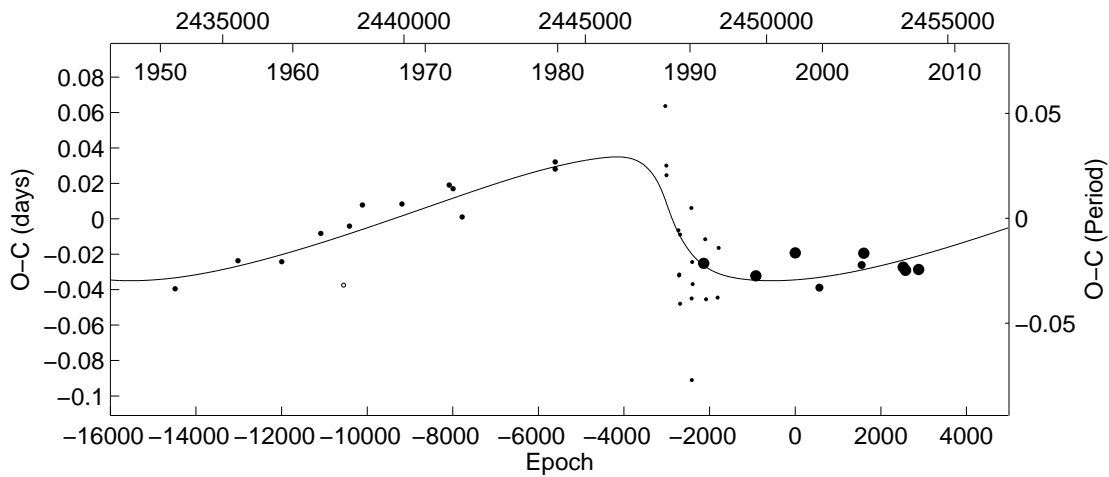}}
  \caption{An $O-C$ diagram of BF~CMi. The description is the same as
  in the previous figure.}
  \label{FigBFCMi}
\end{figure}

There were 39 times of minima collected from the published literature, resulting in an $O-C$ diagram shown in
Fig.\ref{FigBFCMi}. The parameters of LITE are in Table \ref{TableBFCMi}. As is evident from the abrupt period
jump near 1990, the eccentricity of the orbit should be rather high. Note also quite large scatter of recent
photoelectric and CCD observations since 1990, which is much larger, than one could expect from these kind of
measurements.

If one assume the masses of the individual components $M_1 + M_2 =
(1.9 + 0.9)$\Mo, the minimal mass of the predicted third component
is about 2.1\Mo, which is rather high value and would dominate in
the system. This hypothesis could not be proved until the detailed
analysis of the system is performed. Regrettably, neither
photometry nor spectroscopy was carried out. The abrupt changes in
period could be also caused by the two period jumps near 1988 and
1991, produced by some mass-transfer phenomena in the system.

\section{RW~Cap}

The eclipsing binary RW~Cap (AN~21.1910, BD-18~5641, HD~192900) is a system with an orbital period of about
3.4~days. Its spectral type was classified as A3+A4 (according to \citealt{Budding1984}) and its apparent
brightness is about 10.3~mag in V filter. The depth of primary minimum is about 1.2~mag. Str\"omgren photometry
of the system by \cite{1983WolfKern} agrees with its spectral type.

\begin{figure}[b!]
  \centering
  \scalebox{0.75}{
  \includegraphics{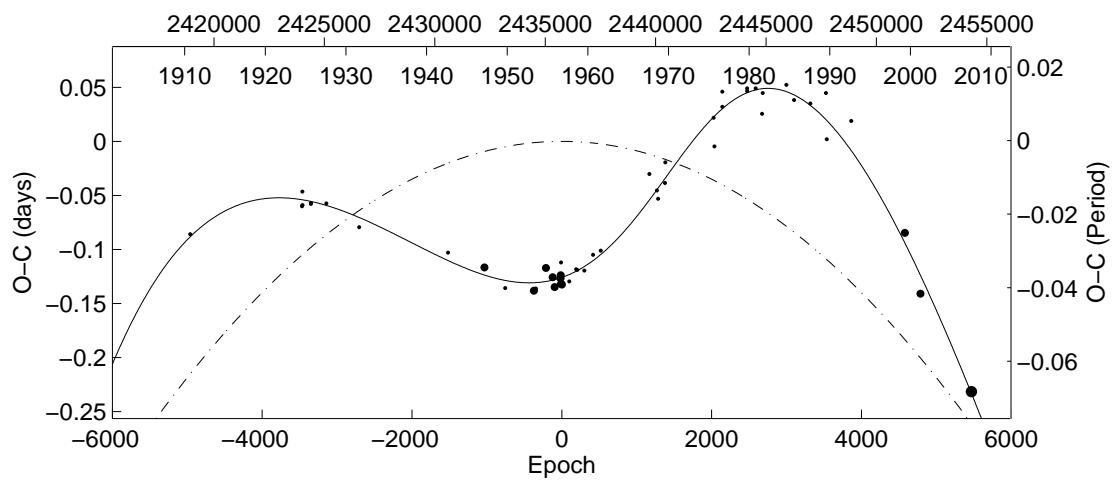}}
  \caption{An $O-C$ diagram of RW Cap. The description is the same as
  in Fig.\ref{FigRYAqr}, the dash-dotted line represents the quadratic term.}
  \label{FigRWCap}
\end{figure}

Its photometric variability was discovered by \cite{Pickering1910}
and until now there were 52 observations of times of minima
obtained. \cite{Tsesevitch1957} collected all available minima and
proposed an abrupt period jump near 1920. After then
\cite{Kreiner1971} compiled large set of times of minima and
already in this paper is evident that there could be some periodic
variation in $O-C$ diagram. Regrettably during the last two
decades only a few times of minima were obtained, so the LITE
hypothesis is still not very conclusive. Two of last three data
points in the $O-C$ diagram are only a mean values from the
automated surveys ROTSE (see \citealt{ROTSE}) and ASAS (see
\citealt{ASAS}). But the last one was obtained and kindly send by
A.Liakos from Athens university in July 2007.

\begin{table*}[t]
 \caption{The final results: RW~Cap.}
 \label{TableRWCap} \centering
 \scalebox{0.77}{
 \begin{tabular}{c c c c c c c c c | c c }
 \hline \hline
   Parameter& $JD_0 $    &      $P$      &  $p_3$   &  $T_0$    & $\omega$ &   $e$    &  $A$      &        $q$       &  $f(M_3)$ & $M_{3,min}$\\
   Unit     &$[$HJD$]$   &   $[$day$]$   & $[$yr$]$ & $[$HJD$]$ & $[$deg$]$&          & $[$day$]$ &$[10^{-10}$ day$]$&  [\Mo]    & [\Mo]     \\\hline
   Value   &$2435750.857$&  $3.3923745$  & $80.1$   & $2440900$ & $ 0 $    &  $0.23$  & $0.130$   &    $87.4$        &  $1.91$   &  $5.9$    \\
   Error   &$\pm 0.014$  &$\pm 0.0000055$& $\pm 4.6$& $\pm 2800$& $\pm 42$ &$\pm 0.16$&$\pm 0.011$&     $0.2$        & $\pm 0.26$& $\pm 1.0$ \\
 \hline \hline
 \end{tabular}}
\end{table*}

For the times of minima analysis the LITE and the quadratic term
was used. This means, during the computation process, altogether 8
parameters ($JD_0, P, q, p_3, A, T_0, \omega, e$) were adjusted,
resulting in a set of parameters written in Table
\ref{TableRWCap}. Because the system is semidetached, the
mass-transfer hypothesis could play a role. The quadratic term
coefficient $q = (87.4 \pm 0.2) \cdot 10^{-10}$~day leads to the
period change about $1.88 \cdot 10^{-6}$~day/yr. From this value
the conservative mass transfer rate could be derived $\dot M = 9.4
\cdot 10^{-8}$~\Mo/yr. Regrettably, these values cannot be proved
by some other independent method.

Applying the LITE hypothesis to the system one gets the 80-years variation (see Fig.\ref{FigRWCap}), but as is
evident from Table \ref{TableRWCap}, the orbit is still not very well-defined and the errors of the individual
parameters are large. Resulting mass function is quite high and using $M_{12}=4.59$~\Mo \citep{Brancewicz1980}
one gets the minimal mass of the third body of about 5.9\Mo. Such a body would be dominant in the light-curve
solution as well as in the spectroscopic analysis. Unfortunately there were no such analysis performed. Another
explanation is that the third component is also a binary.

\section{TY~Cap}

The next eclipsing binary with period changes is TY~Cap (AN~243.1932, BD-13~5664, HD~194168). It is an
Algol-type EB with apparent brightness about 10.3 in V filter and spectral type classified as A2/3V (according
to \citealt{Halbedel1984}). Its orbital period is about 1.4 days.

\begin{table*}[b]
 \caption{The final results: TY~Cap.}
 \label{TableTYCap} \centering
 \scalebox{0.87}{
 \begin{tabular}{c c c c c c c c | c c }
 \hline \hline
   Parameter&  $JD_0 $   &      $P$      & $p_3$   & $T_0$    &$\omega$ &   $e$    &  $A$      &  $f(M_3)$ & $M_{3,min}$\\
   Unit     & $[$HJD$]$  &   $[$day$]$   &$[$yr$]$ &$[$HJD$]$ &$[$deg$]$&          & $[$day$]$ &  [\Mo]    & [\Mo]    \\\hline
   Value   &$2444793.489$&  $1.4234574$  &$70.4$   &$2439600$ &$147$    & $0.79$   & $0.045$   &  $0.23$   &  $2.2$   \\
   Error   &$\pm 0.006$  &$\pm 0.0000009$&$\pm 8.6$&$\pm 1200$&$\pm 29$ &$\pm 0.12$&$\pm 0.005$& $\pm 0.02$& $\pm 0.2$\\
 \hline \hline
 \end{tabular}}
\end{table*}

Its photometric variability as well as its Algol-type were
discovered by \cite{Hoffmeister1933}. Altogether 96 times of
minimum light were carried out, only 5 data points were neglected
due to their large scatter. The $O-C$ plot is in
Fig.\ref{FigTYCap}, the curve represents the least-square fit with
the LITE parameters written in Table \ref{TableTYCap}.

Due to missing detailed analysis of the system, our knowledge
about TY~Cap is only limited. According to \cite{Brancewicz1980}
the total mass of the EB system is $M_1+M_2 = (2.5 + 2.06)$\Mo.
With this mass and with the parameters of the LITE from Table
\ref{TableTYCap} one can calculate the minimal mass of the third
body, which results in 2.18~\Mo. Unfortunately there is no
spectroscopic, as well as no light-curve analysis and the star was
not measured by Hipparcos, so the angular separation of the third
component also cannot be derived.

\begin{figure}[t]
   \centering
   \scalebox{0.75}{
   \includegraphics{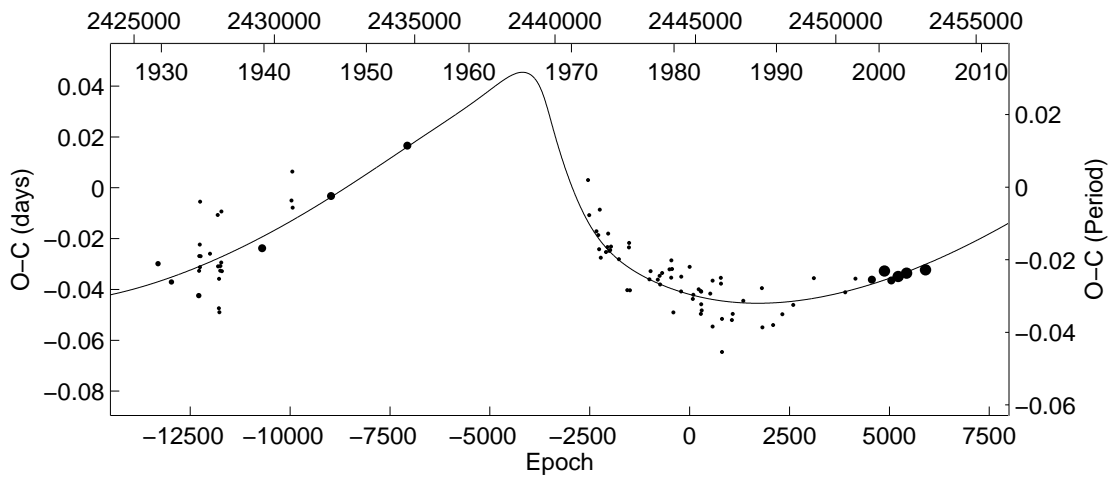}}
 \caption{An $O-C$ diagram of TY~Cap.}
         \label{FigTYCap}
\end{figure}

The eccentricity of the LITE solution is quite high, and could be even higher, but due to lack of data points
near the periastron passage this could not be proved. The next periastron passage will occur about 2035.
Generally the third-body orbit is not covered sufficiently and the parameters of this predicted body have to be
derived by some other independent method.

\section{SS~Cet}

SS~Cet (BD+01~491, HD~17513) is a semidetached Algol-type EB with an apparent brightness of about 9.4~mag in V
filter and spectral type classified as A0+K3III (according to \citealt{Budding2004}). Orbital period is about 3
days.

\begin{table*}[b]
 \caption{The final results: SS~Cet.}
 \label{TableSSCet} \centering
 \scalebox{0.87}{
 \begin{tabular}{c c c c c c c c | c c }
 \hline \hline
   Parameter&  $JD_0 $   &      $P$      & $p_3$   & $T_0$   &$\omega$ &   $e$    &  $A$      & $f(M_3)$  & $M_{3,min}$\\
   Unit     & $[$HJD$]$  &   $[$day$]$   &$[$yr$]$ &$[$HJD$]$&$[$deg$]$&          & $[$day$]$ & [\Mo]     & [\Mo]    \\\hline
   Value   &$2442451.330$&  $2.9739737$  &$20.4$   &$2449500$& $304$   & $0.21$   &$0.014$    & $0.031$   &  $0.75$   \\
   Error   &$\pm 0.002$  &$\pm 0.0000007$&$\pm 0.5$&$\pm 900$&$\pm 73$ &$\pm 0.09$&$\pm 0.002$&$\pm 0.002$& $\pm 0.13$ \\
 \hline \hline
 \end{tabular}}
\end{table*}

Its variability was discovered by \cite{Hoffmeister1934SSCet}. The most detailed analysis was performed by
\cite{Narasaki1994} on the basis of their \emph{BVRI} photoelectric photometry and spectroscopy. This study
results in a semidetached system, with no peculiarities in the light curve, $M_1 = 2.15$\Mo, $M_2 = 0.6$\Mo, and
the spectral observations indicate that no circumstellar matter is presented in the system. On the other hand
the spectroscopic analysis by \cite{Vesper2001} presents evidence for mass transfer in the binary because of the
behavior of the H$\alpha$ emission. The mass transfer was not recognized in our current analysis of the times of
minima.

Since its discovery there were 111 minima observations obtained, but only 95 were used for the period analysis,
because of the large scatter of the first ones from the 1930's. There could be a period jump near 1950, this is
the reason why only the recent data were analyzed. The period changes were firstly noted by \cite{Kreiner1971},
but only with a small set of times of minima, displaying the steady increase. The final LITE curve is in Fig.
\ref{FigSSCet} and the resultant parameters in Table \ref{TableSSCet}.

\begin{figure}[t]
   \centering
   \scalebox{0.75}{
   \includegraphics{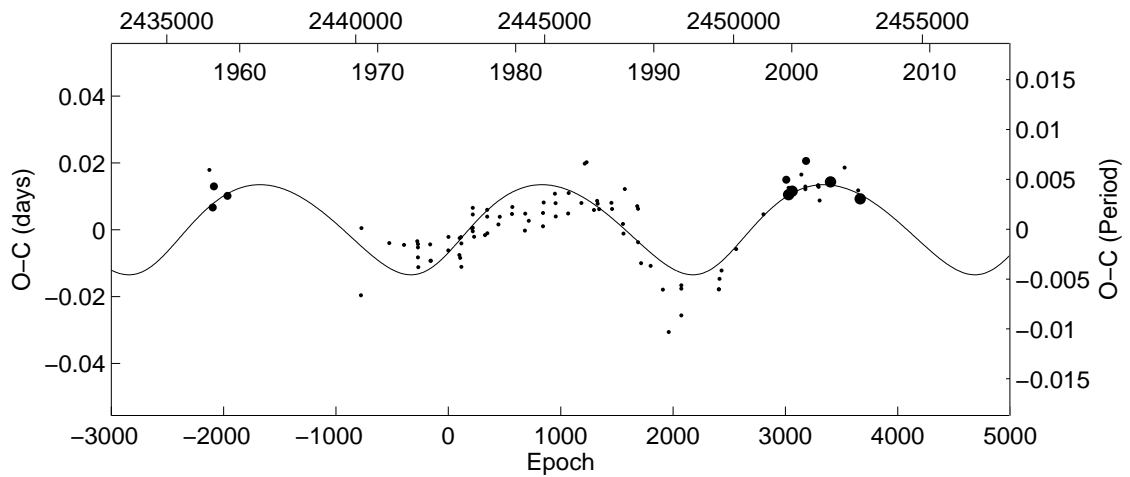}}
   \caption{An $O-C$ diagram of SS~Cet. The description is the same as
   in the previous figures.}
   \label{FigSSCet}
\end{figure}

Using the mass of SS~Cet derived by \cite{Narasaki1994} $M_{12} = 2.75$\Mo, one gets the minimal mass of the
third component about 0.72\Mo. The predicted value of the third light is only about 1 \% and the third component
would be also similar to the secondary component in the spectra. Detection of the third body in the light curve
as well as in the spectra is a difficult task. The distance to the system was derived by \cite{Narasaki1994},
resulting in $d=486$~pc. According to this value and derived parameters of the third body, one could calculate
the predicted angular separation of the third component about only 23~mas and magnitude difference about 5~mag.
Such a large magnitude difference and low separation is hardly detectable, and the third body remains
undetectable also by interferometry.

\section{TY~Del}

Another EB with apparently variable period is TY~Del (AN~141.1935,
BD+12~4539), spectrum classified as B9+G0IV 
(\citealt{Hoffman2006}) and relative brightness about 10.1~mag in
V filter. There is a consensus about the spectral types of the
components of TY~Del, but there is a difference between the
masses. \cite{Brancewicz1980}, and after then also
\cite{Budding1984} and \cite{Budding2004} have presented the
masses $M_1 = 5$\Mo, $M_2 = 2$\Mo, while \cite{Svechnikov1990}
presented $M_1 = 2.8$\Mo, $M_2 = 0.84$\Mo, what is more likely to
the proposed spectral types.

\begin{figure}
   \centering
   \scalebox{0.75}{
   \includegraphics{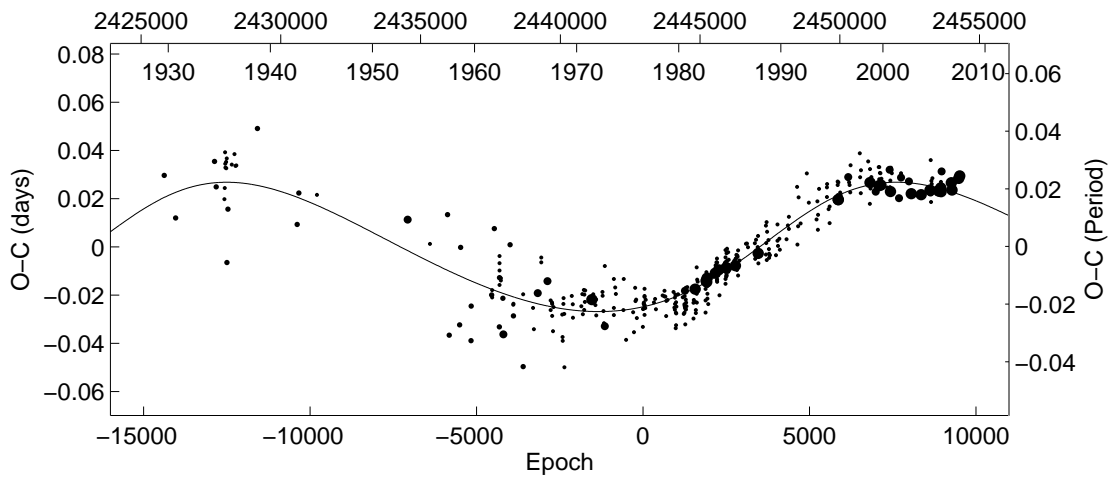}}
 \caption{An $O-C$ diagram of TY~Del. The description is the same as
 in the previous figures.}
         \label{FigTYDel}
\end{figure}

\begin{table*}[b]
 \caption{The final results: TY~Del.}
 \label{TableTYDel} \centering
 \scalebox{0.87}{
 \begin{tabular}{c c c c c c c c | c c }
 \hline \hline
   Parameter&  $JD_0 $   &      $P$      & $p_3$   & $T_0$    &$\omega$ &   $e$    &  $A$      & $f(M_3)$  & $M_{3,min}$\\
   Unit     & $[$HJD$]$  &   $[$day$]$   &$[$yr$]$ &$[$HJD$]$ &$[$deg$]$&          & $[$day$]$ & [\Mo]     & [\Mo]    \\\hline
   Value   &$2442959.471$&  $1.1911264$  & $64.9$  &$2449200$ & $38 $   & $0.22$   &$0.027$    &  $0.025$  & $0.79$    \\
   Error   &$\pm 0.001$  &$\pm 0.0000002$&$\pm 2.3$&$\pm 1000$&$\pm 18$ &$\pm 0.06$&$\pm 0.002$&$\pm 0.001$&$\pm 0.07$ \\
 \hline \hline
 \end{tabular}}
\end{table*}

The star was discovered to be a variable by \cite{Hoffmeister1935}. There was only one attempt to observe the
whole light curve of TY~Del photoelectrically by \cite{Faulkner1983}, unfortunately only about half of the curve
was observed. No analysis of these data was carried out. The star was also studied by \cite{Cook1993} on the
basis of his visual observations for the long-time scale intrinsic variations, but the results are not very
conclusive.

The spectroscopic observations in H$\alpha$ were done by
\cite{Vesper2001}. They conclude that there is no activity in
H$\alpha$ and no evidence for the mass transfer structures was
found in this system.

Altogether 370 times of minima were collected, from which only 5
were omitted due to their large scatter. One period of the third
body is already sufficiently covered by data points, see Fig.
\ref{FigTYDel}, but further observations are still needed. The
last one data point was observed at Ond\v{r}ejov observatory. From
the LITE parameters (see Table \ref{TableTYDel}) and with the
approximate masses of the individual components of the eclipsing
binary $M_1 = 2.8$~\Mo, $M_2 =
0.84$~\Mo~(\citealt{Svechnikov1990}) one is able to derive the
minimal mass of the third component $M_{3,min} = 0.67$~\Mo. Due to
lack of any other observations also this hypothesis cannot be
proved. The spectral types and masses were derived only on the
basis of the photometric indices and are not very conclusive. The
spectroscopic analysis, as well as the analysis of the light curve
of the system is needed, but the third light is undetectable in
the light-curve solution. Regrettably the star was not measured by
Hipparcos, so the distance is not known and one cannot derive the
predicted angular separation of the third component. As one can
see, there is some additional variation besides LITE, which is not
strictly periodic, see Section \ref{Diskuse} for details.

\section{RR~Dra}

Another eclipsing binary which exhibits apparent period changes is
RR~Dra (AN~188.1904). It is an Algol-type semidetached binary,
relative brightness about 9.8~mag in V, spectrum classified as
A2+G8IV \citep{Svechnikov1990}, while \cite{Yoon1994} 
classified a little bit later spectral type of secondary K0.
\cite{Svechnikov1990} presented the masses $M_1 = 2.15$\Mo~ and
$M_2 = 0.6$\Mo. The primary minimum is very deep, about 3.5~mag
and the orbital period is about 2.8 days.

\begin{figure}[b!]
   \centering
   \scalebox{0.75}{
   \includegraphics{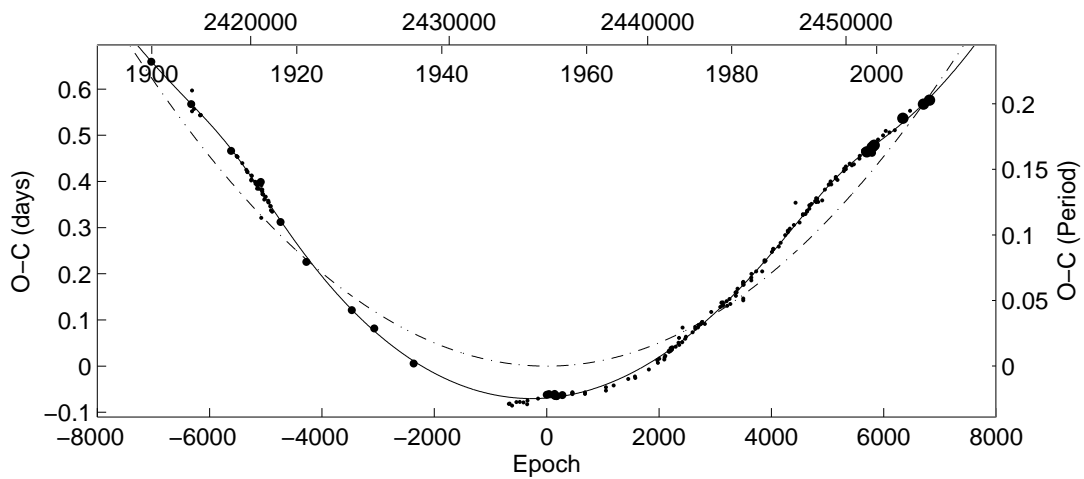}}
   \caption{An $O-C$ diagram of RR~Dra. For the plot where the
   quadratic term was subtracted see Fig.\ref{FigRRDra2}.}
   \label{FigRRDra1}
\end{figure}

The star was discovered to be a variable by Miss
\cite{Ceraski1905}. The minimum is so deep that also visual
observers could provide reliable observations. That is the reason
why most of the collected times of minima are the visual ones (193
out of 219). \cite{Kreiner1971} collected all available minima for
the period analysis. The long-time increase of the period is
evident from his $O-C$ diagram (due to the mass transfer between
the components?). The most recent period study of this system was
performed by \cite{Qian2002}, who considered (besides the mass
transfer) the abrupt period jumps - altogether 8 jumps were
introduced to describe the $O-C$ diagram in detail. Almost the
same goodness of fit could be reached by applying the LITE
hypothesis besides the mass transfer.

\begin{figure}[b!]
   \centering
   \scalebox{0.75}{
   \includegraphics{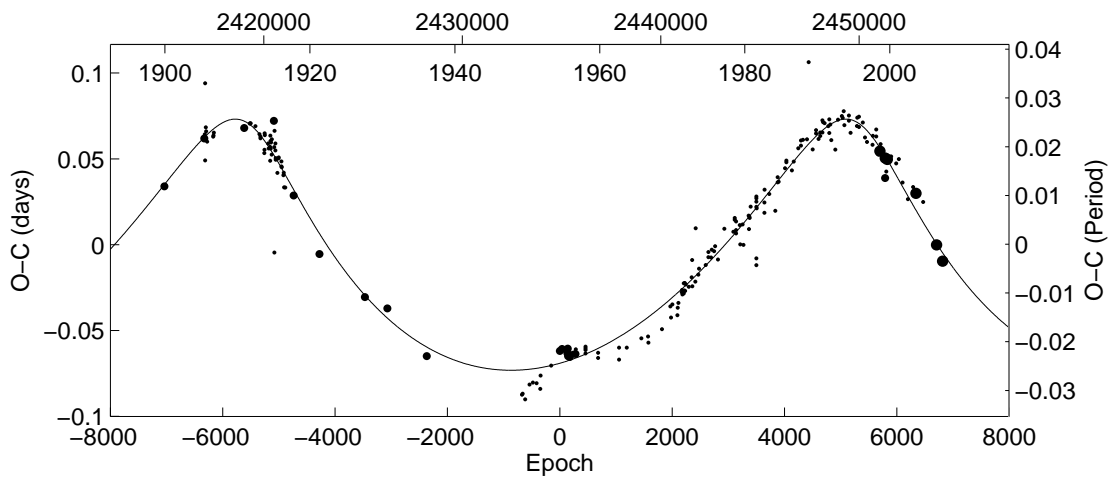}}
   \caption{An $O-C$ diagram of RR~Dra after subtraction of the quadratic term.}
   \label{FigRRDra2}
\end{figure}

\begin{table*}[t!]
 \caption{The final results: RR~Dra.}
 \label{TableRRDra} \centering
 \scalebox{0.77}{
 \begin{tabular}{c c c c c c c c c | c c }
 \hline \hline
   Parameter& $JD_0 $    &      $P$      &  $p_3$   &  $T_0$   &$\omega$ &   $e$    &  $A$      &        $q$       &  $f(M_3)$ & $M_{3,min}$\\
   Unit     &$[$HJD$]$   &   $[$day$]$   & $[$yr$]$ & $[$HJD$]$&$[$deg$]$&          & $[$day$]$ &$[10^{-10}$ day$]$&  [\Mo]    & [\Mo]     \\\hline
   Value   &$2434913.728$&  $2.8312140$  & $84.3$   & $2450100$&$110$    & $0.50$   &$0.073$    &    $-126.2$      & $0.300$   &  $1.85$   \\
   Error   &$\pm 0.022$  &$\pm 0.0000053$&$\pm 0.6$ &$\pm 400 $&$\pm 4 $ &$\pm 0.03$&$\pm 0.002$&     $0.2$        &$\pm 0.002$& $\pm 0.09$\\
 \hline \hline
 \end{tabular}}
\end{table*}

Altogether 219 times of minima were used for the analysis. One new
observation of minimum was obtained by M.Wolf at Ond\v{r}ejov
observatory. The $O-C$ plot is in Fig.\ref{FigRRDra1}, where LITE
and the quadratic term were plotted together. In the next figure
only LITE is shown, see Fig.\ref{FigRRDra2}. As one can see, the
period increase is very rapid, and the amplitude of LITE is still
quite high. This leads to the relatively high mass function, which
results in high predicted minimal mass $M_{3,min} = 1.85$\Mo. This
is larger than the secondary and in the light-curve solution as
well as in the spectrum will be probably observable. Regrettably
no such analysis was performed.

The quadratic term coefficient $q = (126.2 \pm 0.1) \cdot 10^{-10}$~day leads to the period change about $3.26
\cdot 10^{-6}$~day/yr. From this value the conservative mass transfer rate could be derived $\dot M = 3.5 \cdot
10^{-7}$~\Mo/yr. This relatively high value of mass transfer rate arises from the very rapid period change,
which was attributed to the quadratic ephemeris. The spectroscopic observations during the primary eclipse made
by \cite{Kaitchuck1985} indicate the possible presence of a transient accretion disc in the system. The presence
of such a disc also supports the hypothesis of mass transfer in the system.

For the estimation of the angular separation of the third body and the astrometric confirmation of the LITE
hypothesis the distance to the system has to be known. The star was not measured by Hipparcos and the distance
is not known precisely. The only information about the distance is from \cite{Kharchenko2001}, where is
introduced a surprisingly inaccurate value of the parallax $\pi = (0.40 \pm 11.50)$~mas. Distance with such a
large error is useless for the estimation of the angular separation of the predicted component.

\section{TZ~Eri}

The system TZ~Eri (AN~40.1929, BD-06~880) is an EB with an orbital period of about 2.6 days, apparent brightness
of about 9.7~mag in V filter. It has a deep primary minimum (about 2.8~ mag), so the visual observations could
be also reliable.

\begin{figure}[t!]
   \centering
   \scalebox{0.75}{
   \includegraphics{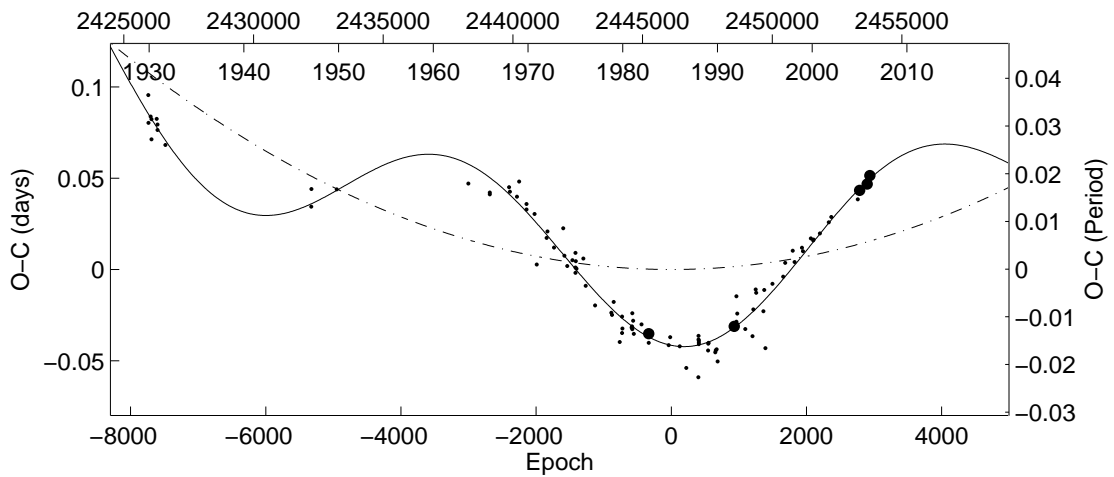}}
 \caption{An $O-C$ diagram of TZ~Eri.}
         \label{FigTZEri}
\end{figure}

Its variability was discovered by \cite{Hoffmeister1929}, who also recognized the system to be an Algol-type.
The spectral type was first classified by Miss \cite{Cannon1934} as F. Later the spectrum was re-classified as
A5/6~V (primary) and K0/1~III (secondary) by \cite{Barblan1998}. In this later paper the light-curve
observations in the Geneva 7-colour photometric system were analyzed together with the radial-velocity curves of
both components. Wilson-Devinney code was used, resulting in a set of parameters describing both components. For
our analysis are the most important the masses, $M_1 = 1.97$~\Mo~ and $M_2 = 0.37$~\Mo.

\begin{table*}[b!]
 \caption{The final results: TZ~Eri.}
 \label{TableTZEri} \centering
 \scalebox{0.77}{
 \begin{tabular}{c c c c c c c c c | c c }
 \hline \hline
    Parameter& $JD_0 $   &      $P$      &  $p_3$   &  $T_0$   &$\omega$ &   $e$    &  $A$      &        $q$       &  $f(M_3)$ & $M_{3,min}$\\
   Unit     &$[$HJD$]$   &   $[$day$]$   & $[$yr$]$ & $[$HJD$]$&$[$deg$]$&          & $[$day$]$ &$[10^{-10}$ day$]$&  [\Mo]    & [\Mo]    \\\hline
   Value   &$2446109.730$&  $2.6061129$  &$48.8$    &$2451100$ & $ 0 $   & $0.01$   &$0.042$    &    $-18.0 $      & $0.165$   &  $1.3$   \\
   Error   & $\pm 0.009$ &$\pm 0.0000034$&$\pm 6.8$ &$\pm 2300$& $\pm 40$&$\pm 0.10$&$\pm 0.014$&     $0.2$        &$\pm 0.013$& $\pm 0.1$ \\
 \hline \hline
 \end{tabular}}
\end{table*}

There were also several studies about the presence of the accretion disc in the system (e.g.
\citealt{Kaitchuck1982}, \citealt{Kaitchuk1988}, \citealt{Vesper2001}). This disc as well as mass transfer from
the secondary to the primary is in agreement with our result about the increasing orbital period (see below).
The system was also included in the sample of Algol-type binaries with radio emission \citep{Umana1998}. The
star was also investigated according to the possible connection between the orbital and pulsational periods, see
\cite{Soydugan2006}.

The analysis of the long-term period changes was done with a set of 108 observations (mostly the visual ones).
The resultant $O-C$ diagram is in Fig.\ref{FigTZEri} and the parameters of the predicted LITE are in Table
\ref{TableTZEri}. The minimal mass of the third component results in $M_{3,min} = 1.3$~\Mo, or the spectral type
F6 (according to \citealt{Hec1988}). Such a body could be evident in the light-curve solution as well as in the
spectra of TZ~Eri. Regrettably, there was no attempt to detect such a body during the detailed analysis by
\cite{Barblan1998}. The long-term period increase is due to the mass transfer from the secondary, with the
conservative mass-transfer rate $\dot M = 6.2 \cdot 10^{-8}$\Mo/yr.

Despite the fact the star was not observed by Hipparcos,
\cite{Barblan1998} estimated the photometric distance to $d = (270
\pm 12)$~pc. Assuming the coplanar orbit of the third component,
then $M_3 = M_{3,min}$ and one could calculate the predicted
angular separation of the third body to $a = 77$~mas and magnitude
difference about 1.7~mag. Such a component is detectable with the
modern stellar interferometers.

\section{RV~Per}

The system RV~Per (AN~61.1905, BD+33~805, HD~279552) is an EB with an orbital period of about 2 days, spectral
type classified as A2+G7IV, according to \cite{Svechnikov1990}. Also this star shows deep primary minimum, about
2.4~mag.

\begin{figure}[b!]
  \centering
  \scalebox{0.75}{
  \includegraphics{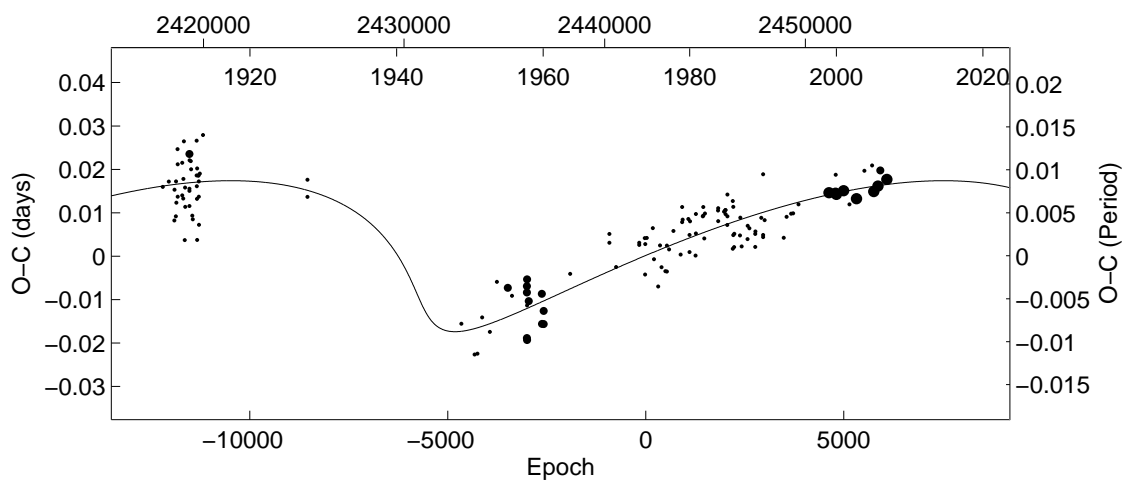}}
  \caption{An $O-C$ diagram of RV~Per.}
  \label{FigRVPer}
\end{figure}

The star was discovered to be a variable by \cite{Blazko1907}. Since then only a few papers on this star were
published, so our knowledge about the system is very limited. According to \cite{Brancewicz1980} the masses of
the individual components are $M_1 = 3.04$~\Mo~ and $M_2 = 0.46$~\Mo. The attempts to prove the existence of the
accretion disc in the system \citep{Kaitchuck1985} were not successful, as well as the presence of the pulsating
component in the system was not confirmed \citep{Kim2003}.

\begin{table*}[t]
 \caption{The final results: RV~Per.}
 \label{TableRVPer} \centering
 \scalebox{0.87}{
 \begin{tabular}{c c c c c c c c | c c }
 \hline \hline
  Parameter&  $JD_0 $  &      $P$      & $p_3$    & $T_0$    &$\omega$ &   $e$    &  $A$      & $f(M_3)$  & $M_{3,min}$\\
  Unit     & $[$HJD$]$  &   $[$day$]$   &$[$yr$]$  &$[$HJD$]$ &$[$deg$]$&          & $[$day$]$ & [\Mo]     & [\Mo]  \\\hline
  Value   &$2442046.920$&  $1.9734888$  &$99.8$    &$2431600$ & $210$   & $0.79$   &$0.017$    & $0.007$   & $0.47$  \\
  Error   & $\pm 0.002$ &$\pm 0.0000003$&$\pm 18.8$&$\pm 7200$& $\pm 44$&$\pm 0.46$&$\pm 0.003$&$\pm 0.002$&$\pm 0.07$\\
 \hline \hline
 \end{tabular}}
\end{table*}

The period changes were first studied by \cite{Wood1950}, but the data set was not sufficient to do any
satisfactory conclusions. In the present thesis the data set consists of 146 times-of-minima observations (see
Fig.\ref{FigRVPer}). One new minimum was observed by M.Zejda. The parameters of LITE are in Table
\ref{TableRVPer}. As one can see, the period of the third body is not covered by observations yet and the errors
of the individual parameters are high. The value of eccentricity could be even higher, but there are no
observations near the periastron passage, and the next one is predicted to occur near 2040. Only further
observations would confirm or reject the third-body hypothesis.

\section{UZ~Sge}

The Algol-type EB system UZ~Sge (AN~435.1936) has an orbital period of about 2.2 days and spectral type
classified as A3V+G0IV \citep{Svechnikov1990}.

Its photometric variability was discovered by \cite{Guthnick1939}.
Since then there was no attempt to do any detailed analysis,
neither the photometric nor the spectroscopic one. The only
spectroscopic observation was done by \cite{Halbedel1984} for
derivation of the spectral type of primary component. 

\begin{table*}[b]
 \caption{The final results: UZ~Sge.}
 \label{TableUZSge} \centering
 \scalebox{0.87}{
 \begin{tabular}{c c c c c c c c | c c }
 \hline \hline
   Parameter&  $JD_0 $   &      $P$      & $p_3$    & $T_0$    &$\omega$ &   $e$    &  $A$      & $f(M_3)$  & $M_{3,min}$\\
   Unit     & $[$HJD$]$  &   $[$day$]$   &$[$yr$]$  &$[$HJD$]$ &$[$deg$]$&          & $[$day$]$ & [\Mo]     & [\Mo]  \\\hline
   Value   &$2445861.420$&  $2.2157425$  &$47.0$    &$2449200$ & $294$   & $0.28$   &$0.023$    & $0.031$   & $0.65$    \\
   Error   &$\pm 0.002$  &$\pm 0.0000007$&$\pm  2.4$&$\pm 2000$& $\pm 47$&$\pm 0.13$&$\pm 0.003$&$\pm 0.002$&$\pm 0.05$ \\
 \hline \hline
 \end{tabular}}
\end{table*}

Altogether 122 measurements of times of minima were found in literature, but 14 observations were neglected.
Four new observations of minima were obtained (two of them by L. \v{S}melcer, one by M.Wolf and one by author).
If the masses of the individual components were taken from \cite{Svechnikov1990},
$M_1 = 2.05$~\Mo~and $M_2 = 0.29$~\Mo, 
then the minimal mass of the third component results in $M_{3,min} = 0.65$~\Mo. But due to absence of any
detailed analysis of this system, this value cannot be proved.

\begin{figure}[t]
   \centering
   \scalebox{0.75}{
   \includegraphics{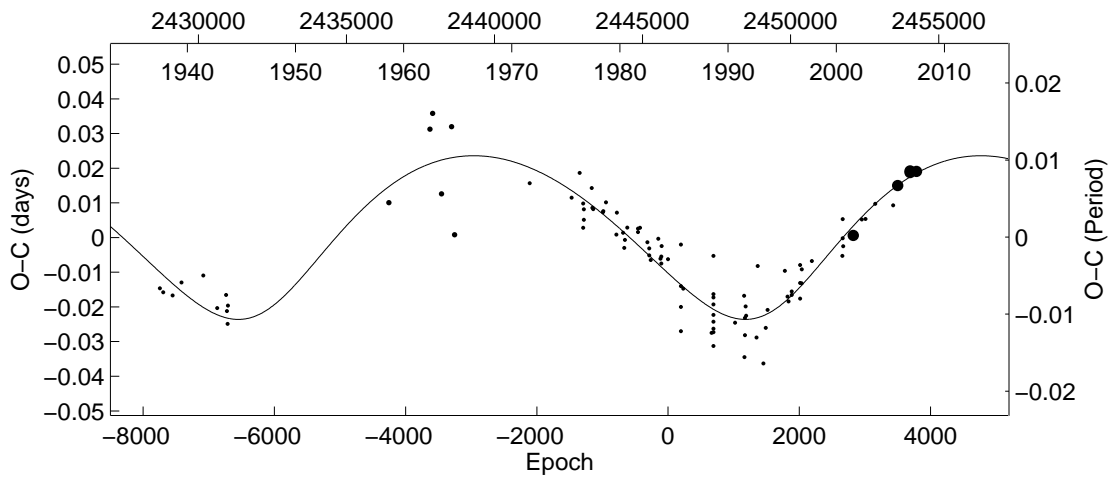}}
   \caption{An $O-C$ diagram of UZ~Sge.}
   \label{FigUZSge}
\end{figure}

\section{BO~Vul}

The last EB system in this thesis which shows long-term period changes is BO~Vul (AN~125.1935, HD~345287). It is
an eclipsing binary with an orbital period of about 1.9 days, apparent brightness 10~mag, depth of primary
minimum about 1.6~mag and spectral type F0+G0IV \citep{Svechnikov1990}.

\begin{table*}[b!]
 \caption{The final results: BO~Vul.}
 \label{TableBOVul} \centering
 \scalebox{0.77}{
 \begin{tabular}{c c c c c c c c c | c c }
 \hline \hline
   Parameter&  $JD_0 $   &      $P$      &  $p_3$   &  $T_0$   &$\omega$ &   $e$    &  $A$      &        $q$       &  $f(M_3)$ & $M_{3,min}$\\
   Unit     &$[$HJD$]$   &   $[$day$]$   & $[$yr$]$ & $[$HJD$]$&$[$deg$]$&          & $[$day$]$ &$[10^{-10}$ day$]$&  [\Mo]    & [\Mo]    \\\hline
   Value   &$2441163.509$&  $1.9458790$  &$42.2$    &$2446900$ &$ 0 $    & $0.33$   &$0.024$    &    $ 25.5 $      & $0.049$   &  $0.73$  \\
   Error   &$\pm 0.002$  &$\pm 0.0000006$&$\pm 1.3$ &$\pm 1800$&$\pm 18$ &$\pm 0.10$&$\pm 0.002$&     $0.1$        &$\pm 0.003$& $\pm 0.04$\\
 \hline \hline
 \end{tabular}}
\end{table*}

The star was observed to be a variable by \cite{Hoffmeister1935}
and the first ephemeris were presented by \cite{Guthnick1936}. The
first observation of the whole light curve was carried out by
\cite{Nassau1939}, where also a brief analysis of the system was
presented. Since then there was no detailed analysis of the system
performed.

The changes of its period were firstly mentioned by
\cite{Ahnert1973}. After then also \cite{Baldwin1996} published
new revised elements for BO~Vul, but without any interpretation of
the period changes.

\begin{figure}[t!]
   \centering
   \scalebox{0.75}{
   \includegraphics{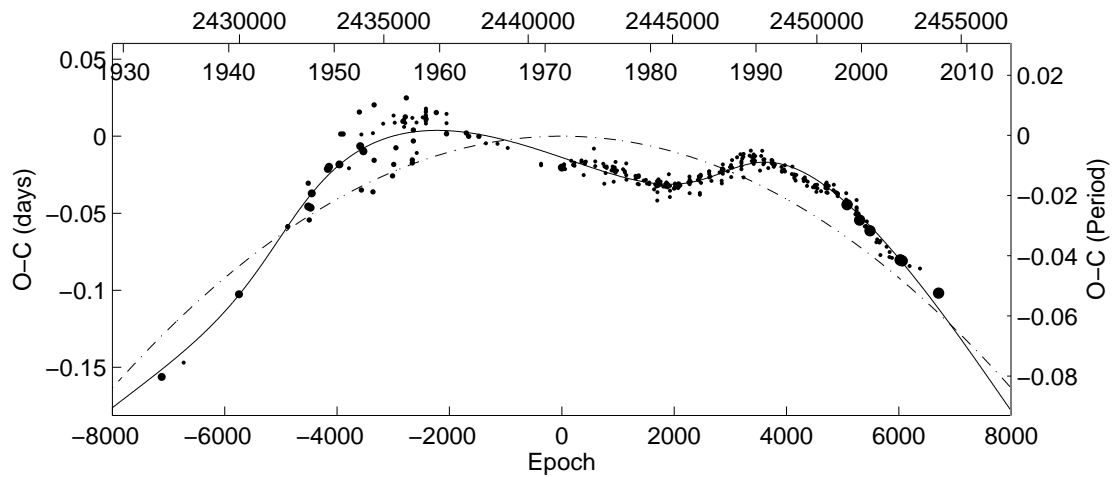}}
   \caption{An $O-C$ diagram of BO~Vul.}
   \label{FigBOVul}
\end{figure}

Altogether 390 times of minima were collected, but only 360 were used for this analysis. One new minimum was
observed by M.Wolf. For the final results see Fig.\ref{FigBOVul} and Table \ref{TableBOVul}. If the masses from
\cite{Svechnikov1990} were taken, $M_1 = 1.45$~\Mo~and $M_2 = 0.64$~\Mo, then the minimal mass of the third
component results in 0.73~\Mo. But as in the previous cases, there is no detailed analysis, which could prove
this result. The quadratic term leads to the conservative mass-transfer rate of about $\dot M = 1.3 \cdot
10^{-7}$~\Mo/yr.

One can also see some additional non-periodic changes, which are evident since 1970's and which could not be
described by applying only LITE and the quadratic term. The amplitude of these variations is smaller than the
amplitude of LITE, but one cannot doubt about their presence nowadays. These could be caused by the abrupt
period changes, or due to magnetic activity cycles presented in the system. See the next section for a brief
analysis.

According to \cite{Brancewicz1980} the distance to the system is about 233~pc, but the star was not measured by
Hipparcos, so this value is not very reliable (there is no information about the error of this value). From the
distance one could estimate the predicted angular separation of the third component, resulting in 63~mas, and
magnitude difference about 3.4~mag. Such body is perhaps marginally detectable by the modern stellar
interferometers.

\section{Alternative explanation}

One can also see additional non-periodic variations in some of the $O-C$ diagrams, which could not be described
by applying only the LITE hypothesis. In Figs. \ref{Figs} there are shown four cases with the most evident
variations. The amplitudes of these variations are usually about 10 minutes in the $O-C$ diagram and are not
strictly periodic (the "periods" are from 5 to 20 years). This could be caused by the presence of stellar
convection zones and magnetic activity cycles in an agreement with so-called Applegate's mechanism, see e.g.
\cite{Applegate1992}, \cite{Lanza1998}, or \cite{Hoffman2006}. The effect could play a role, because the
spectral types of the secondary components in most of the systems are later than F5 (see \cite{Zavala2002} for a
detailed analysis). This explanation would clarify the non-periodicity and the changes in amplitude of such
variation, as well as why in some binaries this phenomena is presented, while in some others not.

Due to missing information about the properties of the eclipsing
components in most of the systems one also cannot estimate the
predicted variation of the quadruple moment $\Delta Q$, which
causes the period variations. This value could be computed from
the equation
$$\frac{\Delta P}{P} = -9 \frac{\Delta Q}{M a^2},$$ where P is the
orbital period of the system, 
$M$ is a mass of the star and $a$ is the separation between the
components, see  \cite{LanzaRodono2002}. $\Delta P$ is the
amplitude of the period oscillation and could be computed from the
LITE parameters from equation
$$\Delta P = A \cdot \sqrt{2[1-\cos(\frac{2\pi P}{p_3})]},$$ see
\cite{Rovithis2000}. The typical values of $\Delta Q$ are of the
order of $10^{51} - 10^{52} \mathrm{g \cdot cm^2}$
\citep{LanzaRodono1999}.

Due to missing light-curve and radial-velocity curve analysis, the value $a$ is missing. For the few cases where
this value is known (RY~Aqr, SS~Cet and TZ~Eri) only RY~Aqr does not satisfy the condition about the limits for
$\Delta Q$ (being about 10 times lower). This result does not indicate that the magnetic activity cycles are not
presented in this system, but only the fact that this effect cannot be used as an alternative explanation of the
period changes. The effect could be present in addition to the light-time effect and describe the non-periodic
variations (shown in Figs. \ref{Figs} after subtraction of LITE).

To conclude, for the better description of the observed period
variations of these systems, the magnetic activity cycles could be
presented together with the LITE. On the other hand one has to
take into consideration that the spectral types of most of these
binaries were derived only on the basis of their photometric
indices \citep{Svechnikov1990} and are not very reliable.

\begin{figure}[t]
 \scalebox{0.88}{
 \includegraphics[16mm,190mm][30mm,270mm] {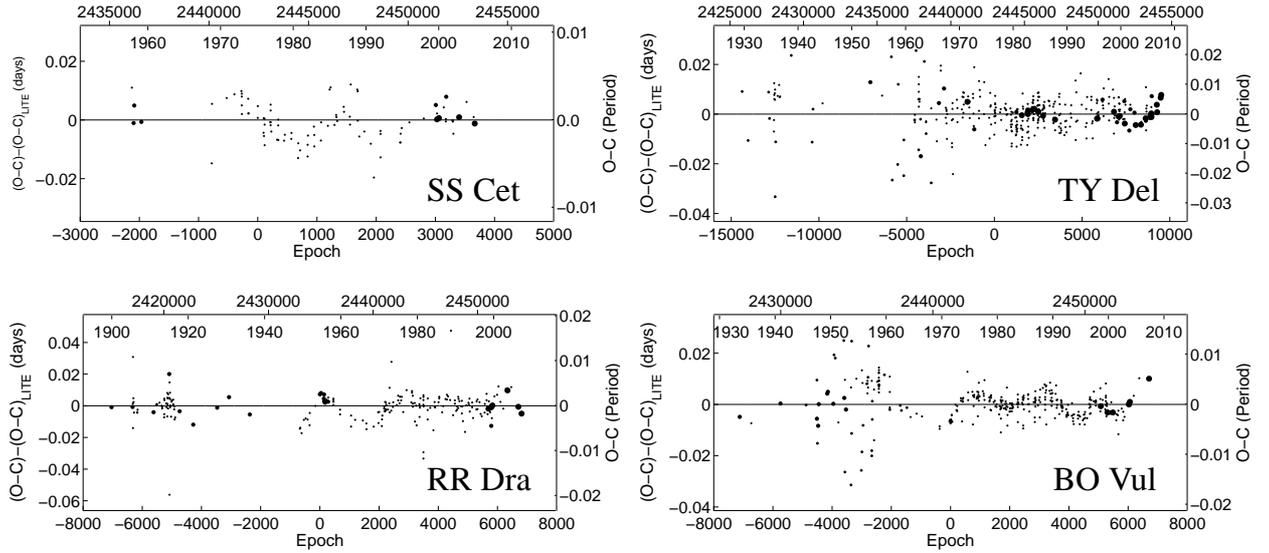}}
 \caption{The $O-C$ diagrams of four systems after subtraction
 of the LITE and the quadratic term (cases RR~Dra and
 BO~Vul). The additional variation is clearly visible.}
 \label{Figs}
\end{figure}

\section{Brief summary} \label{Diskuse}

Eleven Algol-type semidetached eclipsing binaries were analyzed
for the presence of LITE on the basis of their $O-C$ diagram
analysis and the times-of-minima variations. A few new
observations of these systems were obtained. All of the systems
above show apparent changes of their orbital periods, which could
be explained as a result of orbiting the EB around the common
center of mass with the third component.

Such a variation usually has a period on the order of decades. The light-time effect was applied as a main cause
of these changes (as one can see from Figs.\ref{FigRYAqr}--\ref{FigBOVul}). In four cases (RW~Cap, RR~Dra,
TZ~Eri and BO~Vul) also the quadratic term in the light elements was used. This could be described as a mass
transfer between the components, which could play a role, because all of the systems are semidetached ones. In
some cases also the proof of presence of mass-transfer structures or accretion discs were revealed by
spectroscopy.

Regrettably, in most of the systems no detailed analysis (neither the photometric nor the spectroscopic) was
carried out. The spectral types and the masses of the individual components in most of the systems are only
approximate, so the parameters of the predicted third bodies are also affected by relatively large errors. Due
to missing information about the distances to most of these binaries also the prediction about the angular
separation could not be done. As one can see, only further detailed photometric, as well as spectroscopic and
interferometric analysis would reveal the nature of the system and confirm or refute the third-body hypothesis.

\chapter{Systems with combined LITE and astrometry} \label{Ch3}

The crucial part of this thesis was the analysis of the systems,
where the EB pair is a component of spatially resolved binary.
Despite increasing number of the visual as well as eclipsing
binaries, the intersection of these two sets is still only very
limited.

Finding appropriate candidates for this analysis turned out to be quite difficult. One of the problems was the
data set and its quality. Such systems have to satisfy the following adopted conditions: 1. More than 10 times
of minima and more than 10 astrometric observations are available. 2. The observed range of the position angle
in the astrometric measurements is larger than 10$^\circ$. The limit for the number of data points was accepted
because of the number of parameters, which have to be found. There were 5 parameters for LITE and 2 ephemeris,
or 7 for astrometry, it means at least 7 data points (in both methods) is the absolute minimum for the analysis.
And the limit for the range of data points was the accuracy of the fit. This means fitting the linear part of
the astrometric orbit (or the $O-C$ diagram) is useless for the parameter determination.

The astrometric measurements were adopted from "The Washington Double Star Catalog" (hereafter WDS
\footnote{$\mathrm{http://ad.usno.navy.mil/wds/}$}), which incorporates a huge database of astrometric
observations, but only a small one about the properties of the individual components in these systems. There was
a problem with identifying the eclipsing binaries in them. Altogether more than 13800 systems in the northern
and southern sky have been inspected. This was the number of systems in WDS with 10 or more astrometric
observations. From this large number of stars only 31 were eclipsing binaries (according to Simbad database).
And from this 31 stars in most cases there were no or only a little data set in times of minima. Systems with
larger times-of-minima data sets which have been analyzed are presented below in this chapter. In the next
section is the brief survey of other systems.

In a few cases the astrometry and the behaviour of times of minimum light were studied, but these two approaches
were usually analyzed separately. Such systems are for example 44~Boo, QZ~Car, SZ~Cam, or GT~Mus (besides the
systems mentioned and analyzed below). The coverage of the astrometric orbit is very poor for some of them. For
SZ~Cam only a few usable astrometric observations were obtained, but the LITE is well-defined and also the third
light in the light-curve solution was detected \citep{Lorenz1998}. QZ~Car is a more complicated, probably
quintuple system - the bright component of the visual binary consists of two eclipsing pairs (P = 20.7~d and
6.0~d). There are also only a few usable astrometric measurements. Also GT~Mus is a quadruple system, consisting
of an eclipsing and RS~CVn component. In many other cases, only measurements of the times of minima are
available, without astrometry. For some others, astrometry without photometry, is only available. Other systems
where astrometric observations were obtained and the LITE is observable or expected are listed in
\cite{Mayer2004}.

The only paper on combining the two different approaches (LITE and astrometry) into one joint solution is that
by \cite{Ribas2002}, where a similar method (but not the same) as described in this thesis was applied to the
system R~CMa, but where only a small arc of the astrometric orbit was available. Besides the astrometry and LITE
also the proper motion on the long orbit was analyzed. On the other hand one has to note, that in
\cite{Ribas2002} the complete astrometric parameters (with proper motions, parallax, etc.) were used, while in
this thesis only the relative astrometry of the distant body relative to the eclipsing pair was analyzed. From
this point of view such an approach to the combination of LITE and astrometry is unique and has never been
published before. Generally, such a combined approach is potentially very powerful, especially in upcoming
astrometric and photometric space missions.

Three of the systems presented below in this chapter (namely VW~Cep, $\zeta$~Phe and HT~Vir) were selected for
publication, see \cite{ZascheWolfPrint} (in print)

\section{QS~Aql}

The first investigated system is QS~Aql (KUI 93, HD 185936, HR 7486, HIP 96840). It is an Algol-type eclipsing,
and also spectroscopic, binary with a period of about $2.5$ days. Its apparent brightness is about 6.0~mag and
the spectral type was classified as B5V (according to \citealt{Holm}).

The star was recognized to be a variable by Dr.Plaskett from The Dominion Astrophysical Observatory from
photographic plates taken in 1924 and 1925 \citep{Millman1928}. The first photometric observations were obtained
by \cite{Guth}. Surprisingly, only 17 times of minima were recorded since then. This is probably due to the
relatively high brightness of the object, which would saturate most telescopes with CCD detectors (the last one
is from Hipparcos).

\begin{figure}[b!]
   \centering
   \scalebox{0.75}{
   \includegraphics{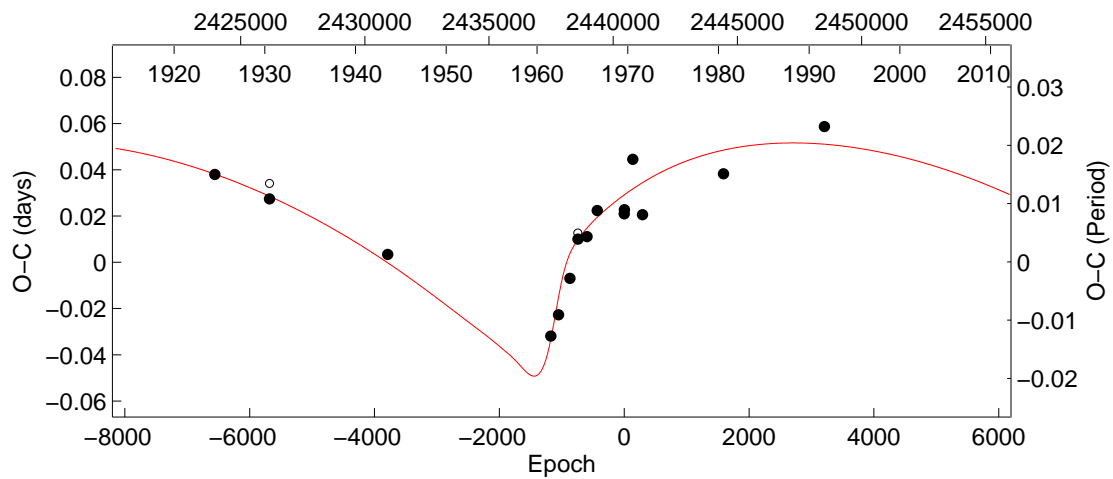}}
   \caption{An $O-C$ diagram of QS~Aql. The individual observations are
   shown as dots (primary) and open circles (secondary) and the curve
   represents the predicted LITE. All of the measurements are
   photoelectric or CCD ones.}
   \label{FigQSAql-OC}
\end{figure}

\cite{Guth} recognized the eclipsing nature of the star. Some 40
years later, \cite{Knipe} discovered a rapid period change, which
occurred at about 1964 (his suggestion) and was caused by the
periastron passage in the wide orbit around the barycenter. The
period change was so rapid that the eccentricity of the wide orbit
must be very high. Unfortunately, during the last decade no
minimum time was obtained, the last one is more than 15 years old.

The first astrometric observations were secured more than 50 years ago, but their accuracy is questionable.
Because both visual components are similarly bright, there could be confusion in the identification of the
primary and the secondary, and some measurements may be shifted for about $180^\circ$ in the position angle.
Especially due to this reason, we have neglected all measurements obtained before 1975. More recent data are
more reliable and more precise (since 1976 the observations are mostly speckle interferometric).

The only paper considering the astrometry together with the LITE
was published by \cite{Mayer2004}. The system QS~Aql is presented
there as one of the systems where LITE could be observed together
with the astrometric orbit. Also the warning regarding the quality
of the old data and their $180^\circ$ ambiguity is given there.

\begin{figure}[t!]
   \centering
   \includegraphics[width=9cm]{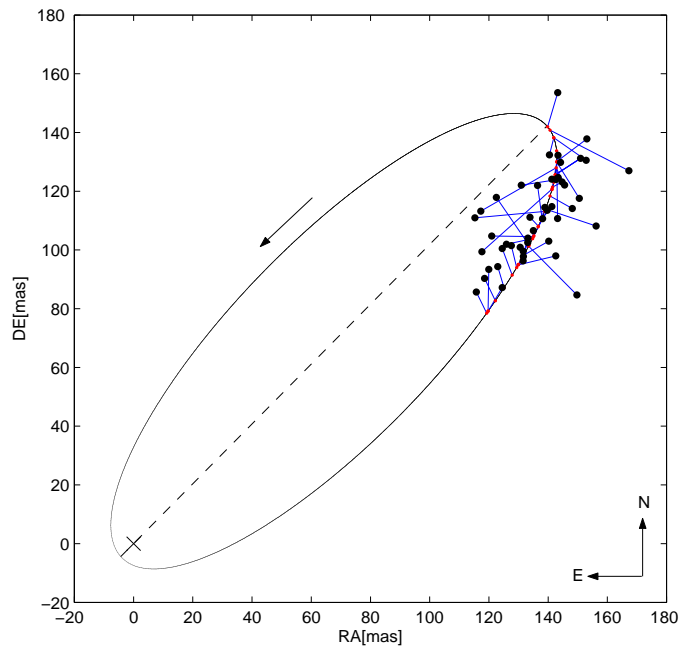}
   \caption{A relative astrometric orbit of a system QS~Aql.
   Older measurements were neglected. The points represent
   individual observations (black dots), while the solid curve corresponds to the
   solution described in the text and parameters in Table \ref{QSAql-Tab}.
   The straight lines connect individual observations with their
   expected positions on the fitted orbit (red dots). The cross indicates
   the position of the eclipsing binary on the sky, the
   arrow indicates the direction of the movement on the third-body orbit, and the
   dashed line represents the line of apsides.}
   \label{FigQSAql-orbit}
\end{figure}

In Table \ref{tabQSAqlMin}, the observed times of minima of QS~Aql
and the corresponding epochs relative to the ephemeris given in
Table \ref{QSAql-Tab} can be found. The algorithm presented in the
introduction was used to analyze this system combining the
astrometry and the times-of-minima analysis. The resultant
parameters of the distant-body orbit are presented in Table
\ref{QSAql-Tab}. In Figs. \ref{FigQSAql-OC} and
\ref{FigQSAql-orbit}, the $O\!-\!C$ diagram of the times of
minima, and the astrometric orbit are shown, respectively. The
curves in both figures show the model fit corresponding to the
resultant parameters given in Table \ref{QSAql-Tab}.

\begin{table}[t!]
\caption{The parameters of QS~Aql.} \label{QSAql-Tab} \centering
\scalebox{0.75}{
\begin{tabular}{c c c c c c c c c c| c c}
 \hline \hline
 Parameter &   $JD_0$     &     P      & $p_3$ &  $A$  &  $T_0$ & $\omega$ & $e$ &  $i$ & $\Omega$& $f(M_3)$ & $M_3$ \\
  Unit     &    [day]     &   [day]    & [yr] & [day]  &  [day]  &[deg]   &   -   & [deg]& [deg] & [\Mo] &[\Mo] \\ \hline
  Value    & 2440443.4680 & 2.51330731 & 82.0 & 0.0516 & 2437313 & 329.8  & 0.940 & 21.0 & 163.7 & 0.535 & 16.5  \\
  Error    &    0.0003    & 0.00000098 & 1.9  & 0.0039 &   24    &  9.6   & 0.008 & 9.9  & 1.5   & 0.425 & $\sb{-12.8}\sp{+77.0}$  \\ 
 \hline \hline
\end{tabular}}
\end{table}

\begin{table}[b!]
\caption{The minimum times of QS~Aql from photoelectric
photometry.} \label{tabQSAqlMin} \centering \scalebox{0.9}{
\begin{tabular}{l c c r c}
\hline\hline
$\!\!\!\!$HJD-2400000 & Prim/Sec & Epoch & Ref. \\
\hline
   $23963.75$   &   Prim  & $-6557.0$ & [1] \\
   $26159.12$   &   Sec   & $-5683.5$ & [1] \\
   $26160.37$   &   Prim  & $-5683.0$ & [2] \\
   $30920.55$   &   Prim  & $-3789.0$ & [3] \\
   $37490.300$  &   Prim  & $-1175.0$ & [4] \\
   $37799.446$  &   Prim  & $-1052.0$ & [4] \\
   $38259.397$  &   Prim  & $-869.0$  & [4] \\
   $38577.35$   &   Sec   & $-742.5$  & [4] \\
   $38578.604$  &   Prim  & $-742.0$  & [4] \\
   $38945.548$  &   Prim  & $-596.0$  & [4] \\
   $39360.255$  &   Prim  & $-431.0$  & [4] \\
   $40443.489$  &   Prim  & $0.0$     & [4] \\
   $40453.544$  &   Prim  & $4.0$     & [5] \\
   $40790.349$  &   Prim  & $138.0$   & [5] \\
   $41182.401$  &   Prim  & $294.0$   & [6] \\
   $44439.6649$ &   Prim  & $1590.0$  & [7] \\
   $48501.190$  &   Prim  & $3206.0$  & [8] \\
\hline
\end{tabular}}
\begin{list}{}{}
\item[Ref.:] [1] - \cite{Guth}; [2] - \cite{GuthnickPrager}; [3] -
\cite{Groeneveld}; [4] - \cite{Knipe}; [5] - \cite{vanderWal}; [6]
- \cite{Knipe72}; [7] - \cite{Skillman}; [8] - \cite{HIP}.
\end{list}
\end{table}

The spectrum of QS~Aql was first classified as B3 \citep{Millman1928}, but already the next spectroscopic
analysis by \citep{Hill1931} indicates a later spectral type B5. In this latter paper also the first
spectroscopic orbit was calculated. The analysis of this SB1-type binary results in $e=0.056\pm0.027$, $K=(47.31
\pm 1.31)~\mathrm{km \cdot s^{-1}}$, $v_\gamma=(-14.21 \pm 0.98)~\mathrm{km \cdot s^{-1}}$. After then a few
spectroscopic investigations of this binary were carried out. The systemic velocities were derived:
$-4.93~\mathrm{km \cdot s^{-1}}$ (low confidence level, no spectroscopic solution, just an estimated mean
velocity of the system, \citealt{Millman1928}), $-13~\mathrm{km \cdot s^{-1}}$ \citep{Lucy1971},
$-15.9~\mathrm{km \cdot s^{-1}}$ \citep{Batten1978}, and the most recent one by \cite{Holm} results in $(-14.8
\pm 0.2)~\mathrm{km \cdot s^{-1}}$. As one can see from Fig.\ref{FigQSAql-RV}, the observed systemic velocities
are almost constant over the period of the third body, which seems unlikely. But it is necessary to take into
consideration the error bars (which are not known for some of these points) and also the very rapid change in
$v_\gamma$ near the periastron and almost constant velocity for a decades.

\begin{figure}[b!]
   \centering
   \scalebox{1.1}{
   \includegraphics[width= 9cm]{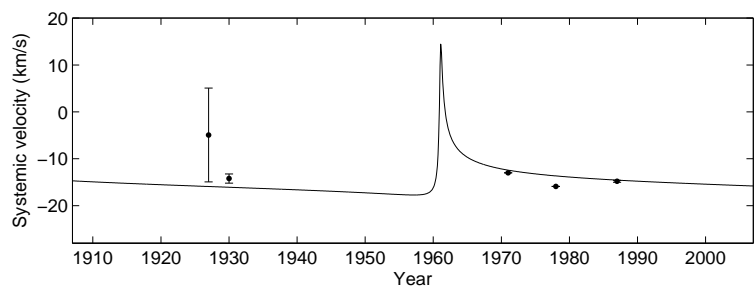}}
   \caption{Systemic velocity variations in QS~Aql. The
   individual points represent derived systemic velocities (see details in text),
   while the solid curve represents the variation on the long orbit described by
   the parameters in Table \ref{QSAql-Tab}.}
  \label{FigQSAql-RV}
\end{figure}

Using the derived parameters and Eq. \ref{eq13}, also the mass of the third component of the system was
computed. If the total mass of the eclipsing binary was assumed to be $M_{12} = 5.9$~\Mo~ (according to
\citealt{Holm}), one obtains the third mass $M_3 = 16.5$~\Mo, large primarily due to the relatively low orbital
inclination of the wide orbit. This means that the third body is much more massive than the individual
components of the eclipsing pair, which seems unlikely. One has to take into account the errors of the resultant
parameters. Due to the relatively large errors of the inclination and mass function, the resultant mass could be
somewhere between 3.7 and 93.5~\Mo, which is a very wide range of masses. The masses in the lower part of this
interval are sufficient to get a reasonable luminosity of the third body.

\cite{Heintze} discussed the spectroscopic observations by
\cite{Holm} and the light-curve observations and also concluded
that the third light is $1.2$ times larger than the combined light
of the eclipsing pair: $l_3 \approx 1.2 \cdot l_{12}$, which means
that in the bolometric magnitude, the third component should be
for about $0.2$ magnitude brighter than the eclipsing binary:
$M_{\mathrm{bol \, 3}} \approx M_\mathrm{{bol \, 12}} -
0.2$~mag. Adopting the spectral types of the primary and
secondary to be B5V and F3 (i.e. $M_\mathrm{{bol \, 12}} \approx
-2.5$~mag), the third body should have the spectral type
B4. If the third star is a main-sequence object, it should have a
mass of about $5.2$~\Mo.

This result lies within the range of the masses received from the
combined analysis. In conclusion, the presented solution is of a
low accuracy, mainly due to a very incomplete coverage of the
astrometric orbit, but leads to an acceptable solution within the
limits of the errors.

\section{VW~Cep}  \label{VWCep}

The eclipsing binary VW~Cep (HD 197433, BD +75 752, HIP 101750) was classified as W~UMa system and in fact it is
one of the most often observed and analyzed system. Its magnitude is about 7.3 in $V$ filter, but its spectrum
is problematic to classify. \cite{Popper1948} and also \cite{Kaszas1998} classified the system as K1+G5, while
\cite{Prib2000} proposed the spectral types G5V + G8V, but \cite{Kaszas1998} noted that the spectral type G5 is
inapplicable and \cite{Hill1989} presented the spectral type K0V. Both components are chromosphericaly active.
VW~Cep is rather atypical, because during the primary (the deeper one) eclipse the less massive star (the hotter
one) is occulted by the larger companion (the more massive and the cooler one).

\begin{figure}[b!]
   \centering
   \scalebox{0.75}{
   \includegraphics{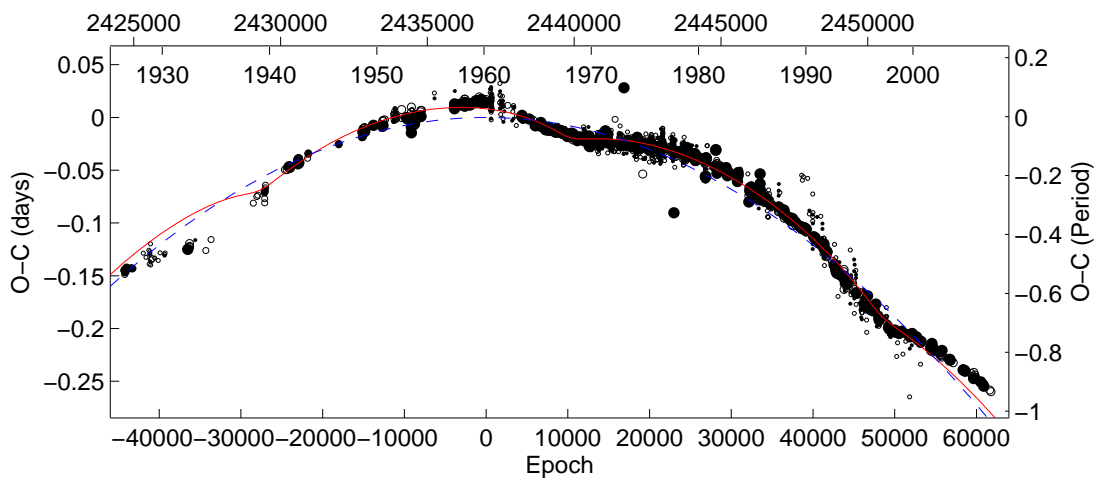}}
   \caption{An $O-C$ diagram of VW~Cep using \emph{Solution I}.
   The description is the same as in Fig. \ref{FigQSAql-OC}.
   The bigger symbols are CCD and photoelectric, while the
   smaller ones are visual. Most of the recent visual
   observations were neglected. The blue dashed line
   represents the quadratic term and the red solid line
   the quadratic plus the LITE caused by the third body.}
   \label{FigVWCep-OC}
\end{figure}

The first observations of its light variations were done by
\cite{Schilt}. Since 1946, a large amount of photoelectric
observations was obtained. However, the observed minima times did
not fit the ephemeris due to the LITE and the mass transfer
between components. There were many light-time effect studies of
this system and \cite{HerSch} proposed the presence of a third
body with an orbital period of 29 years and an angular distance of
the third component between 0.5\arcs and 1.2\arcs.

In 1974, the first successful visual observation of the third
component was obtained and since then, there were 16 observations
of it. Regrettably, the observations near the periastron passage
are missing (the gap in data is from 1991 to 1999). The last two
measurements ($\theta = 231.4 ^\circ$, $\rho = 0.702$\arcs~and
$\theta = 232.6 ^\circ$, $\rho = 0.695$\arcs) were obtained and
kindly sent by Elliot Horch by a speckle camera in April 2007
(priv.comm.).

The most complete set of times of minima is in the most recent
period study of VW~Cep by \cite{Prib2000}. The first times of
minima are from the 1920's and altogether 1907 minima were
collected. From this set of times of minima 313 measurements were
neglected due to their large scatter (mostly the visual ones).
This new minimum-time analysis is based on a larger data set
(about 750 times of minima more than were used by
\citeauthor{Prib2000}), see Fig.\ref{FigVWCep-OC}. Two new CCD
observations of minimum light of VW~Cep were obtained at
Ond\v{r}ejov observatory.

The short-term variations with the period of about two years (see
e.g. \cite{Kwee1966} and \cite{Hendry2000}) are probably caused by
the surface activity cycles on the primary component. Due to this
activity an unique interpretation of the behaviour of period
changes is still missing. \citeauthor{Prib2000} proposed a mass
transfer (the quadratic term) plus the third and the fourth body
in the system (two periodic terms). Nevertheless, they were not
able to explain the $O-C$ diagram in detail.

\begin{figure}[t!]
   \centering
   \scalebox{0.75}{
   \includegraphics{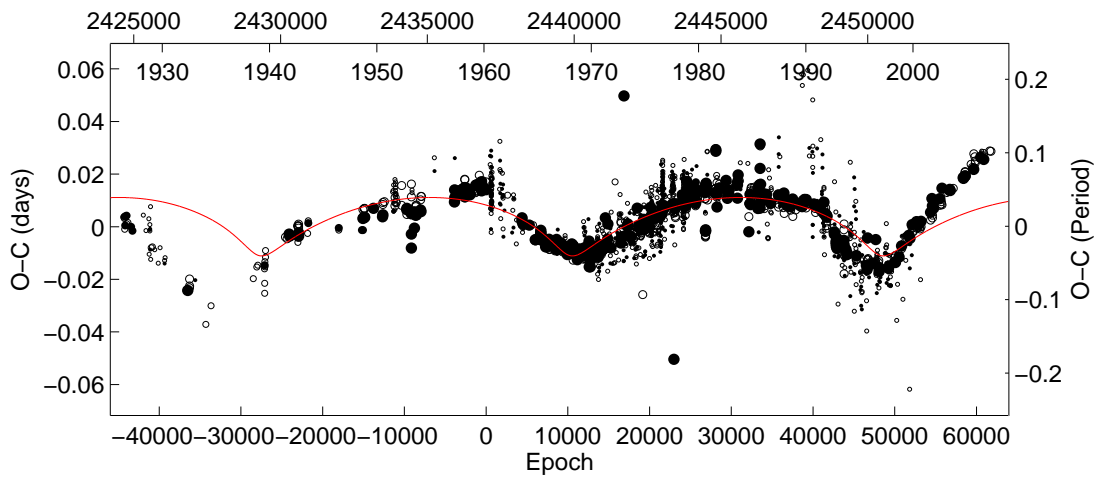}}
   \caption{An $O-C$ diagram of VW~Cep after the subtraction
   of the quadratic term, using \emph{Solution I.} The description is the
   same as in the previous $O~\!-~\!C$ figures, and the red solid line
   represents the LITE caused by the third component in the system.}
   \label{FigVWCep-OC2}
\end{figure}

Another approach was chosen in this thesis. Especially due
to only a few astrometric observations (16 measurements from 1974
to 2007) it was decided to explain only the most significant
effects in the $O-C$ diagram. There were two different approaches
used. In \emph{Solution I.} only the third-body orbit besides the
mass transfer (the quadratic term) was considered, while in
\emph{Solution II.} instead of mass transfer the fourth body on
its very long orbit and the third body was considered. This
approach was chosen especially because of the systemic-velocity
variations, see below. The astrometric variation with a period of
about 30 years has been identified with the $O-C$ variation with
the same period.

\subsection{Solution I.}

The analysis of the times of minima together with the astrometry
led to the parameters shown in Table \ref{TableBig2} and the $O-C$
diagram in Fig. \ref{FigVWCep-OC}. The times of minima, together
with the curve which represents the LITE and the mass transfer,
are shown in this figure. After subtraction of the quadratic term,
one gets Fig. \ref{FigVWCep-OC2}, where only the LITE caused by
the third component is displayed. The quadratic term quotient $q =
-0.756 \cdot 10^{-10}$~day leads to the period change of
about $1.98 \cdot 10^{-7} \mathrm{day/yr}$ and the rate of mass
transfer from the primary component of about $1.30 \cdot
10^{-7}$\Mo/yr (while \cite{Prib2000} derived the value $1.38
\cdot 10^{-7}$\Mo/yr).

The fit is not very satisfactory because of the presence of the
chromospheric activity of the individual components, or due to the
putative fourth component (see e.g. \citealt{Prib2000}). It is
evident, that the recent times-of-minima observations deviates
from the predicted fit. This could be caused by a period jump near
1995. In Fig. \ref{FigVWCep-orbit}, the astrometric orbit of the
binary with the individual measurements and their theoretical
positions is shown. Regrettably, no observations near the
periastron passage are available. The curve represents the
theoretical orbit according to the parameters given in Table
\ref{TableBig2} in agreement with the LITE analysis. Also the
orbit according to the \emph{Solution~II.} is shown, see below.

\begin{table}[t!]
\caption{The final results: VW Cep, Solution I. and II. The table
is divided into three parts, in the first one are eleven computed
parameters, in the second one the values from the literature and
in the last one the quantities computed from the previous parts.
The values of parallax and distance were adopted from the
\emph{Hipparcos} measurements. }
 \label{TableBig2} \centering \scalebox{0.88}{
\begin{tabular}{c c c c}
\hline\hline
Parameter  &  Unit     &    VW~Cep -- Solution I.     & VW~Cep -- Solution II.    \\
\hline
  $JD_0$   & $[$HJD$]$ & $2437001.5289 \pm 0.0034$    & $2437001.4362 \pm 0.0025$   \\
  $P$      & $[$day$]$ & $0.278316241 \pm 0.00000011$ & $0.278315234 \pm 0.00000012$ \\
  $q$      & $[$day$]$ &$(0.756 \pm 0.032)\cdot 10^{-10}$&  --                      \\
  $p_3$    & $[$yr$]$  & $30.04 \pm 0.47$             & $29.99 \pm 0.34$            \\
  $T_0$    & $[$HJD$]$ & $2450402 \pm 91$             & $2450366 \pm 52$            \\
  $\omega$ & $[$deg$]$ & $235.58 \pm 3.01$            & $242.53 \pm 2.89$           \\
  $e$      &           & $0.628 \pm 0.035$            & $0.610 \pm 0.014$           \\
  $A$      & $[$day$]$ & $0.0117 \pm 0.0009$          & $0.0119 \pm 0.0010$         \\
  $a$      & $[$mas$]$ & $447.4 \pm 24.3$             & $451.4 \pm 28.1$            \\
  $i$      & $[$deg$]$ & $30.2 \pm 4.2$               & $28.0 \pm 3.1$              \\
  $\Omega$ & $[$deg$]$ & $200.0 \pm 7.1$              & $192.1 \pm 5.7$             \\
  \hline
  $M_{12}$ & [\Mo]     &  $1.37$                      & $1.37$                     \\
References &           & \cite{Kaszas1998}            & \cite{Kaszas1998}          \\
  $\pi$    & $[$mas$]$ & $36.16 \pm 0.97$             & $36.16 \pm 0.97$           \\
  $D$      & $[$pc$]$  & $27.7 \pm 0.7$               & $27.7 \pm 0.7$             \\
  \hline
  $a_{12}$ & $[$AU$]$  & $4.30 \pm 0.41 $             & $4.59 \pm 0.45 $           \\
  $f(M_3)$ & [\Mo]     & $0.0112 \pm 0.0078$          & $0.0111 \pm 0.0068$        \\
  $M_3$    & [\Mo]     & $0.73 \pm 0.32 $             & $0.80 \pm 0.30 $           \\
\hline \hline
\end{tabular}}
\end{table}

\begin{figure}[t]
  \centering
  \includegraphics[width=9cm ]{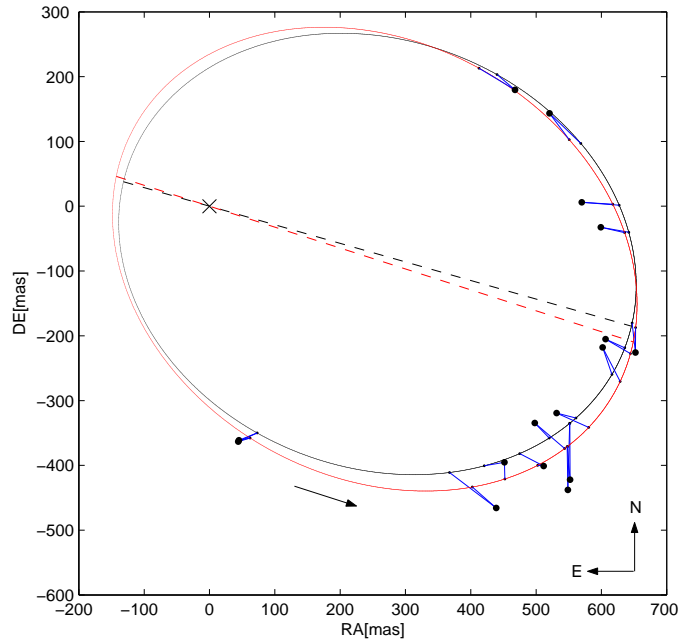}
  \caption{Relative orbit of VW~Cep on a plane of the sky,
  for a detailed description see Fig. \ref{FigQSAql-orbit}.
  Both \emph{Solution I.} and \emph{II.} are plotted, the
  black one for \emph{Solution I.} and the red one for the
  \emph{Solution~II.}.}
  \label{FigVWCep-orbit}
\end{figure}

The parameters describing the LITE and astrometric variation are
in Table \ref{TableBig2} and could be compared to the parameters
derived during the previous analysis by \cite{Prib2000}. Their
values for the third-body orbit are: $p_3 = 31.4$~yr, $e~=~0.77$,
$\omega~=~183^\circ$, and $a_\mathrm{total}~=~12.53$~AU. Our
values are in Table \ref{TableBig2} except for
$a_\mathrm{total}~=~12.35$~AU, and as we can see they differ
significantly in several parameters. This is due to completely
different approach describing the $O-C$ variations. Only the
period and the amplitude of such variation are comparable, but
these are the most important for our combined solution. One has
also to disagree with the result by \citeauthor{Prib2000}, that
the astrometric orbit could not be identified with the LITE$_3$
variation from the $O-C$ diagram. As one can see, our new results
are in agreement with each other without any problems.

Also the astrometric orbit could be compared with the previously
published one. Most recently \cite{Docobo2005} 
published the following parameters of the astrometric orbit:
$p_3~=~31.0$~yr, $a~=~485$~mas, $i~=~39.3^\circ$, and $e~=~0.68$. 
If one compares these values with the new ones (see Table
\ref{TableBig2}), one can see that the differences are slightly
beyond the limits of errors.

If the total mass of the eclipsing binary $M_{12}~=~1.37~$\Mo~ was
taken from \citep{Kaszas1998} and the parallax $ \pi =
36.16~\mathrm{mas}$ (from \citealt{HIP}), the distance to the
system should be only about $27.66~\mathrm{pc}$, which results in
the third-body mass of $M_3~=~0.73~$\Mo. The distant component is
about 2.2 magnitudes fainter than the VW~Cep itself, so its
luminosity and also mass should be much smaller than the mass of
the eclipsing components. Total bolometric magnitude of VW~Cep is
about 4.7~mag, so the magnitude of the third component is about
6.9~mag, which leads to the spectral type of about K3. The typical
mass of this spectral type is about $0.75~$\Mo (according to
\citealt{Hec1988}), which is in an excellent agreement with our
result and within its error limits.

\begin{figure}[t!]
   \centering
   \scalebox{1.1}{
   \includegraphics[width= 9cm]{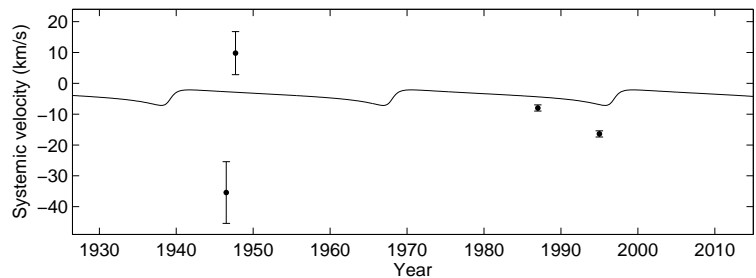}}
   \caption{Systemic velocity variations in VW~Cep, using
   \emph{Solution I.} The individual points represent computed
   systemic velocities (see details in text).}
   \label{FigVWCep-RV}
\end{figure}

Different systemic velocities $v_\gamma$ were found at different
epochs. These values are: $v_\gamma = (-35.4~\pm 10)~ \mathrm{km
\cdot s^{-1}}$ \citep{Popper1948}, $(+9.8~\pm 7)~\mathrm{km \cdot
s^{-1}}$ \citep{Binnendijk1966}, $(-8~\pm 1)~\mathrm{km \cdot
s^{-1}}$ \citep{Hill1989}, and $(-16.4~\pm 1)~\mathrm{km \cdot
s^{-1}}$ \citep{Kaszas1998}. In the time plot (see Fig.
\ref{FigVWCep-RV}) one can see the curve which represents the
theoretical variation of $v_\gamma$ caused by the orbital motion
around the common barycentre. Except for the first one data point
\citep{Popper1948}, the amplitude of the LITE should be much
larger (circa 3 times) than was computed. Keeping the astrometric
amplitude at the value from the fit, this could only be achieved
by decreasing the inclination to smaller values and by modifying
slightly also some other parameters. This speculation could only
be verified after more accurate and larger data set is available.
But considering the spectral analysis and efficiency of the RV
investigations, this result is not very satisfactory, which is the
reason why the different approach was also used, see
\emph{Solution II.} below, which describes better the systemic
velocity variations.

The star was also measured by the \emph{Hipparcos} satellite.
During its 3-yrs mission there were altogether 68 observations
obtained, see Fig. \ref{FigVWCep-orbitHIP}. These are so-called
\emph{abscissa measurements} and are one-dimensional. It means
that only a time of passage through a certain main circle was
measured, but one cannot derive exactly where on this circle the
star really was. The position of these circles (measurements) are
represented by the small abscissae in Fig.
\ref{FigVWCep-orbitHIP}. The observations are connected by a
dotted lines with the theoretical positions on the sky marked as
big points. The theoretical orbit was constructed according to the
parameters from Table \ref{TableBig2}. As one can see, regrettably
only a small part of the orbit was measured, so the Hipparcos
abscissae measurements are not very useful at all.

Another task was to derive the parallax of VW~Cep using this combined approach. Leaving the parallax as another
free parameter, one is able to calculate it from the comparison of the angular and absolute semimajor axis (see
section \ref{dst}). Using this method, most of the relevant parameters remained nearly the same as above, only
the inclination changed a bit, being about $5^\circ$ lower. This led to a higher third mass of $M_3 = 0.99~$\Mo.
The main difference was in the parallax, which decreased from $36.16~\mathrm{mas}$ (Hipparcos) to
$33.63~\mathrm{mas}$. The parallax would shift the distance to $29.74~\mathrm{pc}$. The new value is only about
2 pc higher then the value derived from the Hipparcos measurements. With more precise data points and better
coverage of the orbit also this result would be better. For a comparison with the previously found parallaxes,
see the next section.

\begin{figure}[t]
   \centering
   \includegraphics[width=9cm]{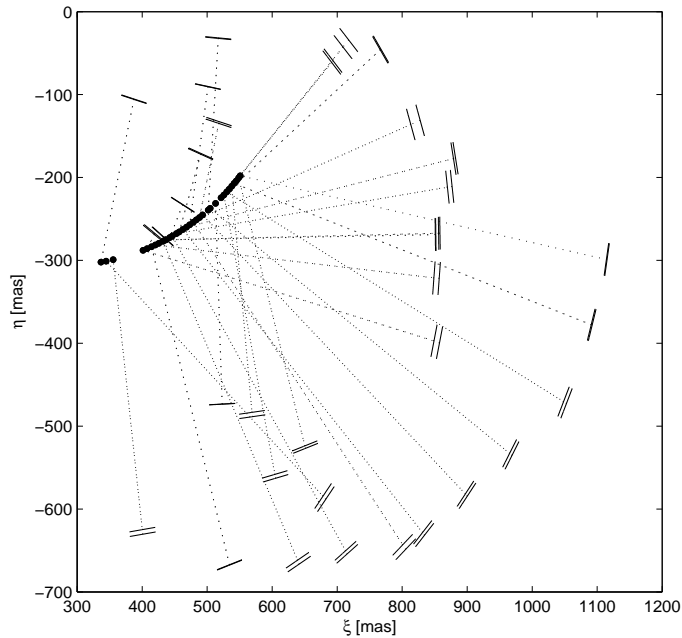}
   \caption{The Hipparcos measurements of VW~Cep, see the text for details.}
   \label{FigVWCep-orbitHIP}
\end{figure}

\subsection{Solution II.}

From the analysis of times of minima and using the long period
perturbation by another distant component instead of the quadratic
term, one gets Fig. \ref{FigVWCep-OC3} and after the subtraction
of the fourth component Fig. \ref{FigVWCep-OC4}. It means during
the computation process altogether 14 parameters
($A,p_3,i,e,\omega,\Omega,T_0,JD_0,P,A_4,p_4,e_4,\omega_4,T_{0,4}$)
were derived minimizing the $\chi^2_{comb}$ value.

Applying this approach to the same data one gets about 12 \%
better result (comparing the sum of square residuals). The
numerical values for the individual parameters are approximately
the same (see Table \ref{TableBig2}). The parameters of the
fourth-body orbit are in Table \ref{VWCep-Tab}, where $M_{4,min}$
denotes for the minimal mass of the fourth body ($i_4 =
90^\circ$). It is obvious that the period $p_4$ is about as long
as our data set. One can judge that this numerical solution is
only an edge-on effect, which fits better the most recent data
points. As one can see, the times of minima since 1995 deviate
from the theoretical prediction and also an additional period jump
should be implemented into the model to describe the data points
in detail. Nevertheless, this combined approach was chosen because
of the RV data (see below). In next few years the behaviour of the
times of minima will decide which solution is the right one. Until
that time this approach is just a hypothesis without any proof,
only for a better description of the systemic-velocity variations.

\begin{figure}[t!]
   \centering
   \scalebox{0.75}{
   \includegraphics{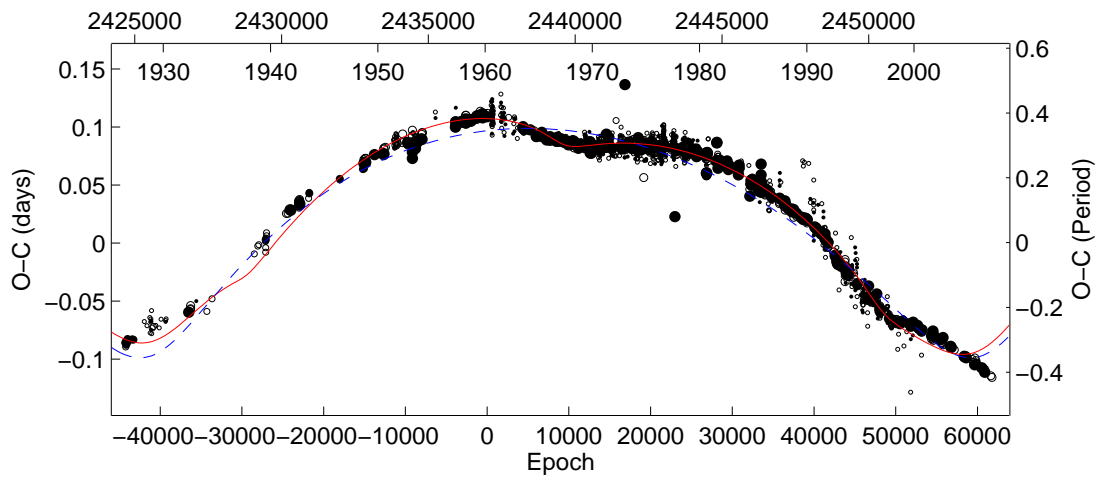}}
   \caption{An $O-C$ diagram of VW~Cep using \emph{Solution II}. The
   description is the same as in Fig. \ref{FigVWCep-OC}. The blue dashed
   line represents the LITE caused by the fourth distant body and the red
   solid line the final fit LITE$_3$ + LITE$_4$.}
   \label{FigVWCep-OC3}
\end{figure}

\begin{figure}[b!]
   \centering
   \scalebox{0.75}{
   \includegraphics{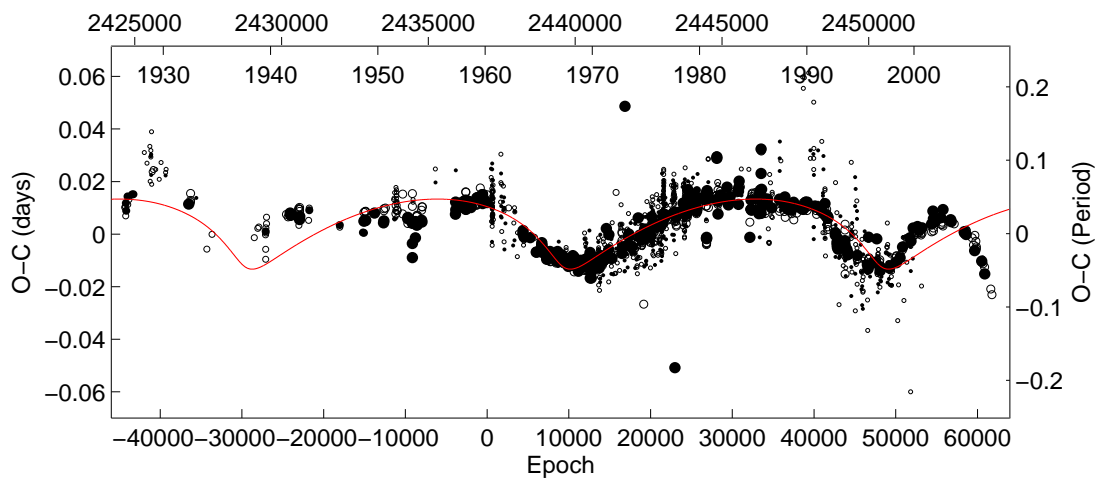}}
   \caption{An $O-C$ diagram of VW~Cep using \emph{Solution II.} after
   the subtraction of the LITE caused by the distant fourth body.
   The description is the same as in Fig. \ref{FigVWCep-OC2}.}
   \label{FigVWCep-OC4}
\end{figure}

The parameters of the third-body orbit according to \emph{Solution
II.} are close to the values from \emph{Solution I.}, that the
theoretical orbit of the binary on the plane of the sky is almost
the same, see Fig.\ref{FigVWCep-orbit}.

\begin{figure}[t!]
   \centering
   \scalebox{1.1}{
   \includegraphics[width=9cm]{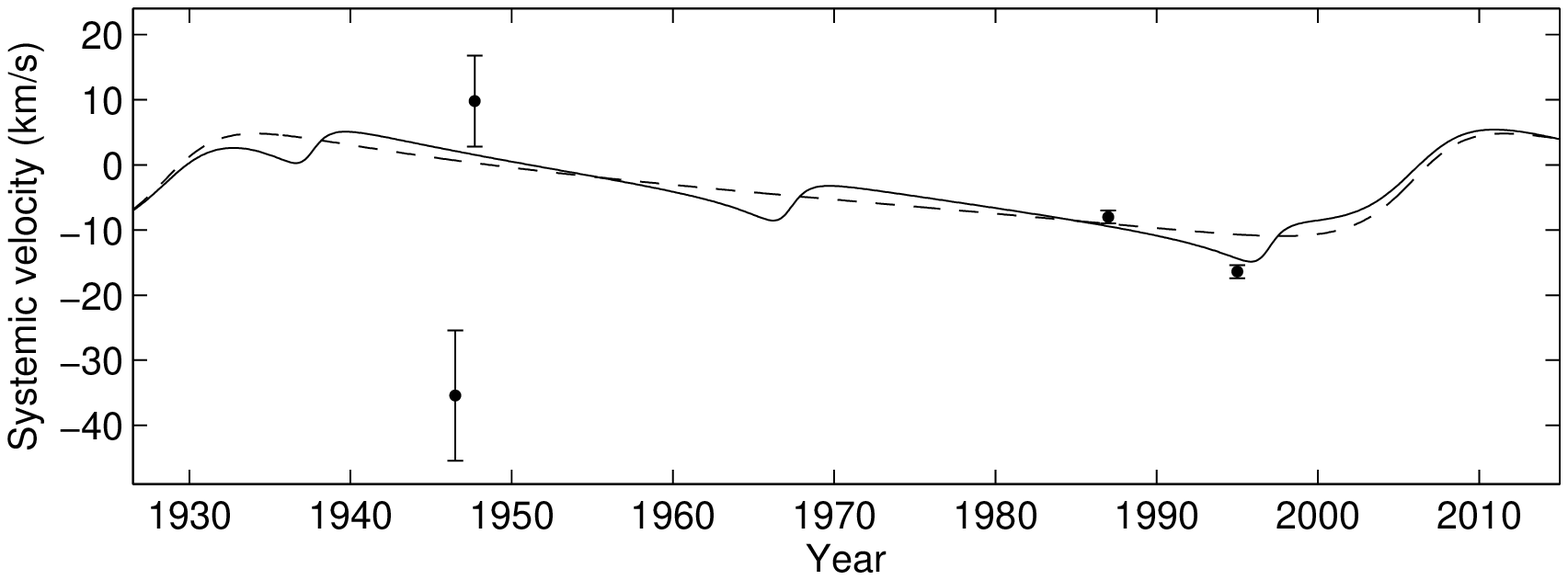}}
   \caption{Systemic velocity variations in VW~Cep, using
   \emph{Solution II.} See details in text.}
   \label{FigVWCep-RV4}
\end{figure}

\begin{table}[b!]
 \caption{The parameters of the fourth-body orbit, VW Cep Solution II.}
 \label{VWCep-Tab} \centering
 \scalebox{0.9}{
 \begin{tabular}{c c c c c c | c c }
 \hline \hline
   Parameter&   $p_4$  &$T_{0,4}$&$\omega_4$ &   $e_4$   &   $A_4$    & $f(M_4)$  & $M_{4,min}$\\
   Unit     & $[$yr$]$ &$[$HJD$]$& $[$deg$]$ &           &  $[$day$]$ & [\Mo]     & [\Mo]    \\\hline
   Value    & $77.32$  &$2397303$& $282.3$   & $0.561$   &  $0.096$   & $0.795$   & $2.64$   \\
   Error    &$\pm 0.04$&$\pm 14$ & $\pm 2.3$ &$\pm 0.008$& $\pm 0.012$&$\pm 0.055$&$\pm 0.45$\\
 \hline \hline
 \end{tabular}}
\end{table}

The only effects which are significantly different are the gamma-velocity variations. It is shown in Fig.
\ref{FigVWCep-RV4}, where the dashed and the solid line represent the LITE$_4$ and LITE$_3$ + LITE$_4$
variations, respectively. As one can see, this approach gives a much better fit. Except for the first data point
\citep{Popper1948} the systemic velocities follow the long-term variation and are almost within its errors near
the theoretical values. The value by \citeauthor{Popper1948} is affected by relatively large error. The scatter
of the individual RV data points by \citeauthor{Popper1948} is larger than those from
\citeauthor{Binnendijk1966}, which could be caused by the combination of two different data sets from different
instruments and obtained after more than 600 orbital revolutions (which could shift the ephemeris).
\cite{PribRuc2006} suggested that the scatter of the systemic velocity data points is instrumental, which seems
unlikely for such a large amplitude. For the final confirmation of $v_\gamma$ variations a more accurate and
larger data set is necessary.

One could also speculate about the possible visual detection of
the fourth component. This suggested body is bright enough to be
visible and its predicted angular separation from the system is
about 1.1\arcs. On the other hand one has to take into
consideration that the period of such a body is not completely
covered by data and could be even higher, as well as the amplitude
could be much higher (and also the angular separation). Nowadays
there is no potential star for this, only the star BD+74~889
shares common proper motion and radial velocity, but it is one
degree distant.

Another task was the distance determination. Due to only slight difference between the parameters of the
third-body orbit from \emph{Solution I.} and \emph{II.}, also the parallax and distance will be approximately
the same. The parallax decreased from $(36.16\pm0.97)~\mathrm{mas}$ (\emph{Hipparcos}) to
$(35.85~\pm~0.37)~\mathrm{mas}$. This parallax would shift the distance from $(27.7\pm0.7)$ pc
(\emph{Hipparcos}) to $(27.90~\pm~0.29)$ pc. Besides the \emph{Hipparcos} value, the most precise parallax was
derived by \cite{Heintz1993} from trigonometry, resulting in $(38.2 \pm 1.9)$~mas. As one can see, the values of
the parallax determined by Solutions I and II are more precise than any of the previously derived parallaxes.
For the summary of the previously derived values see \cite{Hendry2000}.

To conclude, the predicted third body is spectral type K3 with the
mass around $0.73$~\Mo~ (according to the \emph{Solution I.}), or
spectral type K2 with the mass around $0.80~$\Mo~ (applying the
\emph{Solution II.}). It is clear, however, that a more
complicated model will be needed to describe the observed changes
completely. Also new times-of-minima observations would be helpful
to identify the variations in $O-C$ diagram in detail, because
neither the \emph{Solution I.}, nor the \emph{Solution II.} are
able to describe the behaviour of the recent minima observations.
This could be described only applying the hypothesis of an abrupt
period jump. Precise RV investigation (till 2010) would solve the
question about the nature of the variations in gamma velocity.

\section{$\zeta$~Phe}

The system $\zeta$~Phe is the brightest eclipsing binary with two components of early spectral types, exhibiting
total and annular eclipses. This is the only eclipsing binary with an eccentric orbit included in this study.
$\zeta$~Phe (HD 6882, HR 338, HIP 5348) is an Algol-type eclipsing binary. Apparent brightness of the system is
about 4.0~mag in $V$ filter and the spectral types were determined as B6V + B8V (according to \cite{Andersen83},
see also a comment on the spectral types and \emph{Hipparcos} measurements in \citealt{Ling2004}). It is a
visual triple and SB2 spectroscopic binary. The depth of the primary minimum of the eclipsing pair is about
$0.5$~mag, the period of about $1.7$~day.

It is the visual triple system, while the brightest component is the EB, the most distant component is the
faintest (some 6\arcs away and with a magnitude of about $8$, this star is probably not gravitationally bound
with the system). The third component is a $7^\mathrm{th}$-magnitude star at a distance of about
$600~\mathrm{mas}$. This is the astrometric component and this star is supposed to cause also the LITE
variation.

The first astrometric observation of the third component came from
1930's and till now there were collected 14 observations, but two
of them were neglected (see Fig.\ref{ZetaPhe-orbit}).

\begin{figure}[t!]
   \centering
   \includegraphics[width= 9cm ]{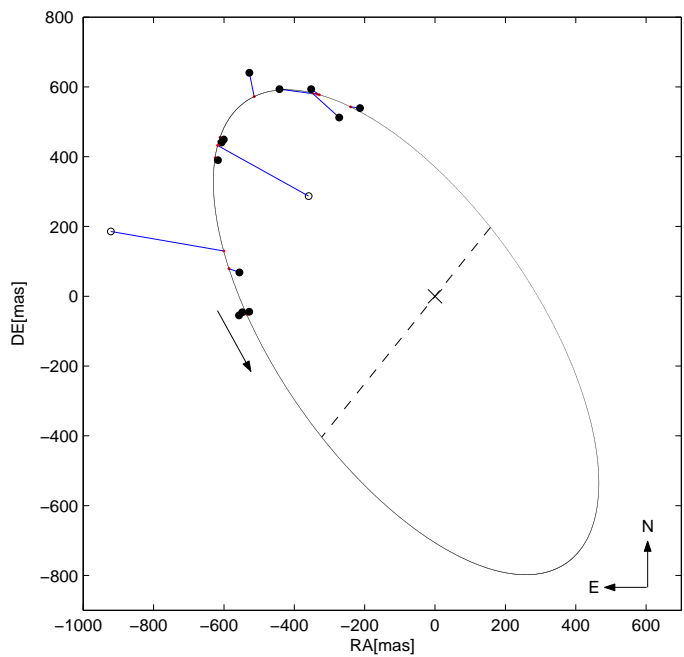}
      \caption{Relative orbit of $\zeta$~Phe on a plane of the sky,
      for a detailed description see Fig. \ref{FigQSAql-orbit}.
      Two measurements (the open circles) were neglected.
              }
   \label{ZetaPhe-orbit}
\end{figure}

The unfiltered light curve was observed in 1950's by \cite{Hogg},
after then by \cite{Dachs} in $UBV$ filters, and the best one by
\cite{Clausen} in $ubvy$ filters. In this latter paper all
relevant parameters of the eclipsing system were derived and also
the third light was computed. Its value changes from $3\% (u)$ to
$8\% (y)$ and the distant component was classified as a spectral
type A7 star (the same result was derived by \cite{Andersen83} on
the basis of his spectroscopic observations).

\begin{figure}[b!]
   \centering
   \scalebox{0.75}{
   \includegraphics{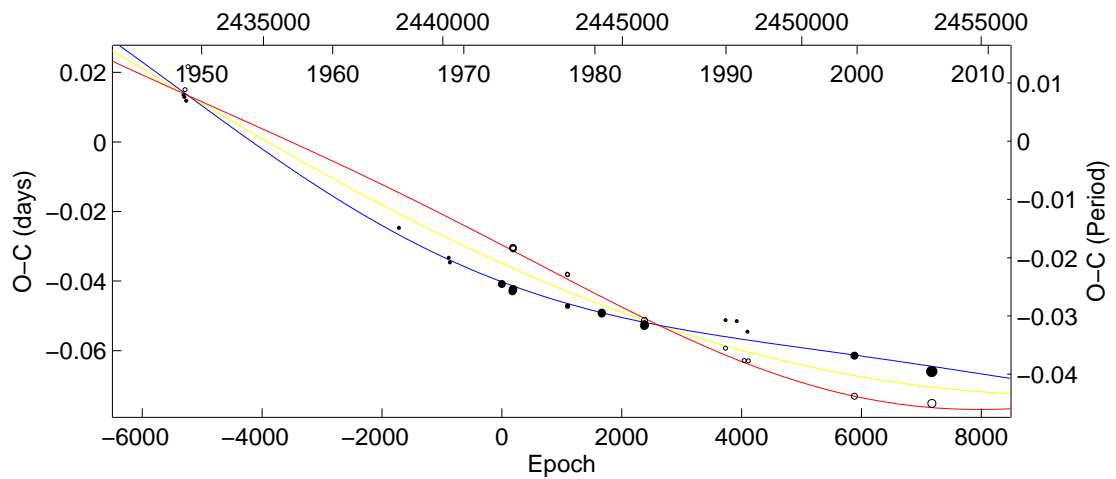}}
   \caption{The $O-C$ diagram of $\zeta$~Phe. The apsidal motion
   curve (the blue one for primary and the red one for secondary)
   is plotted around the (yellow) LITE curve.}
   \label{ZetaPheOC}
\end{figure}

\cite{Clausen} also collected the times of minima obtained before
1975. They concluded that no significant apsidal motion is
observed. The first apsidal-motion study was published by
\cite{Gimen1986}. With an updated list of the times of minima one
is able to conclude that the apsidal motion is definitely
presented. It is clearly visible in the $O-C$ diagrams shown in
Figs.~\ref{ZetaPheOC} and \ref{ZetaPheOC2}. Altogether 36 times of
minima used here came from the paper cited above and from
\cite{Mallama}, \cite{Gimen1986}, \cite{Kviz}. The most recent
ones are taken from \cite{ZascheWolfPrint} (in print).

$\zeta$~Phe has one of the shortest apsidal motions among the
eclipsing binaries (see e.g. \citealt{Claret1993}). Due to a low
eccentricity, the amplitude of the effect is small. For an
accurate calculation of the apsidal motion rate the method
described by \cite{GimenezGarciaPelayo} was routinely used. The
eccentricity of the orbit in the eclipsing binary is $e'=0.0107
\pm 0.0020$, the longitude of periastron $\omega_0 = 12.96^\circ
\pm 5.96^\circ$, and the apsidal motion rate $\dot\omega = (0.028
\pm 0.001) \mathrm{^\circ / cycle} = (6.16 \pm 0.20) \mathrm{^
\circ / yr} $, i.e. the apsidal motion period $U = 58.5$~yr. The
most recent apsidal-motion analysis is more than 20 years old,
made by \cite{Gimen1986}, but with no LITE and with a smaller set
of times of minima. The eccentricity by Gim\'enez was almost the
same, but the apsidal motion rate $\dot\omega$ was
$0.0373~\mathrm{^\circ / cycle}$ and the angle
$\omega_0~=~13^\circ.$

\begin{table}[t!]
\caption{The parameters of $\zeta$~Phe.} \label{ZetaPhe-Tab}
\centering \scalebox{0.75}{
\begin{tabular}{c c c c c c c c c c| c c}
 \hline \hline
 Parameter&   $JD_0$     &     P     &$p_3$&  $A$  & $T_0$ &$\omega$& $e$ & $i$ &$\Omega$&$f(M_3)$&$M_3$\\
  Unit    &   [day]      &   [day]   &[yr] & [day] & [day] & [deg]  &  -  &[deg]& [deg]  & [\Mo]  &[\Mo]\\ \hline
  Value   & 2441643.7382 & 1.6697772 &220.9& 0.0808&2419900& 97.1   &0.366& 64.4& 33.5   & 0.056  &1.73 \\
  Error   &    0.0008    & 0.0000013 & 3.5 & 0.0080& 2500  &  2.2   &0.082& 3.0 & 4.9    & 0.017  &0.26 \\
 \hline \hline
\end{tabular}}
\end{table}

The approach presented here was a combination of the two different
effects. The behaviour in $O-C$ diagram was described as a sum of
apsidal motion and LITE contribution $(O-C) = (O-C)_{apsid} +
(O-C)_{LITE}$, distinguishing the primary and secondary minima. It
means the least-squares algorithm was minimizing the
$\chi^2_{comb}$ with respect to 12 parameters in total
($A,p_3,i,e,\omega,\Omega,T_0,JD_0,P,\dot\omega,\omega_0,e'$).

\begin{figure}[t!]
  \centering
  \scalebox{0.75}{
  \includegraphics{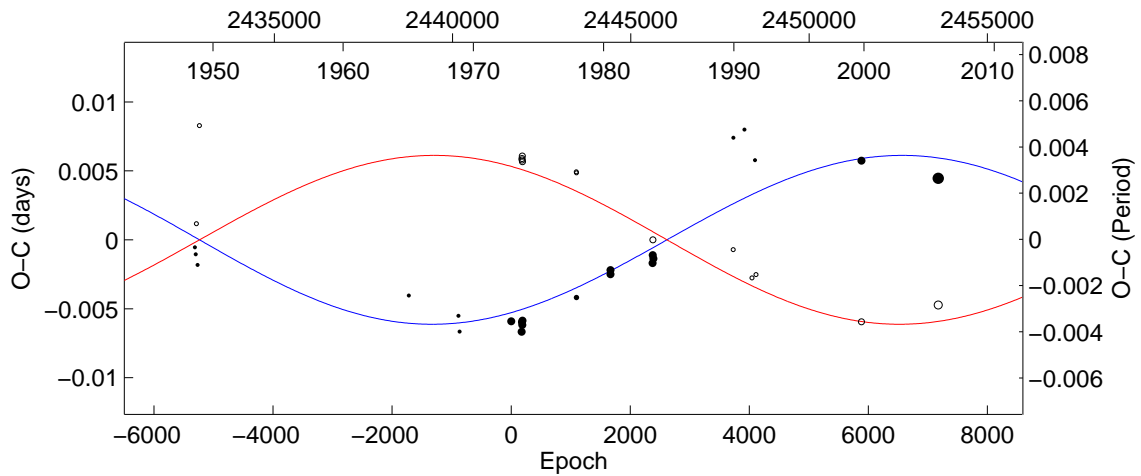}}
  \caption{The $O-C$ diagram of $\zeta$~Phe after subtraction
  of the LITE. Only apsidal motion curve is shown (the blue one
  for primary and the red one for secondary).}
  \label{ZetaPheOC2}
\end{figure}

The astrometric solution based on the combined approach is satisfactory, while the older measurements have
larger scatter then the recent ones (the old ones are visual and the modern speckle-interferometric). Two
measurements were neglected, because of their large scatter (see Fig.\ref{ZetaPhe-orbit}). The solution led to
the parameters listed in Table~\ref{ZetaPhe-Tab}. One could compare this new orbit with the previously found
one, the most recently \cite{Ling2004} reported the parameters: $p_3~=~210.4$ yr, $e~=~0.348$, $a~=~804$~mas,
$i~=~61.9^\circ$, $\Omega~=~33.5^\circ$, $\omega_3~=~271.7^\circ$. It is evident that the new parameters are in
very good agreement with these ones. The new values imply the mass function of the distant body $f(M_3) =
0.056~$\Mo~and with the masses of primary and secondary component of the eclipsing binary $M_1 = 3.93~$\Mo~and
$M_2 = 2.55~$\Mo \citep{Andersen83}, the mass of the astrometric third body was derived $M_3 = 1.73~$\Mo. This
value corresponds to a spectral type around A7, which is in excellent agreement with the photometric analyses by
\cite{Clausen} and \cite{Andersen83}, which result in A7.

There were also 2 RV investigations by \cite{Popper1970} and \cite{Andersen83}, but with only 2 values of the
$v_\gamma$ velocity one cannot do any reliable analysis. In \cite{Andersen83} is also presented that the lines
of the third component are also observable in the spectrum of $\zeta$~Phe, but these lines are hardly separable
from the binary lines.

To conclude, $\zeta$~Phe shows astrometric as well as LITE variations, which are in agreement with each other.
Regrettably, the period of the third-body orbit was not sufficiently covered by the data yet, only about $1/4$
of the orbit is covered in both methods. Only further precise measurements would prove the third-body hypothesis
with higher certainity.

\section{V505 Sgr}  \label{V505Sgr}

Another EB system with apparent changes of the orbital period is V505~Sgr, where the third body has been known
for more than a decade. It is an Algol-type eclipsing binary with a period of about 1.2 days. V505~Sgr
(HD~187949, HR~7571, HIP~97849) was classified as A2V+G5IV spectral types (according to
\citealt{Chambliss1993}), with magnitude of about 6.5 in V filter.

\begin{figure}[t!]
   \centering
   \scalebox{0.75}{
   \includegraphics{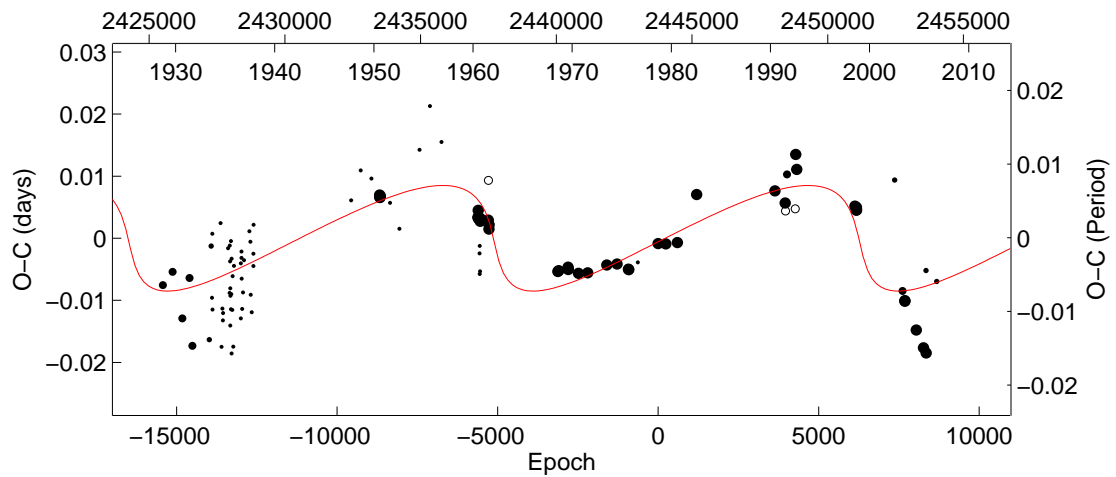}}
   \caption{The $O-C$ diagram of V505 Sgr.}
   \label{V505SgrOC}
\end{figure}

The star was discovered to be an eclipsing binary by
\cite{Hoffmeister1934}. Since then a lot of light-curve
measurements and analyses were carried out (for example
\citealt{Lazaro}). The last one by \cite{Iban2000} indicated that
the mass ratio of the binary is about 0.5 and the contribution to
the total light of the binary by the third component is about
2.6\% in \emph{B} and 3.6\% in \emph{V} filter. This analysis also
yielded the spectral type of the distant component to be roughly
F6, which is in good agreement with the previous analysis by
\cite{Tomkin1992}, which results in A7. The spectroscopic nature
(SB2) was discovered by \cite{Popper1949}.

\begin{table}[b!]
\caption{The parameters from the combined solution of V505 Sgr.}
\label{V505Sgr-Tab} \centering \scalebox{0.75}{
\begin{tabular}{c c c c c c c c c c| c c}
 \hline \hline
 Parameter&  $JD_0$    &    P    &$p_3$& $A$  & $T_0$ &$\omega$&$e$ &  $i$ &$\Omega$&$f(M_3)$&$M_3$\\
  Unit    &  [day]     &  [day]  &[yr] & [day]& [day] & [deg] &  -  & [deg]& [deg]  & [\Mo]  &[\Mo]\\ \hline
  Value   &2443750.6866&1.1828688&36.86&0.0085&2451197& 184.1 &0.802&195.6 & 22.3   & 0.0109 &2.76 \\
  Error   &   0.0004   &0.0000002& 0.09&0.0005&  19   &  3.2  &0.008& 2.4  & 1.9    & 0.0024 &0.98 \\
 \hline \hline
\end{tabular}}
\end{table}

In 1985 an astrometric component was found by \cite{McAlister1987}
by speckle interferometry. The body was 0.3\arcs away from the
eclipsing pair and after a few years a few measurements (16 till
now) was obtained. Nowadays it is evident that the distant
component is moving on its orbit around the EB pair. At the same
time also \cite{Tomkin1992} found the third component lines in the
spectrum of V505 Sgr.

\begin{figure}[t!]
   \centering
   \includegraphics[width= 9cm ]{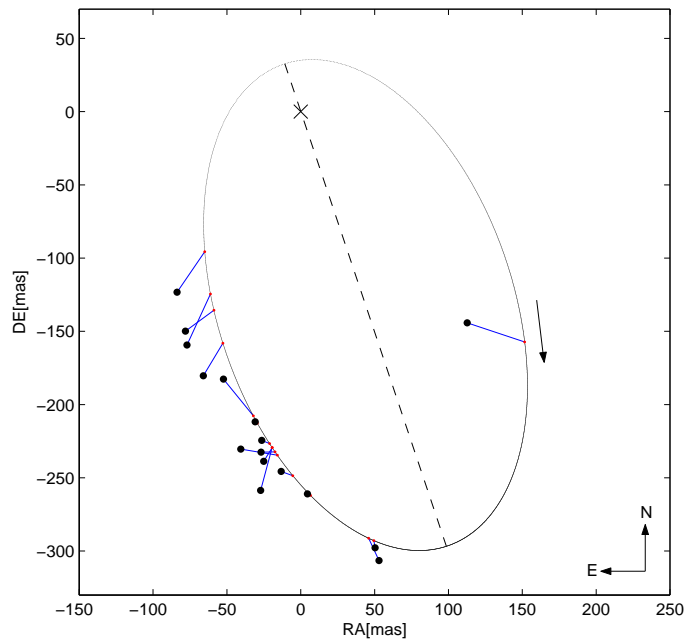}
   \caption{Relative orbit of V505Sgr on a plane of the sky,
   for a detailed description see Fig. \ref{FigQSAql-orbit}.}
   \label{V505Sgr-orbit}
\end{figure}

There are several analyses of its apparent orbital period changes
interpreted as the LITE due to the third body (e.g.
\citealt{Rovithis1991}). The only paper which compares the
astrometry and a period analysis of $O-C$ deviations from the
constant orbital period was published by \citet{Mayer1997}.
Despite existing astrometric measurements, there were no attempts
to combine these two methods together. The results from different
approaches were just compared to each other. The main reason why
such a combined solution is missing, are the differences in
parameters, which result from separate solutions (see below).

The $O-C$ diagram is in Fig. \ref{V505SgrOC} and the astrometric
orbit in Fig. \ref{V505Sgr-orbit}. The last one astrometric
measurement was obtained on the 18$^{th}$ October 2005, using
3.6-meter CFHT on Hawaii Islands, resulting in $\rho =
0.183(4)$\arcs, $\theta = 218(2) ^\circ$ (kindly sent by Theodor
Pribulla). As one can see, this point does not fit the theoretical
orbit well, but one cannot ignore this data point, because it is
the only measurement during the last decade and it is as precise
as the previous ones (the position was obtained after averaging 5
frames).

As one can see from Fig. \ref{V505SgrOC} also the last times of minima in the $O-C$ diagram do not follow the
theoretical curve, and it is really necessary to observe at least one precise minimum of V505~Sgr (the last one
is taken from the VSNET database and is not very precise, derived only from 9 points).

The star was also measured by the \emph{Hipparcos} satellite. Altogether 50 measurements were obtained (see Fig.
\ref{FigV505Sgr-orbitHIP}). Regrettably, at that time the star was not near its periastron, so only a small arc
of the orbit is covered. It is similar as in the case of VW~Cep, also the description of the figure is the same.

The diagrams were plotted according to the parameters from the
combined solution introduced in Table \ref{V505Sgr-Tab}. It is
obvious that the fit to the individual data points is not very
satisfactory. This is due to inconsistency of the two separate
solutions. Only LITE solution leads to the 41-years orbit, while
the astrometric to 33yr. The angle $\omega$ differs about
60$^\circ$ and the amplitude of the astrometric variation is about
2 times larger than one would expect from the LITE analysis. These
are the principle reasons why there is a doubt of identifying the
astrometric and LITE variation to be caused by the same body.

\begin{figure}[t!]
   \centering
   \includegraphics[width=9cm]{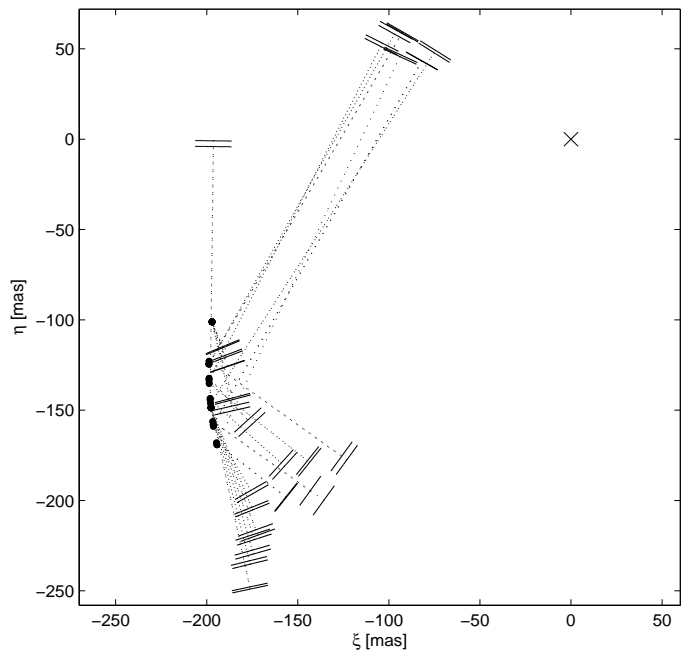}
   \caption{The Hipparcos measurements of V505~Sgr, see
   Fig.\ref{FigVWCep-orbitHIP} and text for details.}
   \label{FigV505Sgr-orbitHIP}
\end{figure}

Because the third body is also visible in the spectra of V505~Sgr, different radial velocities of the third
component were measured from 1979 to 1989 (see \citealt{Tomkin1992}). These measurements together with the
predicted variation based on the parameters from Table \ref{V505Sgr-Tab} are shown in Fig. \ref{V505Sgr-RV}. As
one can see, there is some systematic increase in the radial velocities, but due to only a small part of the
period covered, this is not very conclusive result (regrettably all the measurements were obtained near
apastron).

From the combined solution, together with the parallax from the
\emph{Hipparcos}, one is able to derive the mass of the third
body. This results in 2.76~\Mo, or the spectral type of about B8
(according to \citealt{Hec1988}). This result is in contradiction
with the previous spectral analysis, which indicates a spectral
type of about F6 (see e.g. \citealt{Tomkin1992}), with its typical
mass about 1.3~\Mo. The only acceptable explanation could be that
the third component is also a binary.

\begin{figure}[b!]
  \centering
  \scalebox{1.1}{
  \includegraphics[width=9cm]{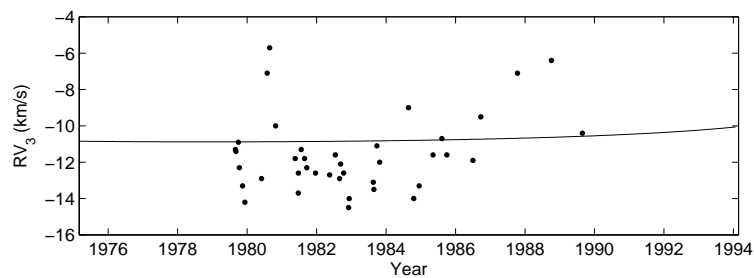}}
  \caption{The radial velocity variations of the third component in V505~Sgr. The solid
  line represents the variation caused by the third body
  according to the parameters from Table \ref{V505Sgr-Tab}.}
  \label{V505Sgr-RV}
\end{figure}

In conclusion, the combined analysis of V505~Sgr leads to results which are not in agreement with previous
analyses. The spectral type and the mass of such a body is in contradiction with the spectral analysis by
\cite{Tomkin1992}. This is especially due to the inconsistency of the results from the separate LITE and
astrometry. Only one astrometric measurement was obtained during the last decade, which is not sufficient for
the precise determination of the orbit. Also in the $O-C$ analysis new precise times of minima are needed to
prove the LITE variation. The similar situation also apply with the radial-velocity measurements. Obtaining the
spectra and the radial velocity of the third component would be very helpful.

\section{HT~Vir} \label{HTVir}

One member of the visual binary STF~1781 is the eclipsing binary system HT~Vir (ADS~9019, HD~119931, HIP~67186,
BD+05~2794). HT~Vir is a contact W~UMa system, with a period of about $0.4$ days and the depths of minima of
about $0.4$~mag. Both visual components have almost equal brightness. The third component of the system is
brighter than the eclipsing binary HT~Vir during its eclipses and fainter than it during its maxima. The system
is apparently about 7.2~mag bright in $V$ filter and the spectral type was classified as F8V (according to
\citealt{LuRucinski}).

According to \cite{WalkerChambl} the distant astrometric component
was discovered by Wilhelm Struve in 1830 at a separation of about
1.4\arcs and position angle $240^\circ$. Since then, numerous
astrometric observations were obtained (altogether 277, from which
275 were used in our analysis) and the orbit is almost completely
covered by the observations (see Fig. \ref{HTVir-orbit}).

\begin{figure}[t]
   \centering
   \includegraphics[width=9cm]{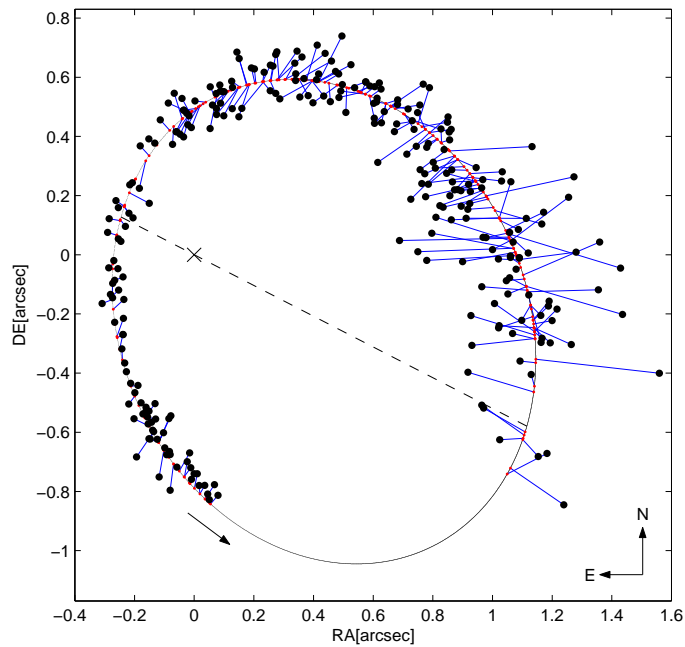}
   \caption{Relative orbit of HT~Vir on a plane of the sky,
   for a detailed description see Fig.\ref{FigQSAql-orbit}.}
   \label{HTVir-orbit}
\end{figure}

\cite{Baize} suggested that the star might be variable. After
then, \cite{WalkerChambl} obtained a complete light curve of
HT~Vir and did the first analysis. It indicated that both
components of the eclipsing pair are almost identical and in
contact. The temperatures of both components are about
$6000~\mathrm{K}$ and the spectral type is estimated as F8V. The
same (combined) spectral type was derived from the spectral
analysis by \cite{LuRucinski}, but with a strong contribution of
the third component. The total mass of the eclipsing pair is
$M_{12}~=~2.3~$\Mo \citep{dAngelo}.

\cite{LuRucinski} discovered that the distant component is also a
binary. They have measured the spectra of HT~Vir eclipsing pair,
and discovered also the lines from the third component in the
spectra and their RV variations with a period of about $32.45$
days. We therefore deal with a quadruple system.

Despite the spectral analysis and a large set of astrometric
observations, there were only a few times of minima published
during the last few decades. The main reason is the relatively
recent discovery of the photometric variability of HT~Vir. The
first times of minima come from 1979. Since then, there were only
31 observations obtained (see Fig. \ref{HTVirOC}). Four new
observations were obtained, two of them were observed at
Ond\v{r}ejov observatory, one by L.Br\'{a}t and the last one by
R.D\v{r}ev\v{e}n\'y. One unpublished observation by M.Zejda was
also used and four times of minima by M.Zejda published in
\cite{Zejda2004IBVS5583} were recalculated, because the
heliocentric correction was wrongly
computed. 

\cite{WalkerChambl} published the first rough estimation of the
proposed amplitude of LITE from the parameters of the astrometric
orbit. Their value (0.18 day) is not too far from the present one
(0.13 day).

\begin{figure}[t!]
  \centering
  \scalebox{0.75}{
  \includegraphics{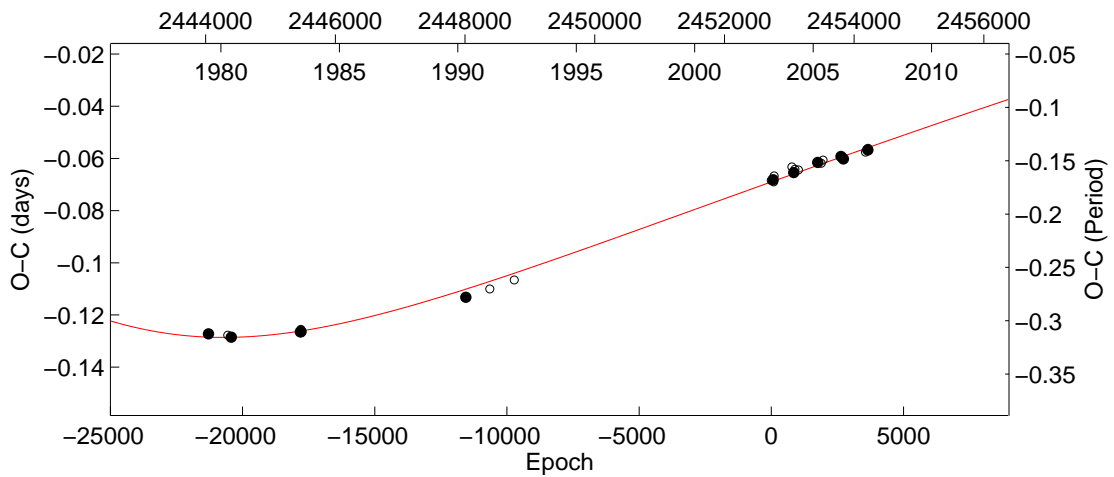}}
  \caption{An $O-C$ diagram of HT~Vir. The description is the
  same as in the previous $O-C$ figures, all minimum times are
  the photoelectric or CCD ones.}
  \label{HTVirOC}
\end{figure}

The final plot of the relative astrometric orbit of HT~Vir is in
Fig. \ref{HTVir-orbit}. The results, the parameters of the orbit
around the common barycenter of the system, are given in Table
\ref{HTVir-Tab}. The values of these parameters
($A,p_3,i,e,\omega,\Omega,T_0,JD_0,P$) were obtained minimizing
the $\chi^2_{comb}$.

The new elements for the astrometric orbit could be compared to
these by \cite{Heintz1986}, which are the following: $p_3~=~274.0$
yr, $e~=~0.638$, $a~=~1010$~mas, $i~=~42.7^\circ$,
$\Omega~=~176.4^\circ$, $\omega_3~=~250.0^\circ$. As one can see,
the period of the new orbit is a bit shorter, but the main
differences in these values are the angles $\omega$ and $\Omega$.
The same fit to the astrometric data could be reached with
simultaneously transformed values $\omega_3 \rightarrow \omega_3 +
180^\circ$ and $\Omega \rightarrow \Omega + 180^\circ$. This only
means the interchange of the role of the two components. This
result therefore indicates the incorrect identification of the
variable HT~Vir in the system in our analysis (the variable was
supposed to be the
component A) and also in the WDS catalogue, see WDS notes \footnote{$\mathrm{http://ad.usno.navy.mil/wds/wdsnewnotes\_main.txt}$}. 
While \cite{PribRuc2006} correctly identified the variable HT~Vir
as a B component and A as a single-lined binary.

If one adopts these parameters to estimate the mass function of
the distant pair (mass function of the whole pair, not the
individual components), one obtains $f(M_3)~=~0.17~$\Mo. This is
quite a high value, dictated by the large amplitude of the LITE.
With the total mass of the primary and secondary $M_{12} =
2.3$~\Mo~one gets the third mass of $M_3 = 2.14~$\Mo. The mass of
the distant pair is quite high (\cite{dAngelo} derived the mass
for some 50\% lower, $M_3=1.15~$\Mo), but note that also this
object is a binary and we do not know the individual masses. From
the spectroscopic observations (to remind, it is a SB1-type
binary), one is only able to estimate the mass function of the
components, or some upper limit for one of them (we do not know
the inclination). The present result $M_3$ is the total mass of
the SB1 pair $M_{3,1} + M_{3,2}$; the limit for the
invisible-component mass $M_{3,2}~\cdot~\sin(i') = 0.075~$\Mo. If
the the coplanar orbit is assumed, high difference in masses would
arise, one component should be much more luminous and also more
luminous than the eclipsing pair itself, which is not the case. In
fact the whole system is not coplanar (see e.g. $i=315.5^\circ$
and the inclination of the EB close to $90^\circ$). If one assumes
two approximately equal masses, there is a problem with the
luminosity, because the distant pair has to be roughly as luminous
as the eclipsing pair. This could only be satisfied if one
component is underluminous or degenerate.

\begin{table}[t!]
\caption{The parameters from the combined solution of HT~Vir.}
\label{HTVir-Tab} \centering \scalebox{0.75}{
\begin{tabular}{c c c c c c c c c c| c c}
 \hline \hline
 Parameter&  $JD_0$    &    P    &$p_3$& $A$  & $T_0$ &$\omega$& $e$ & $i$ &$\Omega$&$f(M_3)$& $M_3$ \\
  Unit    &   [day]    &  [day]  &[yr] &[day] & [day]  &[deg] &   -  &[deg]& [deg]  & [\Mo] &[\Mo] \\ \hline
  Value   &2452722.5040&0.4076696&260.7&0.1274&2442832 &250.9 & 0.640&45.4 & 180.8  & 0.169 & 2.10 \\
  Error   &   0.0050   &0.0000025& 0.5 &0.0026&  61    & 0.7  & 0.005&3.7  & 2.6    & 0.009 & 0.11 \\
 \hline \hline
\end{tabular}}
\end{table}

One has to take into consideration also the comment on the
light-curve solution by \cite{WalkerChambl}. Using the Wood's
model, they discovered that if the third light $L_\mathrm{3}$ is
fixed to be the equal to the light from the distant visual
component (it means $L_\mathrm{3}=0.5$), the solution of the light
curve is unrealistic. To conclude, the system could be much more
complicated than the approach that was used here. There may be
some additional component(s) or the distant pair is composed from
evolved stars, away from the main sequence. Especially because of
the resultant mass and luminosity of the distant pair, the body
causing the astrometric variation is probably different from the
one causing LITE, but this conclusion will be proven only if also
the nonlinear part of the $O-C$ diagram is covered.

\section{The problematic case: V2388 Oph}

Another system where the astrometric orbit is known and also the set of times of minima is available is
V2388~Oph. The contact eclipsing binary system V2388~Oph (FIN~381, HD~163151, HR~6676) is $\beta$~Lyrae type
(according to the Simbad database), or more likely W~UMa type (according to \citealt{Rodrig1998}). Its orbital
period is about $0.8$ days, apparent brightness about 6.3~mag in \emph{V} filter and the spectral type was
classified as F5Vn (according to Hipparcos Catalogue, \citealt{HipCatal1993}), or F3V according to
\cite{Rucinski2002}. The depths of its minima are 0.3 and 0.25~mag for primary and secondary, respectively.

\begin{figure}[t!]
   \centering
   \includegraphics[width=9cm]{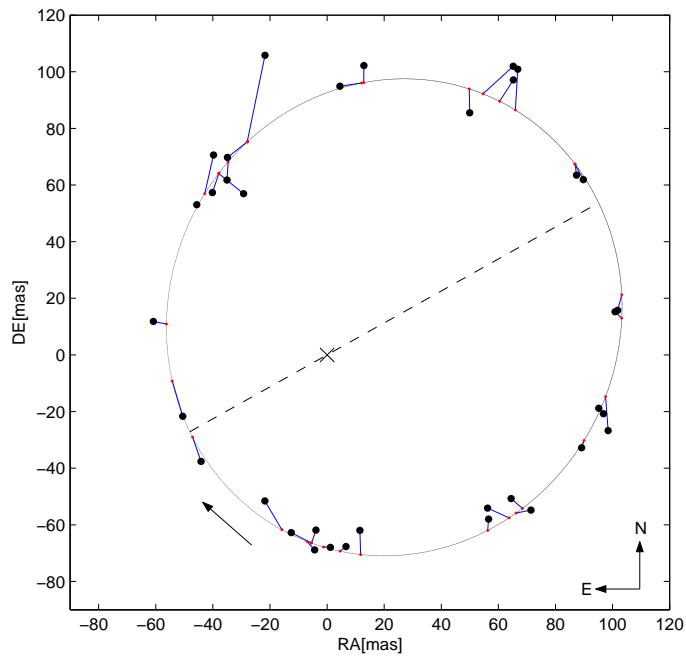}
   \caption{Relative orbit of V2388~Oph on a plane of the sky,
   \emph{Solution I.} Three measurements were neglected due to their
   large scatter.}
   \label{V2388Oph-orbit}
\end{figure}

The first astrometric observation of the third component was obtained in 1959 \citep{Fin1963a} in a distance of
about only 100~mas. During the next decades there was recognized rapid movement of this component around the
primary. A preliminary orbit was calculated by \cite{Baise1988} with a period of about 8.3 yr, semimajor axis
0.09\arcs and eccentricity 0.29. Nowadays orbit is a little bit different ($p_3 = 8.9$ yr, $a=0.09$\arcs,
$e=0.33$). It is evident, that the movement is very rapid and since its discovery the body revolved 5 times
around the eclipsing pair.

The photometric variability was discovered by \cite{Rodrig1998},
but the variability is also evident from the Hipparcos
observations. The $ubvy$ light curves were obtained and analyzed.
With the RV analysis made by \cite{Rucinski2002} one is able to
get the complete picture of the system. The mass ratio from the
photometry is 0.27, but more precisely from spectroscopic analysis
$q=0.186$, the minimum mass $(M_1+M_2) \sin^3 i = 1.93$~\Mo.
According to $ubvy$ analysis the system was classified as F5Vn,
but \citeauthor{Rucinski2002} on the basis of their spectroscopic
observations suggested slightly earlier spectral type F3V. The
magnitude difference between astrometric components is $\Delta m =
1.80$~mag and the contribution of the third component to total
luminosity is circa 20\%. Mass of the third component is
$M_3=1.36$~\Mo $\,$ according to \cite{dAngelo}. The EB system is
SB2-type (according to \citealt{Rucinski2002}) also with the third
component visible in the spectra. The mean radial velocity of the
third component $V_{\gamma,3} = 30.64$ km$\cdot$s$^{-1}$
significantly differs from the center-of-mass velocity of the
binary, $V_0 = 25.88$ km$\cdot$s$^{-1}$, which could be caused by
the motion on the 9-yrs orbit. In the same paper was mentioned an
unexplained \emph{cross-talk}, which means variation in radial
velocities of the distant companion in phase with the period of
the eclipsing binary. Having only this one RV data point one could
not determine any of the relevant parameters of the long orbit.

\begin{figure}[t!]
   \centering
   \scalebox{0.75}{
   \includegraphics{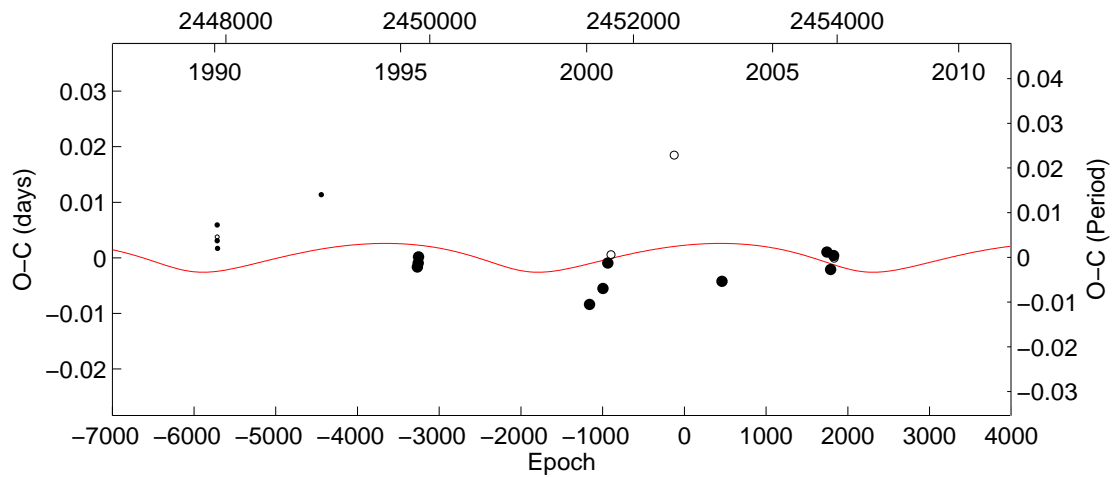}}
   \caption{An $O-C$ diagram of V2388~Oph, \emph{Solution I.} All
   minima times are the photoelectric or CCD ones (the first 5 of
   them are from the \emph{Hipparcos} mission).}
   \label{V2388OphOC}
\end{figure}

Rough estimation of the proposed $O-C$ variation and its magnitude
was done by \cite{Rucinski2002} from the astrometric orbit. Their
analysis results in $A=0.011$ day. Since the discovery of the
photometric variability of V2388~Oph the minimum light of the star
was observed only 18 times. But the individual data points show
different variation, with quite different amplitude, but mainly
with very different period.

The astrometric orbit is plotted in Fig.\ref{V2388Oph-orbit} and
the final fit is satisfactory, with the resultant parameters given
in Table \ref{TableBig2388} -- \emph{Solution I.} The $O-C$
diagram is in Fig.\ref{V2388OphOC}. As one can see, the fit is
unacceptable, but the method applied here was the same as in the
previous cases.

\begin{table}[t!]
\caption{The final results: the case of V2388~Oph, Solution I. and
II. The description is the same as in Table \ref{TableBig2}.}
 \label{TableBig2388} \centering \scalebox{0.88}{
\begin{tabular}{c c c c}
\hline\hline
Parameter  &  Unit      & V2388~Oph -- \emph{Solution I.} & V2388~Oph -- \emph{Solution II.} \\
\hline
  $JD_0$   & $[$HJD$]$  & $2452500.3829 \pm 0.0066$   & $2452500.3799 \pm 0.0009$  \\
  $P$      & $[$day$]$  & $0.8022986 \pm 0.0000025$   & $0.8022995 \pm 0.0000037$ \\
  $q$      & $[$day$]$  & $0.0$                       & $-3.472 \cdot 10^{-10} \pm 0.010$ \\
  $p_3$    & $[$yr$]$   & $9.01 \pm 0.28$             & $9.01 \pm 0.29$       \\
  $T_0$    & $[$HJD$]$  & $2549594.9 \pm 60.7$        & $2549666 \pm 77$     \\
  $\omega$ & $[$deg$]$  & $243.6 \pm 1.7$             & $301.6 \pm 1.6$     \\
  $e$      &            & $0.329 \pm 0.002$           & $0.318 \pm 0.003$    \\
  $A$      & $[$day$]$  & $0.0026 \pm 0.0015$         & $0.00001 \pm 0.00012$ \\
  $a$      & $[$mas$]$  & $88.2 \pm 67.4$             & $85 \pm 1080$       \\
  $i$      & $[$deg$]$  & $156.7 \pm 2.9$             & $180.00001 \pm 1.8$  \\
  $\Omega$ & $[$deg$]$  & $181.0 \pm 4.8$             & $240.5 \pm 2.3$      \\
  \hline
  $M_{12}$ &  [\Mo]     & $2.14$                      & $2.14$               \\
References &            & \cite{Yakut2004}            & \cite{Yakut2004}    \\
  $\pi$    & $[$mas$]$  & $14.72 \pm 0.81$            & $14.72 \pm 0.81$   \\
  $D$      & $[$pc$]$   & $67.9 \pm 3.7$              & $67.9 \pm 3.7$      \\
  \hline
  $a_{12}$ & $[$AU$]$   & $1.14 \pm 0.63$             & $0.56 \pm 1940 $      \\
  $f(M_3)$ &  [\Mo]     & $0.0012 \pm 0.0011$         & $0.00000002 \pm 0.00000001$\\
  $M_3$    &  [\Mo]     & $0.50 \pm 0.47 $            & $0.23 \pm 714.00 $     \\
  \hline
  Data set &            & 35a + 18m                   & 35a + 17m        \\
\hline \hline
\end{tabular}}
\end{table}

Where the problem could be? The crucial point at the first time is to compare the results from these different
approaches and take it into the consideration. If these two results are incompatible (as in this case) this
combined approach is unusable. And this is the case of V2388~Oph.

\begin{figure}[b!]
   \centering
   \scalebox{0.75}{
   \includegraphics{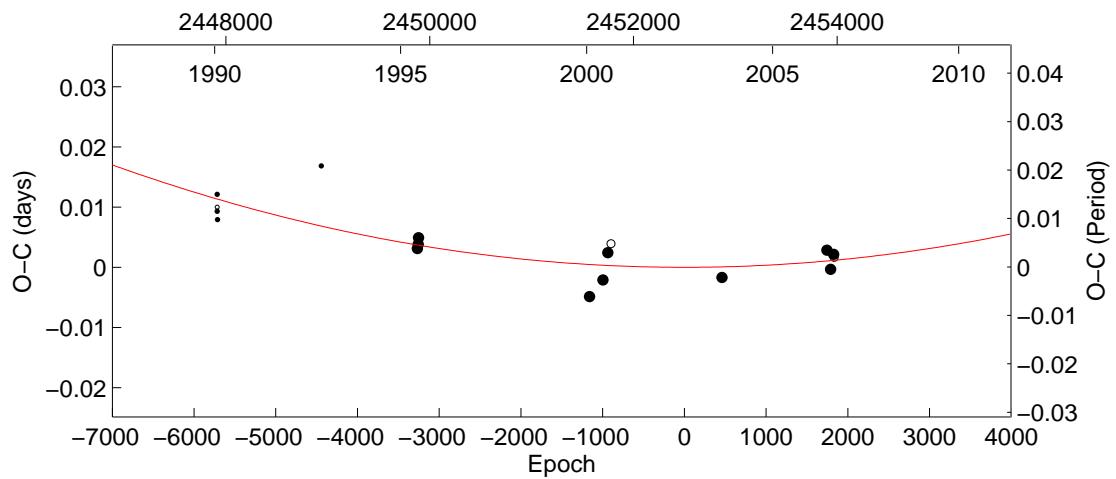}}
   \caption{An $O-C$ diagram of V2388~Oph, \emph{Solution I.} -- only
   the quadratic term was used.}
   \label{V2388OphOC0}
\end{figure}

Using only $O-C$ analysis to the set of times of minima, one gets
the period $p_3=5.1$~yr, $e=0.56$ and $\omega=102.4^\circ$. It is
evident that the astrometry leads to the different set of
parameters and the combined solution should be nonsense. Another
solution could be the very short one, with the period
$p_3=2.99$~yr, but this is only the hypothesis, because the
variation in $O-C$ diagram is not covered very well and this is
only a \emph{sampling} frequency of the individual data points.

The astrometric orbit is well defined, but there is a question
about the accuracy of the individual times-of-minima data points.
Without the input data (the rough photometry), one can doubt, if
all the measurements are accurate enough or some of them could be
neglected. Interesting is the sequence of 4 times of minima, which
are rising up near the epoch -1000 (3 primary and 1 secondary). Is
this the real effect in $O-C$ diagram, or is it just the real
scatter of the measurements? The difference is about 0.01 day, or
circa 14 minutes. This is quite large to be only a scatter, but
one does not know the conditions during the observation, etc.

\begin{figure}[t!]
   \centering
   \includegraphics[width= 9cm ]{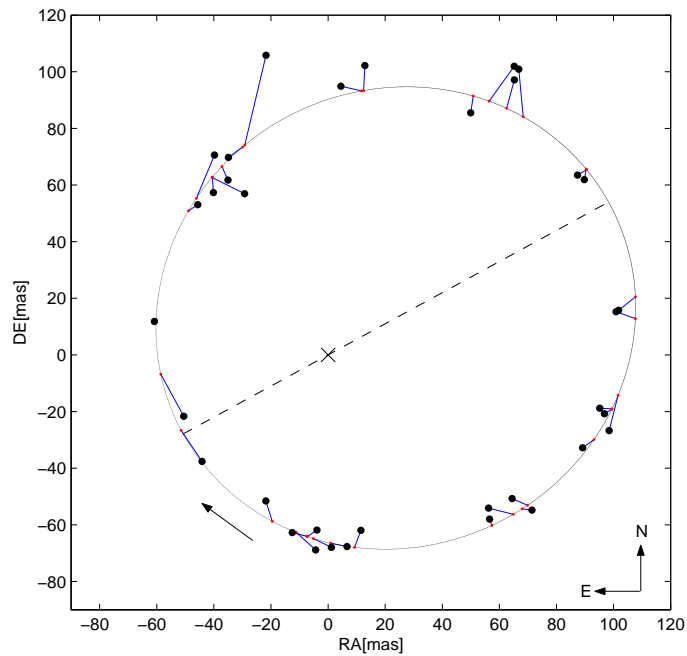}
   \caption{Relative orbit of V2388~Oph on a plane of the sky,
   \emph{Solution II.}}
   \label{V2388Oph-orbit0}
\end{figure}

If one decide to neglect one data point -- the secondary minimum
time near the epoch 0, and include also the quadratic term in the
ephemeris, one will get \emph{Solution II.} -- see Fig.
\ref{V2388OphOC0}. Using the combined approach also the
astrometric orbit could be plotted (see Fig.
\ref{V2388Oph-orbit0}) and as one can see, the difference between
Figs.\ref{V2388Oph-orbit} and \ref{V2388Oph-orbit0} is not so
significant. The main difference is in the $O-C$ diagram, where
only the quadratic term arises and no LITE is presented. This
means that the orbit is just \emph{face-on}, i.e. the inclination
is very close to $180^\circ$. See the resultant parameters of such
fit in Table \ref{TableBig2388}. The parameters of LITE are not
very convincing, because of inclination is almost $180^\circ$, but
the LITE is necessary to compute, because the amplitude of
astrometric variations is computed from the amplitude of LITE.
This means the mass of the third component was derived precisely,
while the mass function of such a body is very inaccurate.

To conclude, it is difficult to decide which solution is the right one. Only further data points, especially
times of minima, would confirm the 9-yrs variation in the $O-C$ diagram. Due to very short period of the third
body the shift in the times of minima should be evident after a few months of observations. Also measuring the
RV curve and the analysis of the systemic velocity could be very helpful, because the period is short and the
last one was carried out more than 6 years ago.

\section{Other systems} \label{Catal}

In this section are presented the systems which were found to be EBs as well as members of the visual binaries.
The limitation about the number of times of minima, which was presented in Introduction to Chapter \ref{Ch3},
does not play a role. Only the systems analyzed in detail above were omitted. The systems presented here were
found by scanning the objects in the WDS catalogue and trying to identify the EBs in this sample of stars. This
survey is slightly following the paper on "Eclipsing binaries in multiple-star systems", \cite{Chambliss1992}.
The number of such systems has grown rapidly since then, but the main difference is the selection criterion. In
\cite{Chambliss1992} are presented all of the multiple systems with eclipsing binaries, which were known for the
author. \citeauthor{Chambliss1992} mentioned that 80 EBs are known to be components of the multiple-star systems
and 37 of them were presented in more detail. For this thesis there were selected only these systems, which were
discovered to be EBs and astrometric variables after \cite{Chambliss1992}, the systems which have sufficiently
large data set in both methods to do the simultaneous analysis, or the systems for which the new astrometric
orbit was calculated for the first time. Some of the presented systems were also included due to their
misidentification as EBs.

\subsection{HD 123}
HD~123 (V640~Cas, HR~5, STF~3062AB) is an eclipsing binary which
spectral type was classified as G5V. Its V magnitude is of about
5.93, but there were only a few times of minima found in the
published literature, no photometric analysis was found. The
eclipse observations are questionable and recent measurements
indicate possible misidentifications of the star as eclipsing
binary. On the other hand the astrometry covers whole orbit.
Altogether 572 data points were obtained during 170 years.
\cite{Soderhjelm1999V640Cas} computed the orbital parameters, the
period about 107~yr and angular semimajor axis about 1.4\arcs.

\subsection{HD 1082}
HD~1082 (V348~And, A~1256AB, HIP~1233) is an Algol-type EB, which
spectral type was classified as B9V. Its apparent magnitude is
6.76 in V filter. The same situation as in the previous case also
apply here, there were neither no times of minima nor the
photometric analysis found in literature. The astrometric orbit is
covered by 61 data points obtained during 93 years and covering
the range from 4 to 223 degrees in $\theta$. From these data the
orbit was calculated by \cite{Olevic2002V348And}, resulting in
$p_3 = 138$~yr and $a=150$~mas.

\subsection{HD 4134}
HD~4134 (V355~And, STF~52AB, HIP~3454) is also an Algol-type EB
with spectral type classified as F5 and the magnitude $V =
7.69$~mag. No times of minima were obtained. Astrometry covers
only 20$^\circ$ with 51 measurements observed in the last 170
years, the orbit was not computed. 

\subsection{HD 10543}
HD~10543 (V773~Cas, BU~870AB, HR~499) is an Algol-type EB with
spectral type A3V and the magnitude $V = 6.21$~mag. The orbital
period of the eclipsing pair is about 1.3 days, but only one time
of minimum was observed. The astrometry covers about
80$^\circ$ with 79 observations made during 120 years. 
The astrometric orbital parameters were calculated by
\cite{Popovic1995V773Cas}, resulting in period about 304~yr and
semimajor axis about 1\arcs.

\subsection{HD 12180} 
HD 12180 (AA~Cet, ADS~1581~A, HIP~9258) is W~UMa type EB, sp F2V,
$V=7.22$~mag, and orbital period about 0.54~d. There were more
than 200 times of minima obtained during the last 40 years, but
with no significant LITE variation. Also the astrometric
observations, which were obtained during more than 200 years, do
not show any evident variation and any orbital solution could be
found from this data set.

\subsection{HD 14817} 
HD~14817 (V559~Cas, STF~257AB, HIP~11318) is one component of the visual binary STF~257AB. It is the eclipsing
binary of Algol-type, as well as spectroscopic binary, spectrum classified as B8V, apparent brightness of about
7.02~mag in \emph{V} filter and orbital period of about 1.58~day. There were 7 times of minima observed since
1971 to 1991. Due to its very long orbital period, about 836~yrs (see e.g. \citealt{Hartkopf2001Catalog}), only
about one third of the orbit is covered by the observations (101 observations and the change in $\theta$ is
about 100$^\circ$). The astrometric measurements are available since 1830 and the periastron passage occurred in
1932, so the part of the orbit near periastron is sufficiently covered. Regrettably, in that time the minima
times are missing.

\subsection{HD 18925}
HD~18925 ($\gamma$~Per, 23~Per, HJ~2170A, HR~915) is an Algol-type
EB with spectral type of about G8III and the magnitude $V =
2.95$~mag. Astrometric observations (altogether 67) were obtained
during 65 years and the orbit was calculated.
\cite{Pourbaix2000GamaPer} published the parameters, $p_3 =
14.6$~yr and $a=144$~mas, and the inclination $i = 90.6^\circ$, so
the orbit is just edge-on and the astrometric observations are
only "in the line". The position of the orbit indicates that the
occultations and eclipses may happen. These were predicted and
successfully observed in September 1990 (see
\cite{Griffin1994GamaPer} for details). The system is similar to
$\beta$~Aur (see below). Therefore, this object is not suitable
for the simultaneous analysis.

\subsection{HD 19356} 
HD 19356 (Algol, $\beta$~Per, LAB~2Aa, HR~936) is well-known
prototype of the Algol-type binaries. Its spectral type is B8V and
apparent brightness 2.12~mag in \emph{V}. The time of minimum
brightness was first measured by Montanari on 8 November 1670
(although known from historical times). Nowadays set of times of
minima is really large, about 1400 observations, covers a few
centuries, but the detailed description of the $O-C$ diagram is
still missing. The system is rather complicated, but the distant
component with the orbital period about 1.8~yr discovered firstly
on the basis of the radial velocity variations was found in 1973
by a speckle camera and the orbit of this component is now well
established ($a=94.6$~mas and $e=0.23$, according to
\citealt{Pan1993Algol}).

\subsection{HD 24071}
HD 24071 (DUN~16, HR~1189, HIP~17797) is probably $\beta$~Lyrae type star, its apparent brightness is about
4.2~mag in \emph{V} filter and spectrum classified as B9V. The star is hardly measurable, because there are
together 4 stars very close each other (only 5\arcs distant) and it is not clear, if all these components belong
to the system. The astrometry was obtained in 1826 for the first time, there were 80 measurements in total,
which reveals the change in position angle of about 15$^\circ$. The orbit computed according to these data is
not very conclusive (the orbital period more than 5000~yr).

\subsection{HD 25833}
HD~25833 (AG~Per, STT~71AB, HIP~19201) is an Algol-type EB
spectral type B5Vp and the relative magnitude $V = 6.69$~mag.
There were 101 times of minima, collected from the published
literature. These minima were obtained from 1920's till now.
AG~Per is one of the most typical apsidal-motion systems, which
has been analyzed for the apsidal motion several times (see e.g.
\citealt{Wolf2006AGPer}). One could also apply the hypothesis of
combining the apsidal motion and the LITE into one joint solution
(similar to the $\zeta$~Phe case). Also the precise light curves
were measured and analyzed (see \citealt{Woodward1987}). The main
problem arises with the astrometry. Altogether 38 measurements
cover more than 30 degrees in $\theta$, but there is no evident
periodicity and the orbit could not be constructed from this data set. 

\begin{figure}[b!]
   \centering
   \scalebox{0.72}{
   \includegraphics{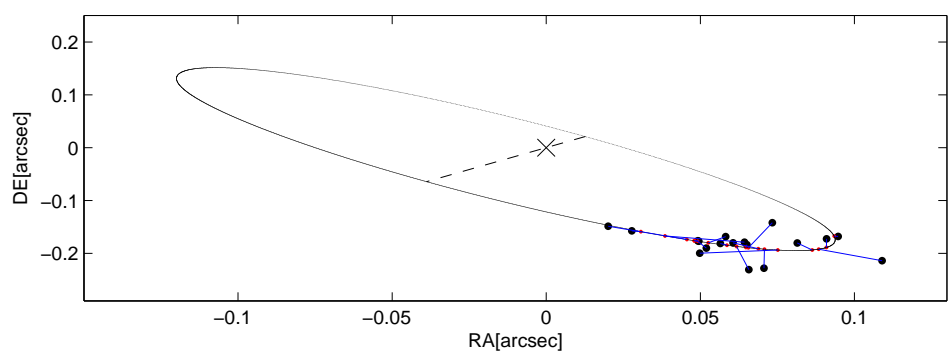}}
   \caption{Relative orbit of V592~Per on the plane of the sky.}
   \label{V592Per}
\end{figure}

\subsection{HD 29911}
HD~29911 (V592~Per, COU~1524, BD+39~1054) is a $\beta$-Lyrae EB
with spectral type classified as F2 and its apparent magnitude $V
= 8.37$~mag. There was only one time of minimum obtained. The
astrometry covers only 19$^\circ$ with 18 data points obtained
during 26 years. The plot is in Fig.\ref{V592Per}, where the
theoretical orbit is also shown. Its period is about 117~yr and
semimajor axis of about 230~mas, it was computed for the first
time.

\subsection{HD 36486} 
HD~36486 ($\delta$~Ori~A, 34~Ori~A, HEI~42Aa, HR~1852) is an eclipsing binary, sp O9.5II, $V=2.23$~mag, and an
orbital period 5.7 days. Only 9 times of minima were found in literature, but these minima do not show any
significant LITE variation (more probably apsidal motion). On the other hand, there is significant motion on the
plane of the sky, while the astrometry was first obtained in 1978 and since then 38 observations were obtained
(see Fig.\ref{DelOri}). The orbit is only a preliminary one and has not been published till yet. The period of
the orbit is about 313~yr and the semimajor axis 280~mas.

\begin{figure}[t!]
   \centering
   \scalebox{0.72}{
   \includegraphics{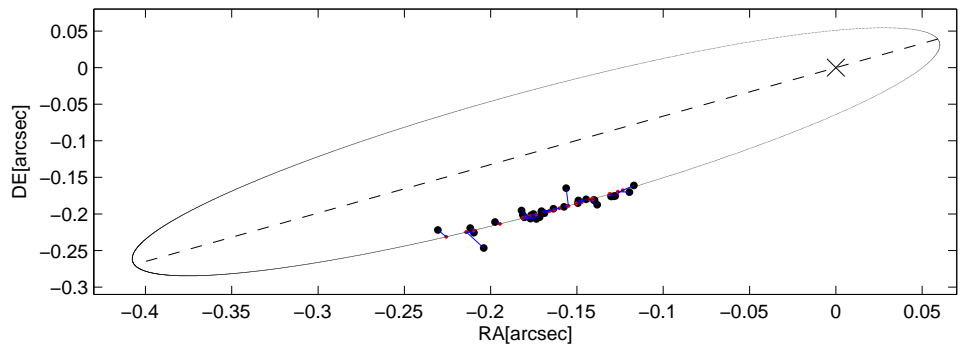}}
   \caption{Relative orbit of $\delta$~Ori on the plane of the sky.}
   \label{DelOri}
\end{figure}

\begin{figure}[b!]
   \centering
   \scalebox{0.72}{
   \includegraphics{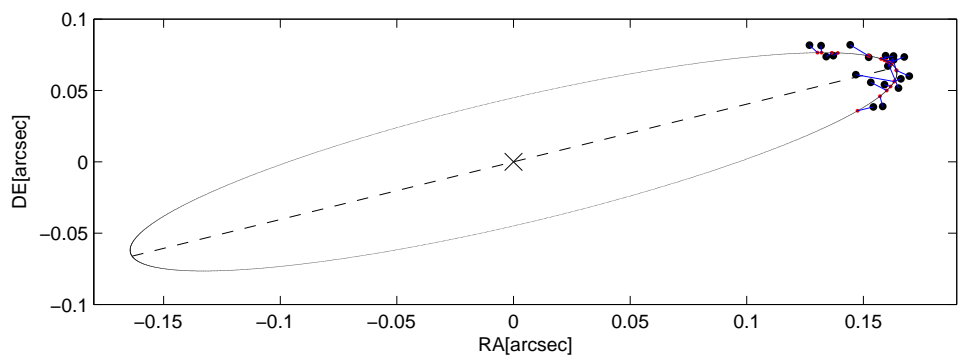}}
   \caption{Relative orbit of V1031~Ori on the plane of the sky.}
   \label{V1031Ori}
\end{figure}

\subsection{HD 38735} 
HD 38735 (V1031~Ori, MCA~22, HR~2001) is an Algol-type detached system, $V=6.06$~mag, sp A4V, period about 3.41
d. There were 9 times of minima found in literature. The orbit of the binary is shown in Fig.\ref{V1031Ori}. It
consists of only 20 observations obtained from 1980 to 1997. This orbit was not published yet and is only a
preliminary one. Its orbital period is about 92~yr and the semimajor axis about 0.18~\arcs. But according to the
RV measurements by \cite{Andersen1990V1031Ori}, the orbit should be much larger, and the period about 3700~yr.
The third-component lines were observed in the spectra of V1031~Ori and radial velocities on the 92~yr orbit
would be much larger than measured. Because the orbit is covered by data only very poorly, only further
astrometric observations, as well as precise radial velocity investigation will reveal the nature of the system.

\subsection{HD 40183}
HD 40183 ($\beta$~Aur, 34~Aur, HR~2088) 
is an Algol-type EB, apparently bright
about 1.9~mag in V filter and its spectrum was classified as A2IV.
It is one of the brightest and nearest spectroscopic as well as eclipsing binaries, but due to its high
brightness only a few observations were done. Altogether 23 times of minima were measured over the whole
century. Photometric (see \citealt{Johansen1971BetaAur}) and also spectroscopic (see
\citealt{Nordstrom1994LCaRVproBetaAur}) analyses were published. Similarly to the previous case $\gamma$~Per,
the astrometric orbit could be identified with the eclipsing binary orbit.
Therefore, this system does not belong to this
survey, but it is of big importance for the present EB knowledge.
Altogether 28 data points 
sufficiently cover the whole 4-day astrometric orbit. 
The EB components could be resolved, because the system is
relatively close (about 24~pc). Detailed description of the
technique used (interferometry with The MARK III long-baseline
optical interferometer on Mount Wilson) and the analysis is in
\cite{Hummel1995}. 
The parameters of the orbit from interferometric measurements were compared by \citeauthor{Hummel1995} with the
previously found values from photometry and spectroscopy. The different approaches lead to the same results
within their respective errors. Also the determination of the distance to this unique binary from four
independent methods gave the comparable results.

\subsection{HD 57061} 
HD 57061 ($\tau$~CMa, 30~CMa, FIN~313Aa, HR~2782) is the brightest star in the open cluster NGC~2362. It is a
$\beta$~Lyrae-type EB, period about 1.28~d. $\tau$~CMa is also a spectroscopic binary with an orbital period of
about 154.9~day and the EB is probably the main component of the SB. This interesting system therefore contains
both the longest period spectroscopic binary and the shortest period eclipsing binary known among the O-type
stars. The system was precisely analyzed by \cite{Leeuwen1997TauCMa}. This triple system is one member of the
visual binary FIN~313Aa, which has been measured 32 times since 1951. The change in position angle is only about
15$^\circ$, so any orbital solution is acceptable.

\subsection{HD 66094}
HD 66094 (V635~Mon, A~1580AB, BD-08~2186) is an Algol-type EB with
primary star classified as a spectral type F5 and apparent
brightness of about 7.31~mag in \emph{V} filter. A lot of times of
minima were collected (altogether 113), but these data points
follow the linear ephemeris without any indication of the proposed
LITE. On the other hand the astrometric orbit is defined very
precisely. 23 data points measured over the century cover about a
half of the orbit, and the analysis results in 160-yrs orbit, with
the semimajor axis of about 280~mas.

\subsection{HD 71581} 
HD 71581 (VV~Pyx, B~2179AB, HR~3335) is an Algol-type EB, spectrum
A1V, brightness 6.58~mag in \emph{V} and orbital period about
4.6~days. There were 7 times of minima found in literature (1976 -
1983), but these data show very long apsidal motion (in order of
decades or centuries). The astrometric orbit is also covered only
very poorly (11 observations obtained during 38~years show the
change in position angle of about 13$^\circ$).

\subsection{HD 74956}
HD 74956 ($\delta$~Vel, HR~3485, HIP~42913) is an Algol-type
eclipsing binary classified as A1V spectral type, with $V = 1.95$~
mag. The star was discovered to be a photometrically variable in
1997 (see \cite{Otero2000} for details), the period of such
variation is about 45 days. Altogether 8 times of minima were
collected, but these data indicates very long apsidal motion (on
the timescale of centuries). Astrometric orbit consists of 37
measurements, which define the orbit with the period of about 142
years and the semimajor axis of about 2\arcs (according to \citealt{Alzner2000DelVel}). 
The whole system is in fact more complicated, consists of two
proper motion pairs (2\arcs and 6\arcs) separated by 69\arcs. Also
the primary component was resolved as a double star
interferometrically. We therefore deal with a quintuple system (at
least 5 components).

\begin{figure}[b!]
  \centering
  \scalebox{0.72}{
  \includegraphics{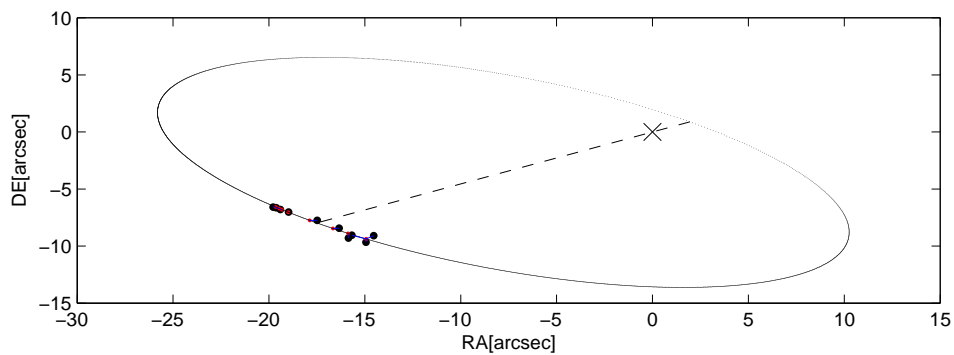}}
  \caption{Relative orbit of AC~UMa on the plane of the sky.}
  \label{ACUMa}
\end{figure}

\subsection{AC~UMa} 
AC~UMa (ARG~21B, BD+65~671B, AG+65~453) is an Algol-type EB,
spectrum classified as A2, brightness 10.3 in V filter. The
orbital period is about 6.85~days. There were 69 times of minima
found in literature and there could be some variation in order of
decades, but this is only hypothesis, larger data set is needed.
The astrometry is shown in Fig. \ref{ACUMa}, only 10 observations
during 106 years were obtained. As one can see, only a linear part
of the orbit is covered by data, so one cannot derive the
parameters of the orbit precisely. This leads to extremely long
period about 1200~yr, which could be even higher.

\subsection{HD 82780}
HD 82780 (DI~Lyn, A~Hya, STF~1369AB, HR~3811) 
is an Algol-type EB, classified as F2V. Its magnitude is $V =
6.76$~mag. There is only one time of minimum measured and the
small arc of the astrometric orbit was observed during 22 years,
covering about 30 degrees. No acceptable solution could be found.

\subsection{HD 91636}
HD 91636 (TX~Leo, 49~Leo, STF~1450AB, HR~4148) is an Algol-type
EB. Its spectrum was classified as A2V and its apparent brightness
is about $V = 5.67$~mag. There were 6 times of minima observed
since 1930. The astrometric data set is much larger, about 132
measurements secured during the last 180 years, but the motion is
undetectable. 

\subsection{HD 101205}
HD 101205 (V871~Cen, I~422AB, HIP~56769) is a $\beta$~Lyrae type EB with its spectrum classified as O8V and the
brightness of about $V = 6.49$~mag. There is a brief paper on the photometric observations of V871~Cen, together
with a minimum time derived (see \citealt{Mayer1992V871Cen}). The astrometry secured during the last 90 years
reveals the change in $\theta$ about 20$^\circ$, but no acceptable solution could be found (these data lead to
an orbit of period about 4500~yr).

\subsection{HD 101379j}
HD 101379j (GT~Mus, 12~Mus, B~1705AB) is an eclipsing binary, with
spectrum classified as G2III and the brightness $V = 5.17$~mag.
The astrometric data were obtained during 60~years and cover about
130$^\circ$ of the orbit. This SB1-type spectroscopic binary was
analyzed by \cite{Parsons2004}. The orbital period of GT~Mus is
about 56 days, which is different from the period of the SB1
binary. Therefore, the component A is the EB, while B is the SB.
No minima were derived. The astrometric orbit has a period circa
91~years \citep{Parsons2004}. Some observations
indicates that one of the components is RS~CVn-type star. 

\subsection{HD 103483} 
HD~103483 (DN~UMa, 65~UMa~A, HR 4560) is an Algol-type EB, sp
A3Vn, $V=6.54$~mag, orbital period 1.73~days. There were only twelve  
times of minima found in published literature (the first ones from
1979). The astrometric orbit is covered sufficiently, the first
astrometric observation came from 1908, and the parameters of the
orbit are known ($p_3 = 136.5$~yr, $a = 230$~mas, according to
\citealt{Aristidi1999DNUMa}).

\subsection{HD 110317j}
HD 110317j (VV~Crv, STF~1669AB, HIP~61910) is an eclipsing binary with the spectrum classified as F5IV and the
brightness of about $V = 5.27$~mag. There were no times of minima found in the literature. The astrometric data
set consists of 156 measurements secured during 180 years, which yielded a change in $\theta$ of
about only 14$^\circ$. 

\subsection{HD 114529}
HD 114529 (V831~Cen, SEE~170AB, HR~4975) is the $\beta$~Lyrae
system, sp B8V, $V=4.58$~mag, orbital period of about 0.64~d. No
published minima were found. The astrometric orbit was derived
according to 40 observations secured during the last 100~years,
resulting in $p_3 = 27$~yr and $a = 185$~mas (according to
\citealt{Finsen1964V831Cen}).

\subsection{SAO 45318}
SAO 45318 (ET~Boo, COU~1760, HIP~73346) is a $\beta$~Lyrae eclipsing binary, spectral type F8. Its apparent
brightness is $V = 9.09$~mag. There were found a few times of minima, covering the last 5 years. There is
possibly some variation in the $O-C$ diagram, but its amplitude is only about 0.001 days and the period about
1.25 years. On the other hand the astrometric measurements were obtained since 1978 till 1999, altogether 20
observations show the change in $\theta$ about 40$^\circ$. The orbit was derived by \cite{Seymour2001ETBoo},
resulting in period about 113~yr and angular semimajor axis 261~mas.

\subsection{HD 133640}
HD 133640 (i~Boo, 44~Boo, STF~1909AB, HR~5618) is a well-known EB of W~UMa type, spectral type G0Vnv and
brightness of about $V = 4.76$~mag. It is quite a complicated system, consists of more than three components.
Many times of minima were observed during the last 90 years, but the detailed description of the behaiour of
these minima is still missing (mass transfer + LITE ?). It was found to exhibit flares as well as to be an X-ray
binary and also many analyses in this part of spectra were obtained. The large astrometric data set consists of
753 observations secured during the last 223 years, and covers the range of position angle from 240 down to 57
degrees. The orbit has period of about 206~yr and semimajor axis 3.8\arcs (see \citealt{Soderhjelm1999V640Cas}).

\subsection{HD 148121}
HD 148121 (V1055~Sco, B~872AB, HIP~80603) is $\beta$~Lyrae EB with
spectral type classified as G3V and brightness $V = 8.64$~mag.
There were no times of minima found in the literature. Astrometric
measurements were obtained 12 times during the last 70 years
covering about 15$^\circ$ in position angle. The orbit was not
derived.

\subsection{HD 157482}
HD 157482 (V819~Her, MCA~47, HR~6469) is an Algol-type EB,
spectrum analyzed as F9Vn and its magnitude is about 5.57 in V
filter. The EB pair is orbiting around the common center of mass
with the third component on the 5.5 years orbit with eccentricity
0.67, LITE is evident. This is the only system where the LITE was
analyzed together with the other methods, namely the
interferometry and RV (see \citealt{Muterspaugh2006}).

\begin{figure}[b!]
   \centering
   \scalebox{0.72}{
   \includegraphics{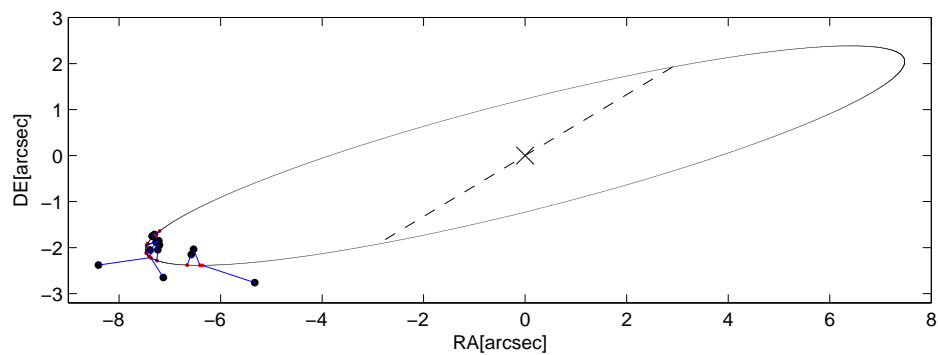}}
   \caption{Relative orbit of V1647~Sgr on the plane of the sky.}
   \label{V1647Sgr}
\end{figure}

\subsection{HD 163708} 
HD 163708 (V1647~Sgr, HIP~88069) is an Algol-type EB with the
spectrum classified as A3III and the relative brightness $V =
6.8$~mag. A few dozens of times of minima are available, showing
very slow apsidal motion (in order of centuries). The astrometric
measurements were obtained 15 times during 170 years and covering
about 14 degrees in position angle, see Fig.\ref{V1647Sgr}. The
orbit was computed first time and was not published yet, but the
result is not very convincing due to poor coverage of the orbit by
data points. The period is about 1200~yr and semiamplitude about
7.7\arcs.

\subsection{HD 174932} 
HD 174932 (COU~510, BD+24~3555, SAO~86519) is a member of visual
binary COU~510. The system also does not belong to this list,
because the star was incorrectly classified as an eclipsing binary
by \cite{Couteau1972JZHer} and designated as JZ~Her. HD~174932
itself is not a variable star 
it was mixed up with the close eccentric eclipsing variable HS~Her. The wrong comment in the WDS notes will be
soon corrected.

\subsection{HD 178125} 
HD~ 178125 (18~Aql, Y~Aql, HEI~568AB, HR~7248) is a spectroscopic variable,
spectral type B8III, brightness about 5.07~mag in \emph{V} filter. The
astrometric orbit of Y~Aql leads to the period of about 58~yr.
In fact, the star does not belong to this survey, because it is probably not eclipsing, but rather an
ellipsoidal variable (recent observations and also data from Hipparcos indicate this possibility).

\subsection{HD 184242}
HD 184242 (V2083~Cyg, A~713AB, HIP~96011) is an Algol-type EB,
spectral type A3, apparent brightness $V = 6.88$~mag and orbital
period about 1.9 days. There were no times of minima found in
published literature. The astrometry covers
about 70$^\circ$ during the last century. 
The orbit was computed by \cite{Seymour2002V2083Cyg}, resulting in
period about 372~yr and angular semimajor axis about 498~mas.

\subsection{HD 195434}
HD 195434 (MR~Del, AG~257AB, HIP~101236) is an Algol-type EB, spectral type classified as K0 and $V =
11.01$~mag. There were found a few times of minima, covering 2451700-2452100 HJD and astrometry from 1902 to
2001, with the change in position angle of about 15$^\circ$. This leads to an orbit with period of about
6500~yr.

\subsection{HD 201427}
HD 201427 (BR~Ind, HU~1626AB, HIP~104604) is an Algol-type EB,
spectral type F8V and with its apparent brightness of about
7.1~mag in \emph{V} filter. Astrometry was obtained since 1914
till 2001, when the position angle has changed from 208 down to
124 degrees. These data lead to the orbital parameters $p_3 =
167$~yr and $a = 894$~mas (according to
\citealt{Seymour2002V2083Cyg}). No times of minima were found in
literature.

\subsection{HD 217675}
HD 217675 ($o$~And, 1~And, BLA~12Aa+WRH~37AB, HR~8762) is a pulsating
Be star, as well as a shell star. 
Its apparent brightness is $V = 3.63$~mag and spectrum was
classified as B6IIIpe. The system is more complicated, consisting
of at least 4 components (see e.g. \citealt{Pavlovski1997}), while
the visual triple consists of Aa-B components. Both orbits (the
longer one with period 68.6~yr, according to
\cite{Hartkopf1996oAnd} and the shorter one with period 8.9~yr,
according to \citealt{Olevic1999oAnd}) were observed and derived.
Some authors (see e.g. \citealt{Schmidt1959}) published the light
curves with the possible eclipsing behavior of the star, but
nowadays it is rather improbable for the star to be an eclipsing
binary. The photometric variability is probably due to the
variability of the shell around the star.

The survey of 37 systems with eclipsing components in visual
binaries is not a complete one. On the other hand it could be
taken as a representative sample of the most interesting ones,
because these systems have the largest data sets in astrometry and
some of them have also the times-of-minima observations.

Scanning the WDS catalogue and trying to find the eclipsing
binaries in this sample there were found a lot of \emph{variable
stars} (according to Simbad catalogue). From this sample of
variable stars there could be a significant number of eclipsing
binaries, but only future photometric observations would reveal
the nature of this variability. Also a few systems mentioned above
are classified in Simbad as \emph{variable stars} or
\emph{ellipsoidal variables}, although they are EBs. Some of the
binaries were also find to be wrongly identified.

\chapter{Discussion and conclusions}

The method of period analysis of eclipsing binaries and its modifications were presented. The method of $O-C$
diagram analysis is not new, but new aspects were also included into the code. The possibility that the third
body resulting from the LITE analysis is also detectable via astrometry was discussed. With this assumption the
modified algorithm of simultaneous solution of LITE and astrometry was presented.

The theoretical explanation of the effect and the method used is
presented in chapter \ref{Ch1}. It deals with the relative
astrometry only and also the parallax of the system was assumed as
a priori known (mainly from Hipparcos satellite). The distance is
needed for the transformation between the angular and absolute
semimajor axis in both methods. Also the principal limitations for
both methods are presented in chapter \ref{Ch1}. These limitations
have to be considered, especially when one has only very poor data
in one of the methods.

\underline{LITE systems:} In chapter \ref{OnlyLITE} there was presented the application of the LITE analysis on
eleven particular systems. These systems have never been studied for the presence of a third component and LITE
hypothesis is able to describe their long-term minimum times behavior. On the other hand there were neither
detailed spectroscopic, nor photometric analyses of these systems and the third body hypothesis presented here
cannot be proven.

Although the number of systems, where the astrometric orbit together with LITE is known, is growing steadily, in
most cases only very limited coverage of the orbit, both in astrometry and times of minima is available.
Especially due to this reason the combined analysis of these systems is still difficult. There were found only a
few appropriate candidates for such an analysis. These cases were studied and discussed in detail and the
principal limitations of the method were pointed out. On the other hand the method itself is very powerful and
efficient. It could be even modified for the estimation of the distance to the suitable kind of binaries.

During the last decade a few papers combining the approach of simultaneous solution of radial velocities,
spectral analysis, astrometry, \emph{Hipparcos} measurements or LITE were published. Besides the systems
mentioned in the introduction (44~Boo, QZ~Car, SZ~Cam, GT~Mus) there were also the analysis of V1061~Cyg
(combining the light curve analysis, radial velocity analysis, light-time effect and \emph{Hipparcos}
measurements, see \citealt{PaperV1061Cyg}), papers where radial velocity measurements and astrometry were
combined (see \cite{Muterspaugh2006} for the solution of the LITE system V819~Her, or \cite{Gudehus} for
$\mu$~Cas), the paper on HIP 50796 combining the radial-velocity measurements with the \emph{Hipparcos} abscissa
data (see \citealt{Torres2006}), or the paper on $\delta$~Lib comparing the results from the period analysis,
light-curve analysis, spectral analysis, radio emission and astrometry, respectively; see
\cite{Budding2005-DelLib}.

Such a combined analysis is very important, and the individual
methods could be tested. Their independent results have to be in
agreement with each other. The method presented here is also the
combination of the two independent methods into the one joint
solution. It was never been done before in this way. Various
modifications were presented, but most similar was the analysis of
the LITE together with the Hipparcos observations, and the
absolute astrometry by \cite{Ribas2002}.

The code itself is presented in section \ref{program}. It could be
downloaded from the web sites and it is ready to be used. Short
description of the code is presented and also the brief manual is
available. Only slight modifications of the algorithm are
necessary before the first run of the code. The numerics and the
computing time required for the code strongly depends on the
initial parameters and the input data (their quality and the size
of the data set), but it could be slightly improved, see section
\ref{numerics}.

A few eclipsing binaries were studied in this thesis. Detailed
analysis was performed for QS~Aql, VW~Cep, $\zeta$~Phe, V505~Sgr,
HT~Vir and V2388~Oph. These systems have relatively best coverage
both in astrometry and LITE variation. This is the crucial part,
as one can see in the case of VW~Cep, which is the most suitable
system for the simultaneous analysis.

If precise measurements and good coverage of at least one period
of the distant body in both methods are available, the presented
method is very powerful and the parameters of the distant-body
orbit could be derived very precisely. Even the distance of the
system could be computed with high confidence level. The limiting
factor is mainly the coverage of the orbit. In most of the cases
the orbit was not covered sufficiently with data. Also almost all
the systems included in the catalogue in chapter \ref{Catal} have
only poor coverage of the orbit by data in both methods.

\underline{VW Cep:} The case where both methods have relatively best coverage of the orbit is VW~Cep. In this
case the resultant parameters of the third body satisfies the limit for the luminosity, and also the systemic
velocity variations coincide with our hypothesis. New results are comparable with the previous ones. An
additional fourth body was introduced to describe the long-term variation in times of minima, as well as in
radial velocities. The system is probably more complicated than was assumed (chromospheric activity cycles,
stellar spots and flares), and it was decided to explain only the most pronounced effects in the $O-C$ diagram.
Using the combined approach it is possible to derive the parallax to VW~Cep more precisely than in any other
previous papers, resulting in $\pi=(35.85~\pm~0.37)$~mas. The two different approaches (LITE$_3$ + LITE$_4$ and
LITE$_3$ + mass transfer) were used and their results compared. Both approaches lead to approximately the same
results both in astrometry and times-of-minima analysis. The simultaneous analysis is able to describe the
system in its complexity and one has to disagree with the result by \citet{Prib2000}, that the astrometric orbit
could not be identified with the LITE$_3$ variation from the $O-C$ diagram. As one can see, our new results are
in agreement with each other without any problems. On the other hand only further observations of this system
will decide which approach (Solution I. or II.) is the right one. Two new times of minima were observed at
Ond\v{r}ejov Observatory.

\underline{QS Aql:} In the case of QS~Aql the parameters of the distant-body orbit were mainly derived from the
LITE analysis, because the coverage of the astrometric orbit is very poor and the old data are not very
reliable. Due to this difficulty, the inclination of the orbit could not be derived precisely. The computed
value of the inclination is quite low and the error quite high. Low inclination dictates high mass of the third
body (but with large errors). The new derived mass of the third body is in contradiction with the previous
photometric analysis, but one can get a consistent result within the error of this value.

\underline{$\zeta$~Phe:} The system $\zeta$~Phe displays an apsidal motion together with the LITE and this
explanation fits the $O-C$ residuals quite well. This is the first time when the apsidal motion together with
the LITE hypothesis were applied to this system. The astrometric analysis of $\zeta$~Phe is complicated due to
the fact that the period of the third body orbit is circa 3 times longer than the interval covered by the data.
The time span of the minima measurements is even worse, only about one fourth of the orbit is covered. On the
other hand the powerful combined analysis was able to estimate all of the parameters of the third-body orbit
precisely. This approach lead to the period of the third body about 220~years and the parameters of such a body,
its predicted mass and the spectral type is in an excellent agreement with the previous photometric analysis.

\underline{V505 Sgr:} V505~Sgr is the system, where both the astrometry and also recent times of minima
observations deviate from the predicted trend. Despite the fact the third body was detected more than 20~years
ago, it could be even observable in the spectrum of the system, the complex figure of the system is still
missing. The new result from the combined approach is in contradiction with the previous results from photometry
and also spectroscopy. It indicates that the third body observable in spectra and light curve is different from
the fourth body observable astrometrically. Only further detailed analysis would prove this hypothesis.

\underline{HT Vir:} The eclipsing system HT~Vir is the case where
the new value of mass of the distant body is about 2 times larger
than one would expect. The distant component in the system is also
a double and from the spectroscopy one is able to derive an upper
limit for its mass. Regrettably, our new result is in
contradiction with such a mass. This could be due to only a few
times of minima observed in the linear part of the $O-C$ diagram,
new minima are needed in the next decades. Four new times of
minimum light were observed. On the other hand the astrometric
orbit is well-defined and almost whole orbit is covered.

\underline{V2388 Oph:} The last system is V2388~Oph, where two different approaches were used. The astrometric
variation is rapid and since its discovery the third component has revolved a few times around the primary. On
the other hand, the times of minima were obtained only rarely during the last decade. Due to this reason, the
rapid change in order of 9 years is hardly detectable from the $O-C$ diagram analysis and one could speculate
about the inclination of the orbit. If the orbit's inclination is close to 180$^\circ$, there could be no LITE
evident in the $O-C$ diagram, which was presented as another explanation and the mass transfer was suggested as
an alternative explanation for the $O-C$ diagram. Only further times of minima would prove or refuse this
hypothesis. Another possible explanation is that the system is quadruple and the third body observable
interferometrically is not the one which causes LITE.

The final result is that the method itself is potentially very
powerful but it is also very sensitive to the quality of the input
data, especially if the method is used for determining the
distance of these binaries. It can only be applied successfully in
those cases where the astrometric orbit and the LITE in the $O-C$
diagram are well defined by existing observations and lead to the
approximately same parameters of the distant-body orbit. This is
necessary condition, as one can see for example from the case
V505~Sgr.

\underline{The catalogue:} The catalogue of other suggested systems for the prospective simultaneous analysis
with the introduced algorithm was presented in the chapter \ref{Catal}. The main purpose of the catalogue was to
critically consider the potential objects for such a combined approach and from the eclipsing binaries in the
spatially resolvable systems identify those, which are suitable for the introduced method.

During the inspection of such systems there were found a few
binaries which were often presented as eclipsing binaries, but
which are in fact not. These are for example 18~Aql (=Y~Aql) which
is an ellipsoidal variable, $o$~And which is photometrically
variable, but it is not due to the eclipses, or V640~Cas which is
probably also not an eclipsing variable. Also one
misidentification of the EB was presented (HD~174932).

On the other hand there were found a few systems which are the most suitable ones for the method presented here.
Such systems are for example V348~And, V592~Per, V635~Mon, DN~UMa, V831~Cen, ET~Boo, or i~Boo.

Additional material is also available via the web
pages\footnote{\href{http://sirrah.troja.mff.cuni.cz/~zasche/}{http://sirrah.troja.mff.cuni.cz/ $\sim$
zasche/}}. The code for computing the combined analysis could be downloaded together with the brief manual and
instructions for the user. On the same web pages there are also the complete data files, which were used as the
input files for the analysis.

\bibliographystyle{aa}
\bibliography{Dezert}

\end{document}